\documentclass[12pt]{article}

\usepackage[a4paper, margin=2cm]{geometry}

\usepackage[numbers]{natbib}
\usepackage{bbm}

\usepackage{tcolorbox}
\usepackage{algorithm}
\usepackage{times}
\usepackage{soul}
\usepackage{url}
\usepackage[hidelinks]{hyperref}
\usepackage[utf8]{inputenc}
\usepackage[small]{caption}
\usepackage{graphicx}
\usepackage{amsmath}
\usepackage{amsthm}
\usepackage{amsfonts}
\usepackage{dsfont}
\usepackage{booktabs}
\usepackage{algorithm}
\usepackage{algorithmic}
\usepackage{soul}
\usepackage{cancel}

\usepackage{tcolorbox}
\usepackage{algorithm}
\usepackage{times}
\usepackage{soul}
\usepackage{url}
\usepackage{booktabs}
\usepackage{algorithm}
\usepackage{algorithmic}
\usepackage{soul}
\usepackage{cancel}

\usepackage{enumitem}
\usepackage{subcaption}

\usepackage{todonotes}

\usepackage{xspace}

\usepackage{xcolor}

\newcommand{\N}{\mathbb{N}}
\newcommand{\code}[1]{\texttt{#1}\xspace}

\newcommand{\1}{\mathbbm{1}}

\newcommand{\lobgan}{\texttt{LOBGAN}\xspace}
\newcommand{\imb}{\code{BookImbalance}}
\newcommand{\imbI}{$\imb_1$\xspace}
\newcommand{\imbV}{$\imb_5$\xspace}
\newcommand{\vol}{\code{TotalVolume}}
\newcommand{\volI}{$\vol_1$\xspace}
\newcommand{\volV}{$\vol_5$\xspace}
\newcommand{\tradeimb}{\code{TradeImbalance}}

\newcommand{\fscale}{0.6}
\newcommand{\fscales}{0.5}
\newcommand{\fscaless}{0.43}

\begin{document}

\title{Conditional Generators for Limit Order Book Environments:
Explainability, Challenges, and Robustness}

\author{
Andrea Coletta\\
J.P.Morgan AI Research\\
andrea.coletta@jpmchase.com
\and
Joseph Jerome\\
University of Liverpool\\
j.jerome@liverpool.ac.uk
\and
Rahul Savani\\
Alan Turing Institute/\\
University of Liverpool\\
rahul.savani@liverpool.ac.uk
\and
Svitlana Vyetrenko\\
J.P.Morgan AI Research\\
svitlana.s.vyetrenko@jpmchase.com
}

\maketitle

\begin{abstract}\noindent
Limit order books are a fundamental and widespread market mechanism.
This paper investigates the use of conditional generative models for order book simulation.
For developing a trading agent, this approach has drawn recent attention as an alternative 
to traditional backtesting due to its ability to react to the presence of the trading agent.
Using a state-of-the-art CGAN (from Coletta et al.~(2022)), we 
explore its dependence upon input features, which highlights both
strengths and weaknesses.
To do this, we use ``adversarial attacks'' on the model's
features and its mechanism.
We then show how these insights can be used to improve the CGAN, both in terms
of its realism and robustness. 
We finish by laying out a roadmap for future work. 
\end{abstract}

\section{Introduction}\label{sec:intro}
This paper deals with the construction of robust and realistic limit order book (LOB)
environments for the training and evaluation of trading strategies.
LOBs~\citep{gould2013limit} are a fundamental market mechanism,
which are used across a significant proportion of financial markets, including
all major stock and derivatives exchanges. 
The benefits of having robust and realistic simulators for these markets are
numerous. 
For example, they would allow the study of markets under different assumptions, and
the investigation of AI techniques for training trading strategies.
In a LOB market, matched orders result in trades and unmatched
orders are stored in the two parts of the LOB, a collection of buy
orders called \emph{bids} (the bid book), and a collection of sell orders called
\emph{asks} (the ask book).
Typically, each side of the LOB will contains hundreds of
individual orders, and a real market would be updated at micro-second time
resolution, driven by a wide range of market participants and facilitated
by ``high-frequency'' market makers~\citep{savani2012high}.

The development of AI-based automated trading strategies for LOB
markets has been a growth area in recent years, both within academia and
industry, spurred on in part by developments in deep learning and reinforcement
learning. 
Two typical LOB trading problems that have been investigated are
\emph{market making}, where the goal is to provide liquidity to the market by
being continually willing to buy and sell an
asset~(see, e.g.,~\citet{spooner2018market,jerome2022market,GasperovK21}),
and \emph{optimal execution} where the goal is to execute a pre-determined
amount in the most cost efficient manner possible (see,
e.g.~\citet{Nevmyvaka2006,NLJ21}).

The ultimate test of the efficacy of a trading strategy -- be it a hand-crafted one
or one developed using AI -- is to trade it live in a real market~\citep{Pardo08}.
However, this testing approach is rarely feasible for academic researchers, and
even for industry practitioners it is potentially very costly to evaluate
strategies this way.
Thus effective simulation-based evaluation frameworks are highly desirable both
to drive academic and commercial research. 
In Section~\ref{sec:furtherben}, we discuss in detail desiderata that a
simulation-based strategy evaluation framework should have.

By far the most prevalent choice is market replay, but simulators based on this
approach have no reactivity~\citep{BMLHB19}. 
Thus, one implicitly makes the assumption that the financial market does not
react to the presence of the trading agent which is being tested. Needless to
say, this is an unrealistic assumption. 
Conditional generative models provide a solution to the issue of reactivity, and
for this reason -- and others that we discuss in Section~\ref{sec:benefits} --
they have recently attracted a lot of attention in the AI community (see, e.g.,
\citet{LiWLSW20,KuoCLH21,ColettaPCMBMVB21,ColettaMVB22,shi2023neural}). 
With a deep conditional generative model,
a deep neural network is used to generate the next state of 
the world, conditioned upon the current state of the world. 
The two most popular such models are Conditional Generative Adversarial Networks
(CGANs)~\citep{mirza2014conditional} and Conditional Variational Autoencoders
(CVAEs)~\citep{SohnLY15}. 
All of the main experiments in this paper use a specific class of CGANs
introduced in the paper by \citeauthor{ColettaMVB22}~[\citeyear{ColettaMVB22}].
We refer to this CGAN as \lobgan.

Training of the \lobgan model works as follows. 
The \textit{generator} neural network takes as its conditioning input a set of features that describe
both the recent state of the order book and recent trade action, as well as --
as in a non-conditional GAN -- a random noise vector to add variability. 
It then outputs a vector that represents the predicted next order to arrive in the
book. 
Then a \textit{discriminator} neural network takes the synthetic order vector as well as
the true order vector (from the historical training data) and aims to
distinguish (conditional upon the input to the generator) which is real. 
For more details on training conditional GANs see, e.g.,
\citet{goodfellow2014generative,mirza2014conditional,gulrajani2017improved}.

\subsection{Our contributions}

In this paper, we motivate and study the problem of developing realistic and
robust conditional generative models for simulating LOBs.
Throughout the paper we use \lobgan along with high-fidelity  historical order data from LOBSTER~\citep{huang2011lobster}. 
While our exploration uses this specific model family, our novel contributions to
the methodology of evaluating and interpreting generative order flow models applies
more generally.
We next summarise our contributions:

\begin{itemize}

\item We go beyond prior work in demonstrating the benefits of \lobgan 
by doing a \emph{price impact analysis for market and 
limit orders separately}. 
This approach newly reveals a strength, and separately a weakness of 
\lobgan.

\item
We provide a \emph{new technique for analysing the conditioning of generative LOB models}. This technique helps both with explainability and the design of better models.

\item
We develop and test trading strategies to  \emph{study and stress the robustness} of \lobgan. These adversarial strategies include both hand-crafted and reinforcement-learning-based strategies.

\item Using our insights, we develop \emph{new \lobgan models}, which we demonstrate
are better in terms of \emph{both} realism and robustness.

\item We provide \emph{recommendations} for how to use \lobgan (or similar conditional generative models) \emph{in practice}.

\item We provide a detailed \emph{roadmap for future work}. 

\end{itemize}

\paragraph{Remark: our adversarial attacks are not \emph{``market manipulation''}.}
Our goal in this paper is to understand and develop the methodology 
for designing realistic and robust conditional generators.
As such, we design adversarial attacks to exploit and show weaknesses of such models by manipulating the features and mechanism of the models.
The term ``market manipulation'' has a specific meaning and refers to behaviour
such as spoofing and quote stuffing, whereby orders are placed, with no intention
of them being executed, and with the goal of deceiving and manipulating other 
market participants. 
Such market manipulation is typically illegal and markets are carefully monitored
for these  types of behaviours by exchanges.
In the interests of clarity, none of the strategies we present in 
this paper would be considered as market manipulation, but we rather focus on adversarial attacks on the deep neural network model~\cite{szegedy2013intriguing,biggio2013evasion,nguyen2015deep}.  
Furthermore, whenever we place orders as part of an adversarial attack, there is always a reasonable chance that  order will be executed. And, we never cancel an order purely to avoid that order being executed, but instead because that order is in some sense ``stale''.

\subsection{Outline of the paper}
In Section~\ref{sec:preliminaries}, we provide a discussion of key concepts needed 
in this paper.
We start by explaining how a LOB market works; we then discuss
how to measure the realism of a simulator of such a market, where we stress a 
distinction between two types of realism: realism in isolation and interactive
realism; then we discuss a particular type of interactive realism namely the 
price impact of orders.

In Section~\ref{sec:benefits}, we demonstrate the benefit of using
a generative CGAN-based order flow model as an alternative to market replay
(which is also known as ``backtesting'').
A key benefit of generative models is that they can react to incoming order
flow, and therefore generate realistic price impact.
In the literature, the price impact of market orders has been heavily studied;
the price impact of limit orders has been studied much less so. 
In this section we further analyse the price impact of {\lobgan},
separately for market orders and limit orders.
While we find that \lobgan can generate realistic price impact paths, we also
find that the paths show a greater impact of limit orders over market orders,
which is unexpected, and was not studied in the earlier papers on \lobgan.
We investigate the possible reasons in the paper.

In Section~\ref{sec:realism}, we review realism metrics for trajectories
generated sequentially by CGAN-based order flow models.
These metrics have been the primary way that models have been selected and
evaluated to date, where a human in the loop chooses among a set of candidate
CGAN models one that does well across a range of these realism metrics
~\citep{ColettaMVB22,LiWLSW20}). 

In Section~\ref{sec:lobgan_conditioning}, we \emph{introduce a new analysis technique}
for CGAN-based order flow models that explores the dependence of the model's
output dynamics in terms of the individual input features that are used for
conditioning. 
This technique both helps to explain the causal mechanisms in the model, and
improves the robustness of model selection by identifying redundant features
that can be removed, which reduces overfitting and further improves
explainability.

The realism metrics in Section~\ref{sec:realism} focus on realism in
isolation, whereas Section~\ref{sec:benefits} provides a very simple example of
interactive realism (which is similar in spirit to previous work~\citep{ColettaMVB22,shi2023neural}).
In Section~\ref{sec:attack}, we explore interactive realism in depth -- for the first time.
We analyse \lobgan's outputs when it interacts with a range of trading agents.
Some of these agents emulate standard market participants such as market makers,
where an example of unrealistic behaviour of the CGAN that we discover is a market
maker can make consistently large profits, albeit with an extremely unsophisticated
strategy. 
A second type of trading agent that we introduce explicitly tests how
exploitable the causal mechanisms of the model are by placing orders to
``exploit'' features for the purpose of making unrealistic profits.
This is problematic because if one were to use AI methods such as reinforcement
learning for developing trading strategies, such approaches would likely try and
exploit the features of the model to make profits, whilst being potentially unprofitable in real markets.

In Section~\ref{sec:improved_models}, we develop a variety of
solutions to the ``vulnerabilities'' of the original \lobgan model that we identify.
This leads to new variants of \lobgan that are better both in terms of
their realism and robustness.
In Section~\ref{sec:recommendations} we provide recommendations for 
how best to use \lobgan or similar models in practice, based on the 
insights from this paper.

In Section~\ref{sec:relatedw}, we provide an extensive discussion of related work.
In Section~\ref{sec:future_work}, we finish with a discussion of future work and
next steps in this promising research direction, where many challenges remain in
relation to the CGAN design and model selection, 
We lay out a programme for future research, including, for example, a proposed
generative model training loop that incorporates the use of trading strategies
to ensure that the CGAN is not unrealistically exploitable.

\section{Preliminaries}
\label{sec:preliminaries}

\subsection{Generative Adversarial Networks (GANs)}
In recent years, Generative Adversarial Networks (GANs) have been successfully
applied to a wide range of applications, ranging from images to time-series
data. 
GANs generate samples with high quality and diversity by \textit{implicitly}
learning to generate data without the need for an explicit density
function~\citep{goodfellow2014generative}.

In particular, GANs employ two neural networks and an adversarial training
procedure: a generator $G$ and a discriminator $D$ are trained simultaneously to
play the following min-max game:
\begin{equation*}
   \min_G \max_D \, \underset{\textbf{x} \sim p_{\texttt{data}}((\textbf{x}))}{\mathbb{E}}[\log(D((\textbf{x})))] + \underset{\textbf{z} \sim p_{z}(\textbf{z})}{\mathbb{E}}[\log(1 - D(G(\textbf{z})))]\ . 
\label{eq:gan_game}
\end{equation*}
Given a vector $\textbf{z}$ from a prior distribution $p_{z}$ (i.e., typically a
multi-dimensional Gaussian distribution), the generator network $G(z)$ creates a
new realistic sample $x$ that the discriminator examines to estimate whether $x$
is real (i.e., $x$ belongs to the ground truth training set) or fake (i.e., $x$
has been generated by $G$).
Both networks aim at maximizing their own utility function. As the training
advances, the discriminator network $D$ learns to reject unrealistic synthetic
samples generated by $G$, while the generator $G$ learns to generate more
realistic data to fool the discriminator. 
Finally, in deployment, the generator $G$ is used to generate new (hopefully realistic) data samples.  

\paragraph{Conditional Generative Adversarial Networks (CGANs)}
Conditional GANs~\citep{mirza2014conditional} condition the generative process
by using some extra information, which is captured in a feature vector $\textbf{y}$. 
This extra information can represent, for example, the class of the images to
generate~\citep{gulrajani2017improved}, or the market state to generate
appropriate orders~\citep{ColettaMVB22}. 

CGANs feed both the generator and discriminator networks with this extra
information $\textbf{y}$, resulting into the following game:
\begin{equation*}
   \min_G \max_D \, \underset{\textbf{x} \sim p_{\texttt{data}}(\textbf{x})}{\mathbb{E}}[\log(D(\textbf{x}|\textbf{y}))] + \underset{\textbf{z} \sim p_{z}(\textbf{z})}{\mathbb{E}}[\log(1 - D(G(\textbf{z}|\textbf{y})))].
\label{eq:cond_gan_game}
\end{equation*}

\lobgan uses a particular CGAN architecture namely a Wasserstein CGAN,
combined with a gradient penalty~\citep{gulrajani2017improved}. 
The Wasserstein (also called Earth-Mover's) distance has been found to to
improve the stability of the CGAN training process (compared to the
Jensen–Shannon divergence, which was used in the original GAN), as it is
continuous everywhere, and differentiable almost everywhere, under the
assumption that the discriminator lies within the space of 1-Lipschitz
functions.
The Lipschitz constraint is enforced on the discriminator via the gradient penalty,
which penalises the norm of the gradient with respect to its input.
Importantly, Wasserstein CGANs have demonstrated strong results in modelling
discrete data~\citep{gulrajani2017improved}, which is the form of the order flow
features like order type, volume, and depth that are used by conditional models 
for limit order books including \lobgan.

For the Wasserstein CGAN with a gradient penalty the min-max game is defined as follows:
\begin{equation*}    
\min_D \, \underset{\textbf{z} \sim p_z({\textbf{z}})}{\mathbb{E}}[D(G(\textbf{z}|\textbf{y}))] - 
\underset{\textbf{x} \sim p_{data}(\textbf{x})}{\mathbb{E}}[D(\textbf{x}|\textbf{y})] 
+ \underset{\textbf{z} \sim p_z({\textbf{z}})}{\lambda \cdot \mathbb{E}}[(\Vert \nabla D(G(\textbf{z}|\textbf{y})) \Vert _2 -1 )^2]
\end{equation*}
\begin{equation*}    
\max_G \, \underset{\textbf{z} \sim p_z({\textbf{z}})}{\mathbb{E}}[D(G(\textbf{z}|\textbf{y}))]\ ,
\end{equation*}
where, for \lobgan, the weight $\lambda$ for the gradient penalty is set to 10
as in \citep{gulrajani2017improved}.

\lobgan's generator synthesises subsequent order flow according the
recent state of the market, i.e., the features used for conditioning summarise 
the current market state and recent market action.
The features used by \lobgan for conditioning are explained in detail in
Section~\ref{sec:feature_definitions}.

\begin{figure*}
\centering
\includegraphics[scale=0.63]{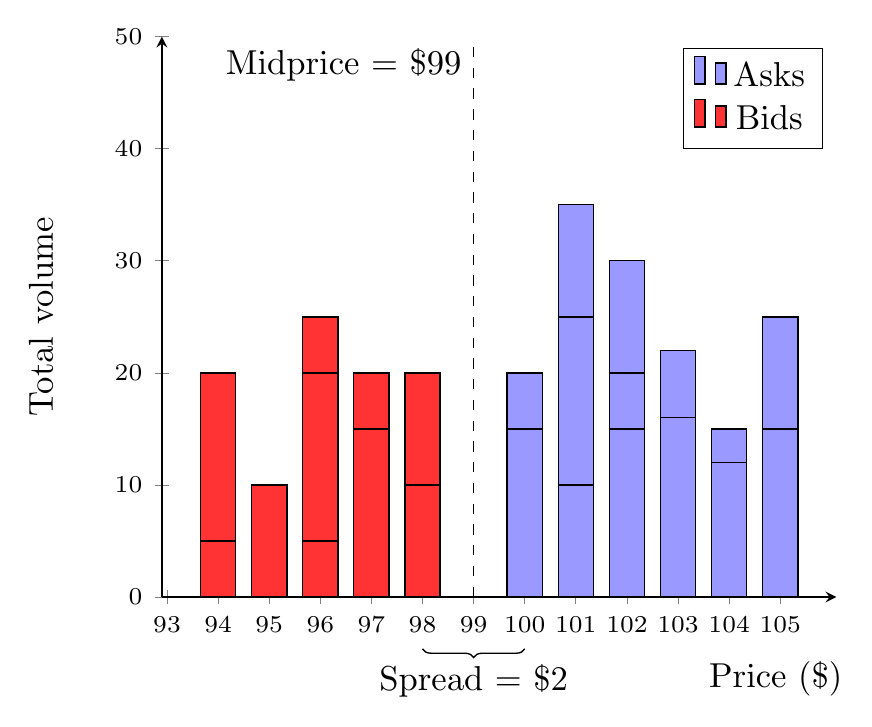}
\includegraphics[scale=0.63]{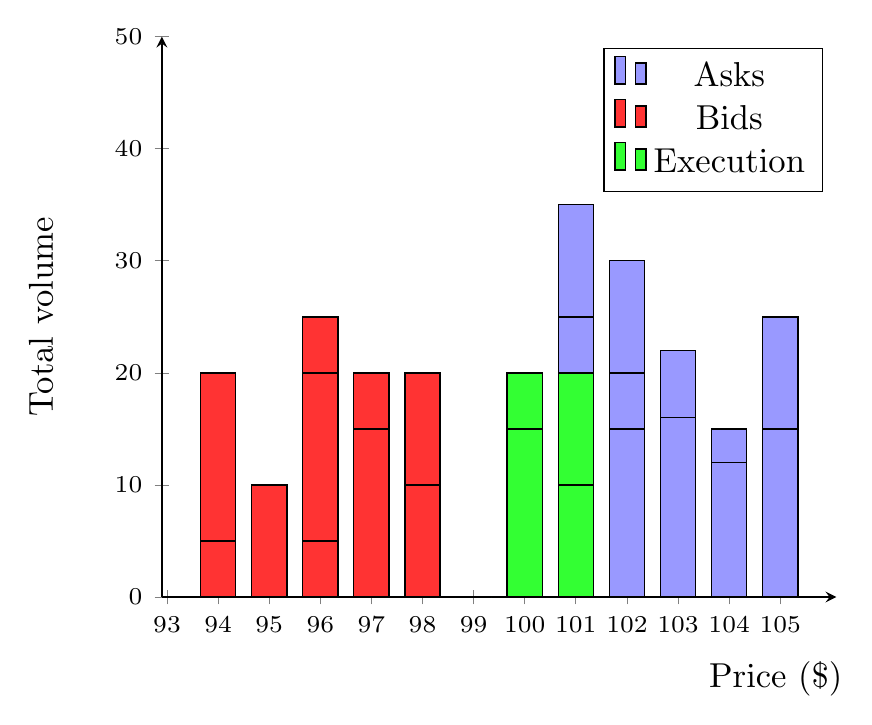}
\includegraphics[scale=0.63]{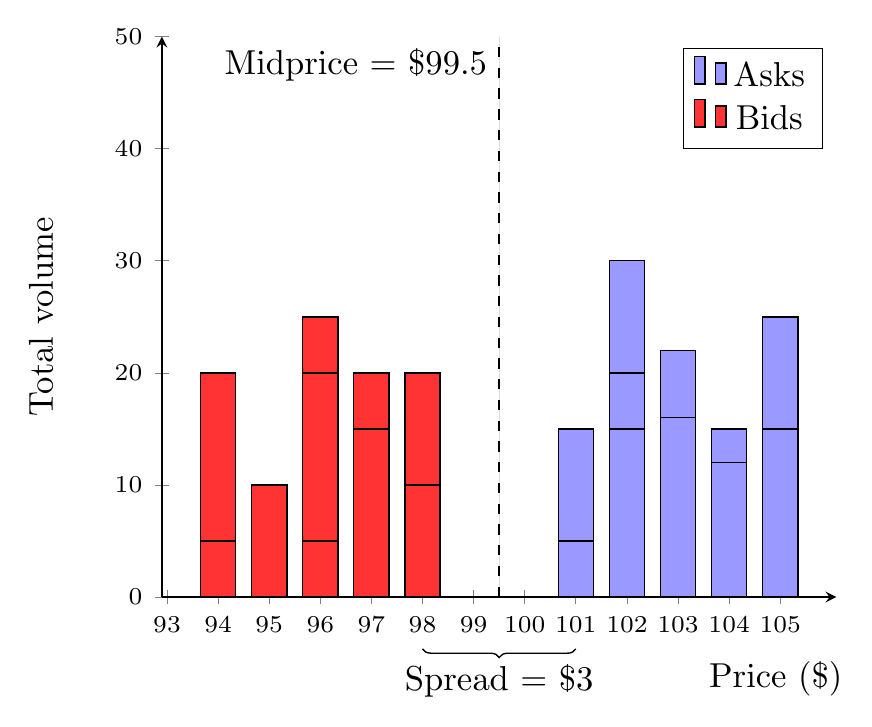}
\caption{In the leftmost subfigure, we see the bid limit orders in blue and 
ask limit order in red. The best bid (ask) price is \$98 (\$100). At a given
price level, orders of different market participants are delimited by vertical
lines, with orders nearer the front of the queue for that price level being nearer
the bottom. In the middle subfigure a buy market order with volume 40 has arrived.
It totally takes out the old best ask and a further volume of 20 from the next best ask
(price \$101). After this execution the spread has widened from \$2 (i.e., 100-98) to \$3 (i.e., 101-98).
}
\label{fig:lobex}
\end{figure*}

\subsection{An illustrative example of limit order book dynamics}
\label{sec:lobexample}

The example given in Figure~\ref{fig:lobex} illustrates both the execution
mechanism of a limit order book and highlights the lack of reactivity in market
replay that we demonstrate further in Section~\ref{sec:price_impact} below. 
The left-hand panel of the figure shows a collection 
of unmatched bid and ask limit orders -- forming the \textit{order book}, with
the bid book (at lower price levels) on the left, and the ask book (at higher price
levels) on the right.
In the middle panel, a buy market order\footnote{A market order can be thought
of as a limit order with an infinite limit price if it is a buy and a limit
price of $0$ if it is a sell.} arrives and is executed against the ask limit
orders. 
It does so according to price-time priority -- first matching against the lowest
price and then matching orders at the best price according to time priority
(i.e., transacting first against limit orders that arrived earlier). 
The right-hand panel shows the state of the order book after the execution of
this market order; the spread has widened and the mid-price has gone up.

In a real market, the key question would be what happens next? For a market
replay environment, the answer is simply \textit{whatever happened in the
historical data}. 
This is entirely the correct answer when simply replaying historical order data,
but ultimately the purpose of a backtest is to evaluate an exogenously defined
trading agent (or perhaps to explicitly train a learning agent). 
Making an assumption that a trading agent can place orders
without affecting subsequent order flow is unrealistic:
The historical orders that are placed after the ``external'' order 
of the trading agent, are being added to an order book that does not look like
the order book in which they were placed historically;  if the external order had 
actually been placed historically then the market would have been
made aware of the presence of the exogenous agent and its orders and would have
reacted accordingly. 
In particular, the traders responsible for subsequent historical order flow 
may have no longer chosen to place the same orders that were historically observed. 
Furthermore, the more that an external agent interacts with the order book, the
more the observed order book differs from the historical order book and the more
unrealistic the assumptions underlying market replay are. 

On the other hand, in a CGAN-based environment, subsequent order flow depends on
the current state of the world, so that when an exogenous agent interacts with
the book and changes it, the CGAN generates orders that are realistic
\textit{conditioned on} the updated order book as well as any trades that ocurred. 
This \emph{reactivity} is a primary advantage of a CGAN-based approach, but it
is not the only one. In Section~\ref{sec:benefits} we discuss some of the many
other benefits.

\subsection{Realism in isolation and interactive realism}
\label{sec:realismtypes}

The example and discussion in Section~\ref{sec:lobexample} explain how 
reactivity in response to incoming orders is a strength of the CGAN approach.
However, so far evaluation of CGAN models has focused primarily on the realism
of the CGAN outputs when the only inputs come from model itself. 
In this paper, we extend the focus to include realism of outputs when at least
one additional trading agent is also in the system.
It is thus useful to introduce terminology to distinguish between these two
types of realism:

\paragraph{Realism in isolation.} This type of realism is where no agents are
added to the simulation, and the desideratum is that the simulation outputs
should look realistic in terms of their statistical properties. 
This is discussed in detail for limit order books in~\citet{VyetrenkoBPMDVB20}. 
In this case of realism in isolation, market replay will -- by definition -- look completely realistic. 
This type of realism has been the primary driver of model selection and evaluation
for existing CGAN LOB models such as \lobgan~\citep{ColettaMVB22} and Stock-GAN~\citep{LiWLSW20}. 

\paragraph{Interactive realism (i.e., realistic reactivity).} 
In addition to realism in isolation, we would like a simulator to also react
realistically when an external agent places orders. 
Again, we would like that the outputs of the model should look realistic given
the behaviour of the interacting trading agent(s).
By definition, market replay cannot do this.
However, a CGAN model can, by conditioning on recent market action, which includes the
actions of external agents. 

\subsection{Price Impact}\label{sec:price_impact}

In financial markets, price impact is the effect that an agent has on the 
midprice due to its trading activity. In particular buy (sell)
orders tend to push the price of the asset up (down).
There is debate as to whether market
impact comes from trades revealing private information about the fundamental
value of the asset or whether it emerges naturally from the limit order book mechanism, but
there is agreement that it is a fundamental aspect of markets to consider when
placing large orders~\citep[Chapter 11.1]{bouchaud2018trades}.
Therefore, a realistic simulator has to show price impact phenomena to properly
capture market response to an experimental strategy. 

When one only has historical market data one is limited in the type of analysis
of price impact that one can carry out. 
In particular, one is not able to do the type of ``A/B'' testing that one can
do when one has a simulator, as we do.
In this case, with a simulator, one can study the market's price evolution in
two situations that only differ according to whether a certain set of orders was
placed or not.
Later in the paper, we do this for the CGAN models that we explore, and compare
the results with earlier findings from the literature.

In more detail, to consider price impact we define the (reaction) \textit{impact
path} \citep[(11.1),(12.2)]{bouchaud2018trades} as the average price dislocation
between the beginning and the end of a metaorder execution (a collection of
smaller orders in one direction):
\begin{eqnarray}\nonumber
    \lefteqn{\mathcal{I}^{\rm react.}_{t+l}\left({\rm exec}_t\,\middle\vert\,\mathcal{F}_t\right)} 
     \\ \nonumber
    &=& \mathbb{E}\left[ P^{\rm mid}_{t+l} \,\middle\vert\, {\rm exec},\mathcal{F}_t \right] - \mathbb{E}\left[ P^{\rm mid}_{t+l} \,\middle\vert\, {\rm no~exec},\mathcal{F}_t \right]\ .
\end{eqnarray}
In particular, we measure the price difference between a simulation with and
without the metaorder in our simulated environments, and then we compare the 
resulting impact paths against the form of these paths that have been found in
the empirical literature on price impact (Figure \ref{fig:price_impact_path}).
As mentioned above, we note that this approach cannot be implemented in market 
replay or in a real financial system as the two situations (i.e., the metaorder 
arriving or not) are mutually exclusive.

We distinguish between three types of price impact:
\textit{temporary}, \textit{permanent}, and
\textit{transient}~\citep{almgren2001optimal,HoltLM90,bouchaud09price}. 
We define \textit{temporary} price impact as the impact occurred during the
execution of the metaorder -- and we further define the \textit{peak} price
impact to be the impact when the whole order has been executed. 
After the metaorder has been executed we have: the \textit{transient} price
impact which is the component of the metaorder's price impact that decays to
zero; and the \textit{permanent} price impact that persists in the market,
formally defined as $\lim_{l\to\infty} \mathcal{I}^{\rm react.}_{t+l}\left({\rm
exec}_t\,\middle\vert\,\mathcal{F}_t\right)$.

\section{Benefits of \lobgan over backtesting}
\label{sec:benefits}

Simulated environments are increasingly used by academic researchers, trading firms, and investment
banks to evaluate and train trading strategies and study market response to order placement.
In this
section, we discuss the benefits of conditional generators over traditional
market replay simulators.

\subsection{Price impact of \lobgan}

Going beyond what one typically finds in the literature, we investigate the 
price impact of market and limit orders on \lobgan separately. 
This allows us to better evaluate and understand \lobgan.

\subsubsection{Impact path of market orders}

We first evaluate the \lobgan and market-replay environments by considering a 
Time-Weighted Average Price (TWAP) agent that uses market orders.
TWAP is a widespread benchmark execution strategy which splits a larger order into 
small parts equally over time (in this case, over a period of 5 minutes). 

\begin{figure}[h]\centering
\includegraphics[width=\fscale\linewidth]{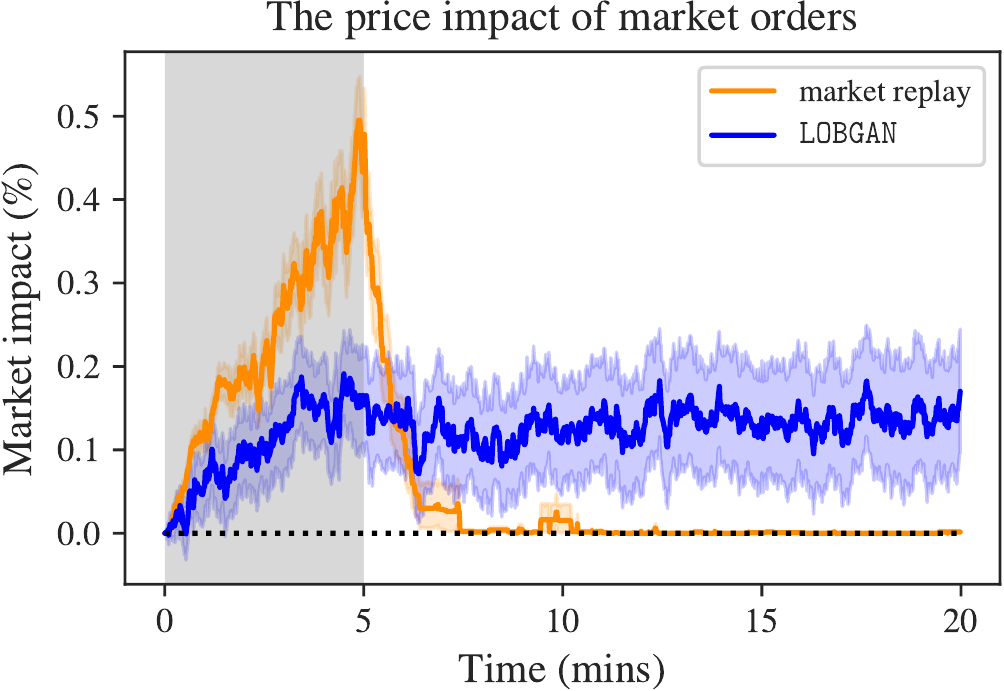}
\caption{The average price impact path of a TWAP market metaorder in the CGAN environment
and in the market replay environment. This TWAP execution consitutes an average of 70\% of the traded Percent of Volume (POV) over its 5 minute execution window (shaded in grey). Error bars represent the 5th-95th\% confidence interval. The start time of the trajectory is given by 30 equally-spaced times for two days for a total of 60 trajectories.}
\label{fig:price_impact_market}
\end{figure}

Figure~\ref{fig:price_impact_market} shows the average impact path when the
TWAP agent is included in the simulation environment. 
It may seem counterintuitive that a market-replay environment exhibits price impact at all -- 
since the subsequent order flow does not \textit{react} to the orders of the the TWAP agent. 
Nonetheless, under market replay the limit order book mechanism will lead to changes in the midprice for sufficiently 
large incoming market orders:
when an arbitrarily large buy market order comes in, this order could take out a large number of ask price levels
and create an arbitrarily large increase in the midprice; if this is an isolated order, then this increase will
just be an instantaneous spike, with the midprice reverting to its old level as soon as the next historical asks
arrive; however, if, as with our TWAP agent, there are persistent incoming market orders over time, then it is 
possible that these incoming agent market orders outweigh the incoming historical asks and so one sees temporally-extended,
albeit temporary, price impact, as we see in Figure~\ref{fig:price_impact_market}.
We see this price impact because market replay is using a limit order book mechanism, so the midprice naturally increases as price levels are taken out by the agent.
Since these price levels that are taken out are not restored as frequently with market replay as in the \lobgan 
(since \lobgan reacts to the incoming market orders),
\footnote{In a real financial market, these levels would be refilled by market-making agents. 
In \lobgan, the order flow for the market as a whole is learnt -- which includes the order flow from market makers.} which has \textit{reactive} order flow, the temporary price impact is larger. 
As soon as the TWAP agent stops interacting with the financial market, the incoming limit orders (that were placed in the historical financial market where the midprice hadn't moved upwards) quickly move the midprice back down to its historical level. 

It is worth noting that this analysis is similar to that performed in
\citet{ColettaMVB22}. However, in \citet{ColettaMVB22}, the authors look at the
price impact of a TWAP agent that, every minute: first, places a limit order on
the touch; then, if the order doesn't get filled, places a market order to hit
their target for that period.

\subsubsection{Impact path of limit orders}\label{sec:market_impact_limit}

Whilst the impact of market orders has been much studied, the impact of limit
orders has received less attention. 
An in-depth study of the
price impact of limit orders is conducted in  \citet{hautsch2012market}. 
The authors find that relatively large limit orders do have a significant market
impact, pushing the price up for large bids and down for large asks.

\begin{figure}[h]\centering
\includegraphics[width=\fscale\linewidth]{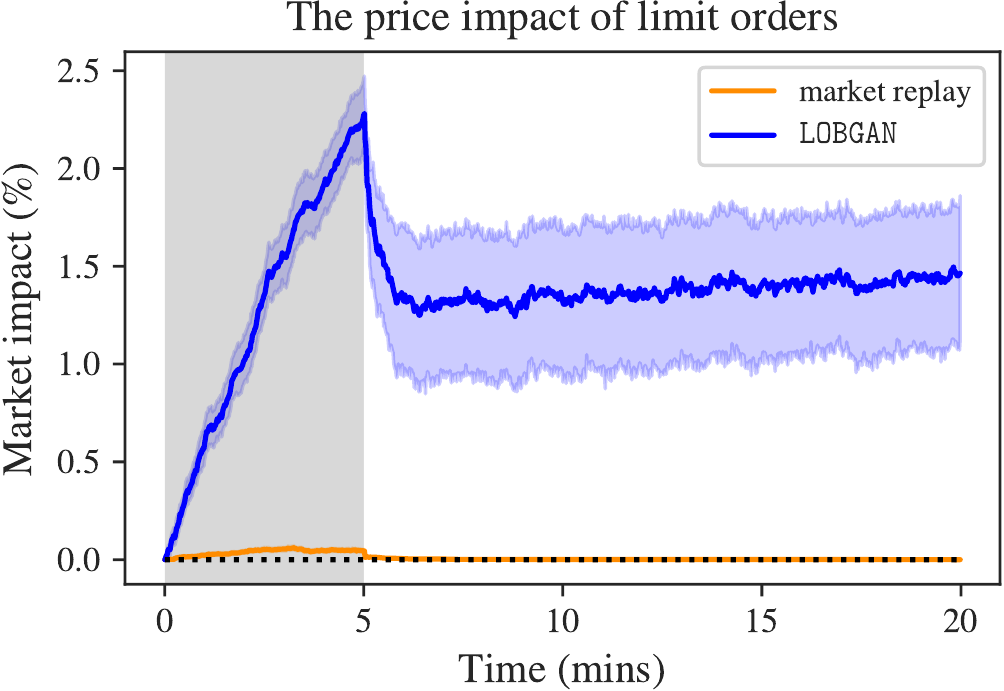}
\caption{The price impact of a an agent placing and maintaining a relatively large limit buy order (of volume 750) 
at the touch over a 5 minute window (shaded in grey). 
The executions from this order constitute approximately 60\% of the traded volume. 
Error bars represent the 5th-95th\% confidence interval. 
The start time of the trajectory is given by 30 equally-spaced times for two days for a total of 60 trajectories. 
Here, the TWAP execution period is shaded in grey.}
\label{fig:price_impact_limit}
\end{figure}

In Figure~\ref{fig:price_impact_limit}, we illustrate the price impact of
limit orders in \lobgan and market replay environments. 
The market-replay simulation produces minimal market impact, since
non-aggressive limit orders (limit orders that do not enter the spread) only affect the price by being filled instead of limit orders deeper in the book that would have been filled had they not been there. 
In particular, the future evolution of order flow is not
affected by their presence leading to a tiny market impact. 
In contrast, these incoming limit orders can impact subsequent order flow in \lobgan, 
and this impact, as shown in Figure~\ref{fig:price_impact_limit}, has same characteristic 
shape in response to a meta-order as found in the literature and shown in 
Figure~\ref{fig:price_impact_path}.
It is worth mentioning that \lobgan exhibits a greater price impact with limit orders than with market orders (see Figure \ref{fig:price_impact_market}). This phenomenon is partially related to the \imbI feature (see Section~\ref{sec:feature_definitions}) which is generally a strong predictor of the sign of future price changes~\citep{bouchaud2018trades}. 
\lobgan is overreliant on this feature and large limit buy orders alter the \imbI and drive the price up. 
We further investigate this effect in Section \ref{sec:book_imb1}.

\subsubsection{Comparison with impact paths from the literature}

\begin{figure}[h]
\centering
\includegraphics[width=\fscale\linewidth]{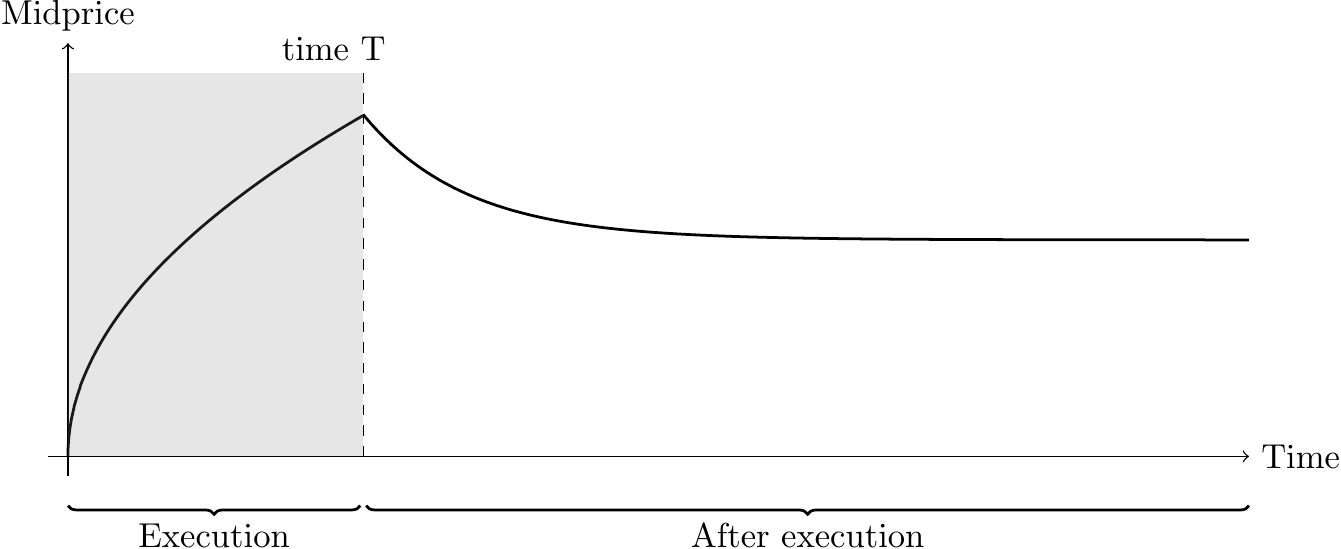}
\caption{The impact of a buy metaorder. 
This form of the impact function represents the
temporary price impact (until time T), the permanent price impact (the
long-run limit level) and the transient price impact (the difference between the
price impact at the peak and the long-run limit level). 
This shape can be seen empirically in Figure 13 of \protect\citet{zarinelli2015beyond} 
or Figure 8 of \protect\citet{bacry2015market}).
}\label{fig:price_impact_path}
\end{figure}

The \lobgan price impact path closely resembles the shape of the price impact path
from the empirical and theoretical literature given in
Figure~\ref{fig:price_impact_path} or \citet[Figure 12.1]{bouchaud2018trades}. 
In particular, {temporary price impact} is increasing in the volume that is executed and concave 
(\citet{bouchaud2018trades} states that it should be approximately square root in volume and \citet{bacry2015market}
find it to be power law with exponent $0.6$). 
Then, there is also clearly a permanent price impact component, as the
midprice mean reverts back to a proportion of its peak impact. Furthermore,
the permanent price impact is even approximately equal to $2/3$ of the peak
price impact (a finding from \citet{zarinelli2015beyond}). 

Apart from the issues relating to the excessive impact of limit orders when compared with market orders, the market impact produced by \lobgan is much more realistic than that produced by market replay. 

\subsection{Further benefits}
\label{sec:furtherben}

We now describe two important further benefits of the CGAN approach which
motivate the challenge of designing robust and realistic models.

The first further benefit is data shareability: Generative
models offer a way for commercial entities to share realistic data with
academia, whereas sharing historical data is constrained by licensing and cost
issues.

The second further benefit is variability of the market scenarios that a
generative model can generate. 
A weakness of backtesting is that there is only one market history, so
backtesting is prone to ``time-period bias'', i.e., in essence, overfitting to
the particular history that was encountered. 
An attractive feature of CGANs is that they can create many possible histories
depending on the randomness that is used every time the CGAN outputs new orders.
While there is only one true market history, with a generative model, you can in
principle generate as much data as you want. 
With an effective generative model, this then allows the use of potentially
sample inefficient machine learning approaches such as Reinforcement Learning,
while avoiding the problems of both ``time-period bias'' and ``data-snooping
bias''~\citep{STW99} (where good results are obtained by luck just because the
same data is used again and again).
        
\section{Realism metrics}
\label{sec:realism}

In Section~\ref{sec:realismtypes}, we introduced the concept of \textit{realism in isolation} as the ability of the CGAN to generate realistic limit orderbook markets when trajectories are rolled out at test time with no other agents in
the system.
In general, a single metric to measure the realism of a synthetic market does not exist, 
and we rather evaluate how well a range of statistical properties align with those of real markets.
These properties are typically highly non-trivial and sometimes counter-intuitive, which 
makes it very difficult, if not impossible, to define a single unified metric~\citep{bouchaud2018trades}.

\begin{figure*}[t]
\centering
\includegraphics[scale=0.40,trim={0 0 0 0}, clip]{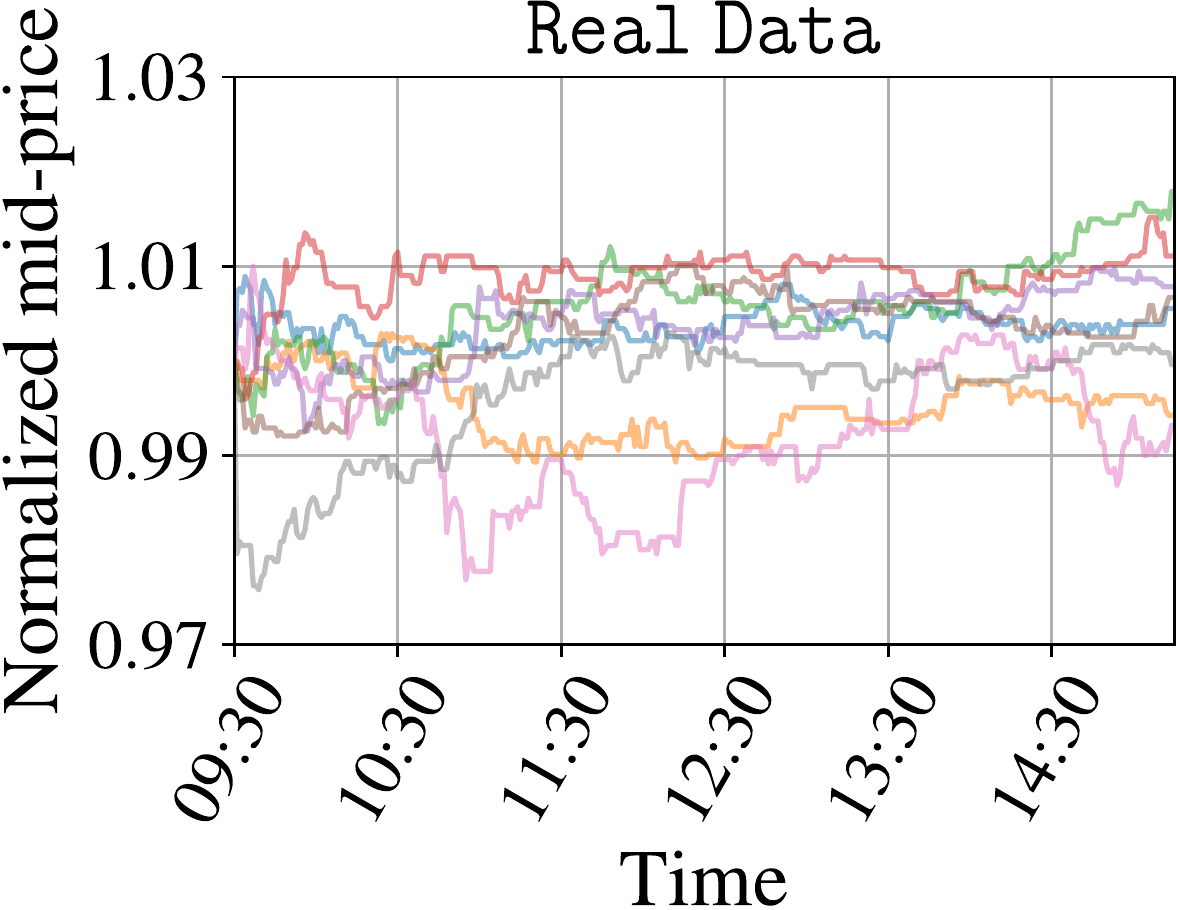}
\hfill
\includegraphics[scale=0.40, trim={0 0 0 0}, clip]{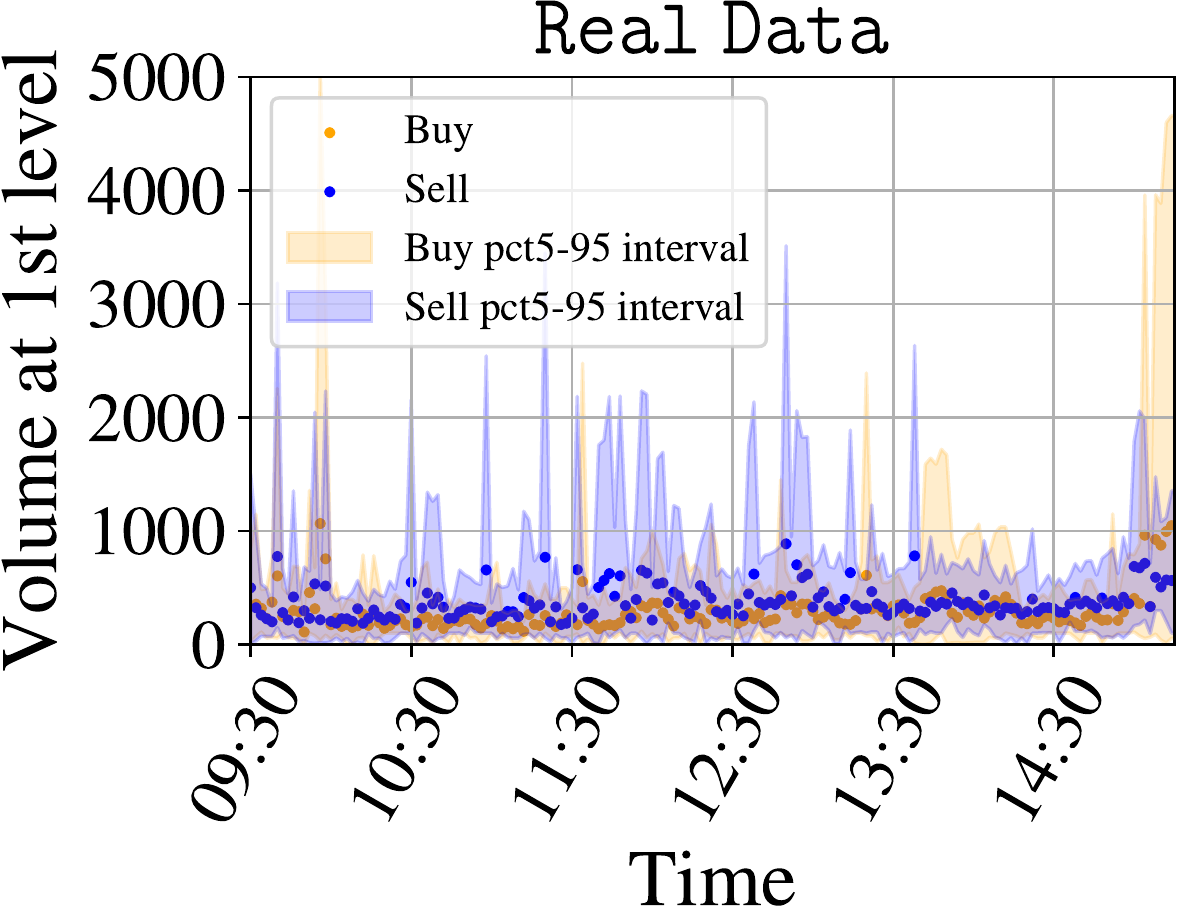}
\hfill
\includegraphics[scale=0.40,trim={0 0 0 0}, clip]{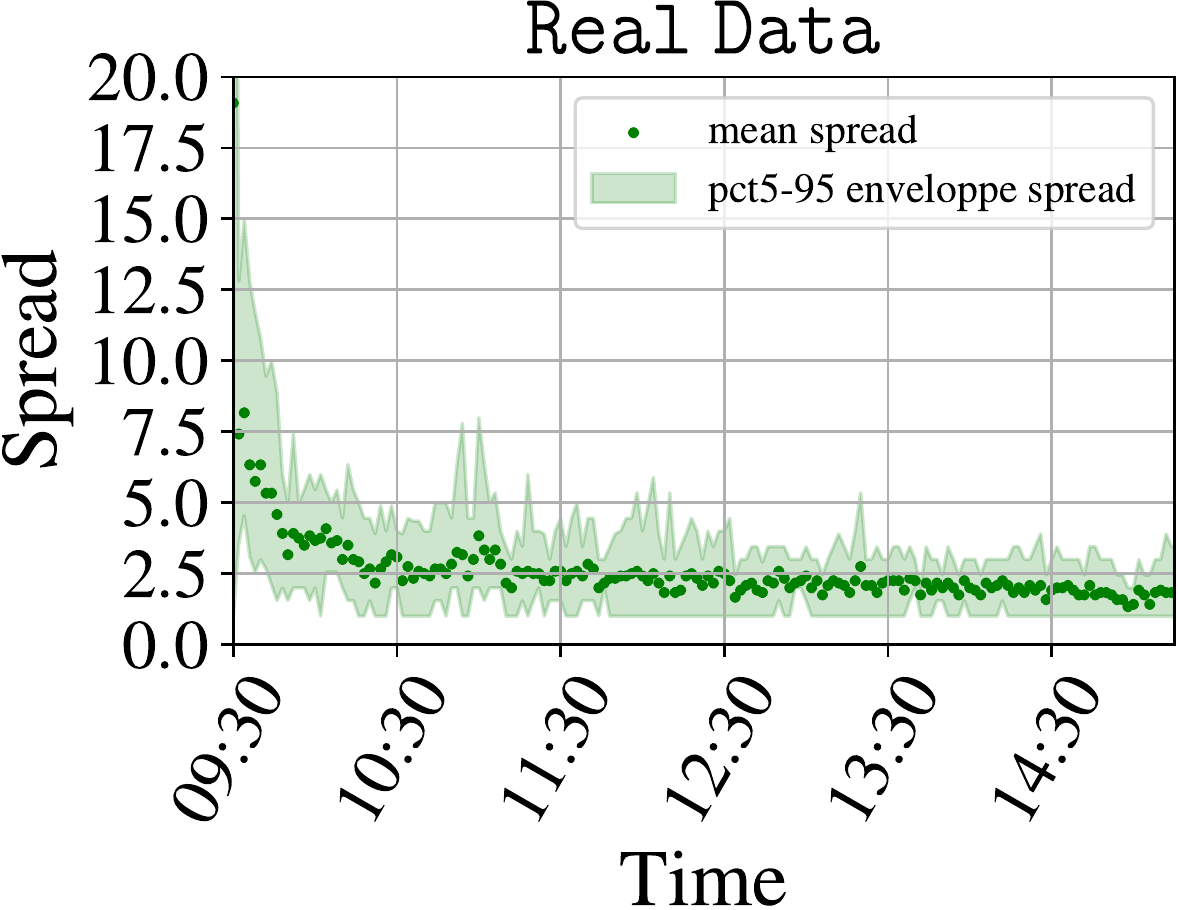}\\
\vspace{0.1cm}
\includegraphics[scale=0.40]{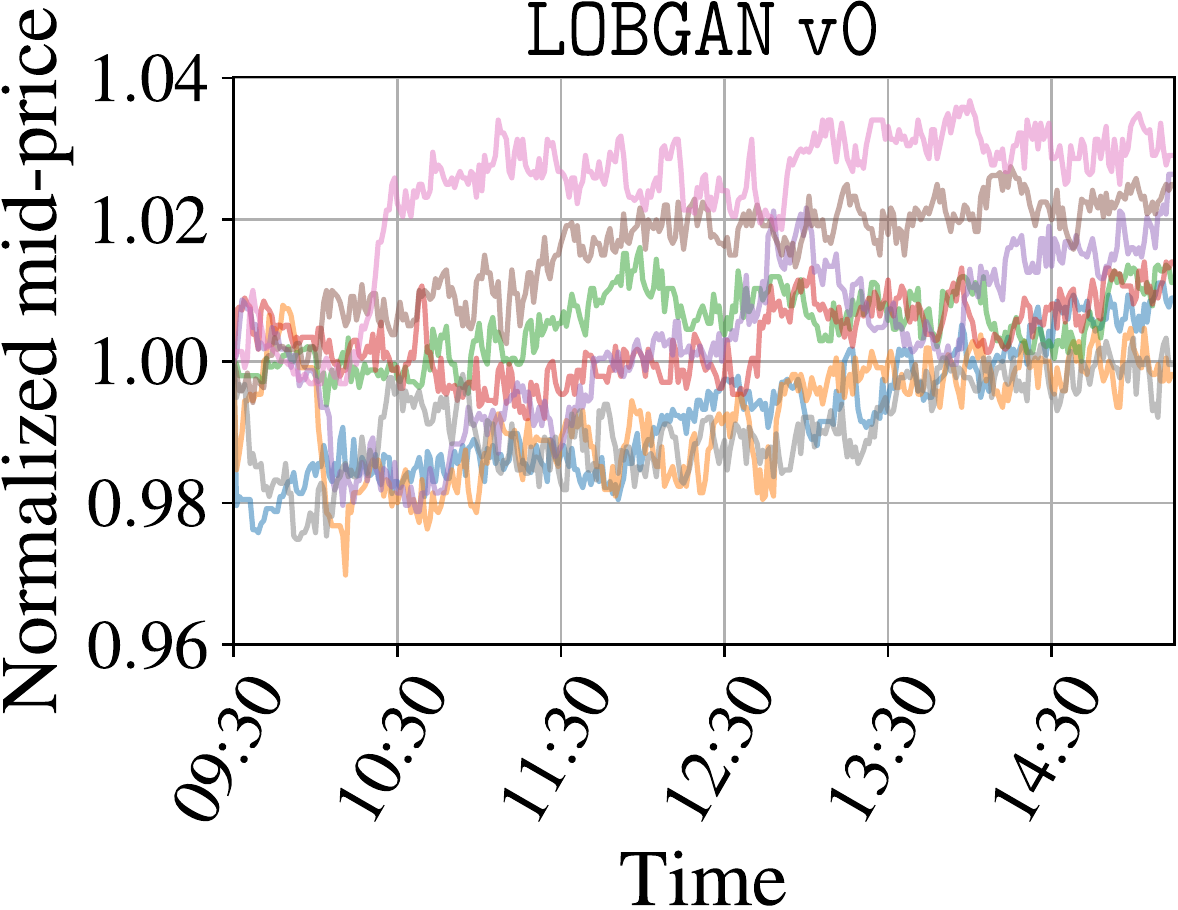}
\hfill
\includegraphics[scale=0.40]{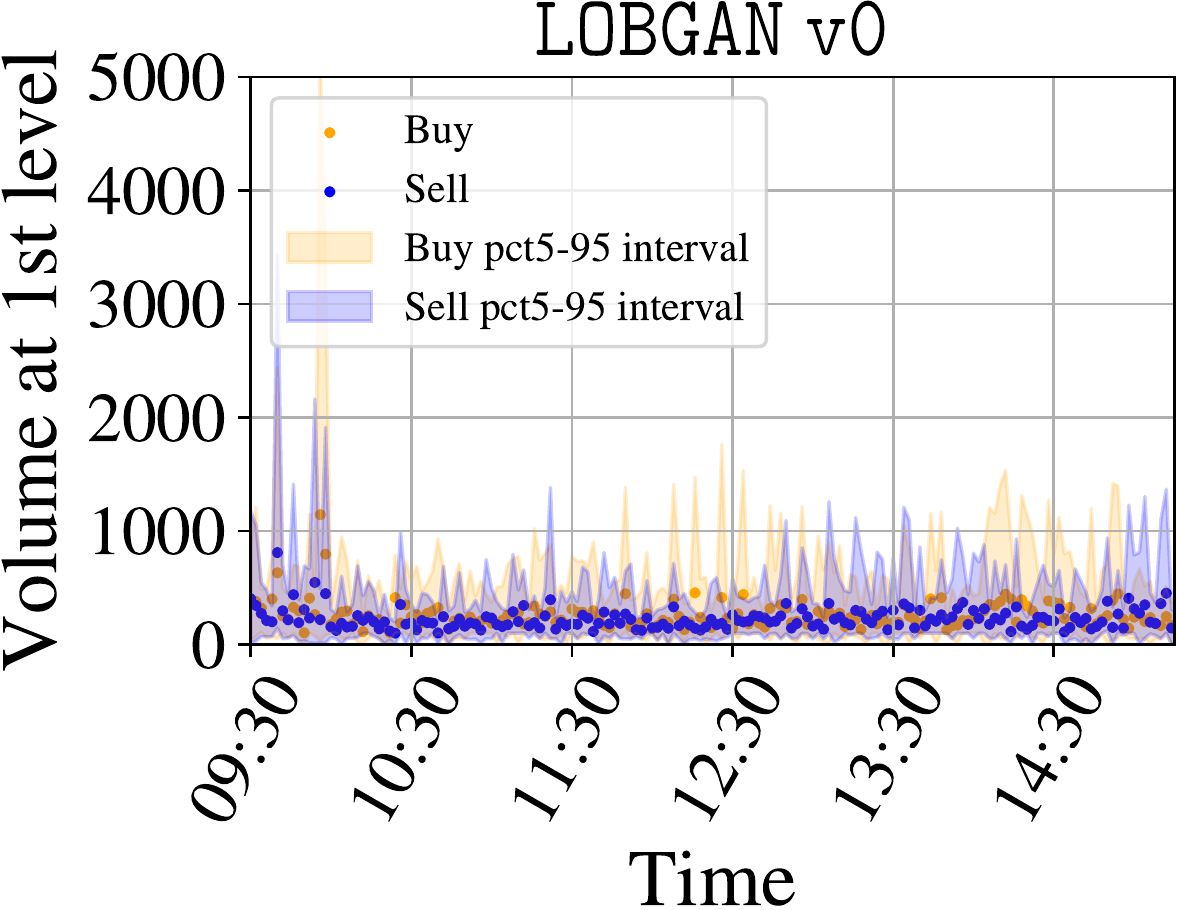}
\hfill
\includegraphics[scale=0.40]{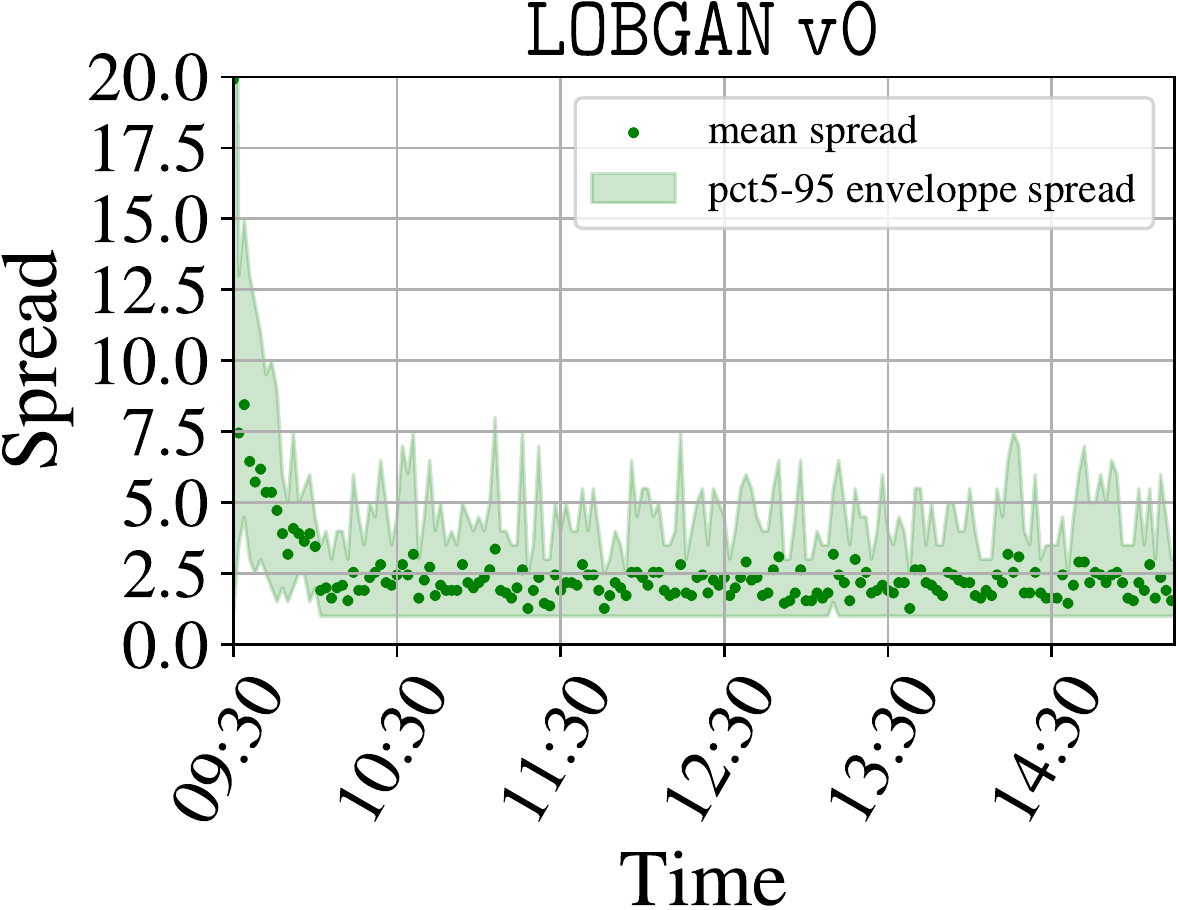}
  \caption{Real vs \lobgan stylized facts.}\label{fig:stylized_lobgan}
\end{figure*}

In this section we briefly review the main statistical properties that 
have been used in the literature to evaluate synthetic LOB markets, 
and we refer to these properties as \textit{stylized facts}~\citep{VyetrenkoBPMDVB20}. 
We use such stylized facts to evaluate the synthetic price series, volumes and order flow. 
We consider: auto-correlations, the heaviness of tails of the
respective distributions, long-range (temporal) dependence, the 
order volume distribution, the time to first fill, 
and the depth and market spread distributions.\footnote
{We refer to~\citet{VyetrenkoBPMDVB20} for a more extensive introduction to stylized facts.}
These stylized facts are used to answer two fundamental questions
during the CGAN training process: 
\begin{itemize}
\itemsep1mm
\item \emph{At which training epoch has the model stabilised?}
\item \emph{Given two trained CGAN models, which one is more realistic?}
\end{itemize}
 
Due to the adversarial training procedure, the CGAN loss is by no means a
perfect indicator of the generator quality~\citep{goodfellow2020generative}. 
Instead, it is common to include humans in the loop (HITL) to evaluate the
realism of generated samples during the training, especially in computer vision
and image generation tasks, where human judgement can reliably reject unrealistic
images~\citep{borji2019pros}. 
In our case, at the end of each epoch we unroll the CGAN in closed-loop
simulation to generate multiple days of synthetic market data. 
We then compute the stylized facts and compare them against those of the real
market. 
As mentioned, we do not have a unified metric but we rather adopt the HITL
approach to distinguish between the properties of synthetic data streams, using
the stylised facts as a basis. 
In particular, considering the HITL approach we restrict the analysis to
\textit{price series}, \textit{volume at best bid and ask}, and \textit{spread}
as shown in Figure \ref{fig:stylized_lobgan}.

We consider them as a prerequisite of a realistic market. 
Importantly, it is the case that humans can easily distinguish between real
stock price series and synthetic price series generated by simple popular stock
price models (members of the Brownian-motion-based midprice processes)
\citep[Chapter 2]{mandelbrot2010mis}. 
Also, since mode collapse is common during training of generative adversarial
networks, one needs to ensure that the generated price and volume trajectories
exhibit significant diversity.

Finally, the best trained model can be tested against all the stylized facts. We
refer to \citet{ColettaMVB22} for the complete evaluation of \lobgan realism,
and in Figure \ref{fig:stylized_lobgan} we report the restricted set of
statistical properties computed in isolation (unrolled in closed-loop) against
real data. 
The Figure also confirms the realism and the ability of {\lobgan} to cope with previously generated
synthetic orders and possible compounding errors during the unrolling of those orders~\citep{suo2021trafficsim}. 

In the original paper Coletta et al. show also that the CGAN produces realistic dynamics in presence of a certain class of experimental order-execution
agents~\citep{ColettaMVB22}. 
In particular, the authors introduce an impact experiment in which an
experimental agent buys (sells) a given amount of volume showing how the
simulation reacts and eventually mean-reverts. They show that, in these cases, the simulator manages to simulate realistic \textit{price impact paths} and \textit{stylized facts}.
In Section \ref{sec:cgan_feature_dependence} we answer a different question:
\textit{can an experimental agent discover weaknesses of \lobgan and drive the market into a desired state?} 

\section{The conditioning of \lobgan}
\label{sec:lobgan_conditioning}

This section explores the conditioning mechanism of \lobgan.
In Section~\ref{sec:feature_definitions}, we define and explain 
a range of relevant features, which are based upon the dynamics of the order book.
These features include all of those that are used by \lobgan, as well as 
some further features that are useful for our analysis and development of \lobgan.
In Section~\ref{sec:features_cgan} we specify the features that are used by
\lobgan.
Then, in Section \ref{sec:cgan_feature_dependence},
we present an analysis
of the feature dependence, highlighting the difficulties in analysing
sequentially generated data.
Finally, in~ Section~\ref{sec:ablation} we perform an ablation study, where 
we train models with and without certain candidate features to understand their
importance.

\subsection{Feature definitions}
\label{sec:feature_definitions}
 
Since extracting features from the raw data is difficult and computationally
expensive~\citep{sirignano2019deep}, \lobgan uses hand-crafted features that have
been successfully used in other parts of the market microstructure
literature \citep{bouchaud2018trades}. 

\subsubsection{Order book features}

Denote the $i^{th}$ best price on the bid and sell sides of the
order book by $P^b_i(t)$ and $P^a_i(t)$ respectively. 
Corresponding to these price levels, let $V^b_i(t)$ and $V^a_i(t)$
denote the volume of limit orders at same $i^{th}$ level of the bid and ask 
sides of the book respectively.
We then define the following order book features:

\begin{itemize}
	\item The \textit{total volume} at the top $n$ levels of the order book at
		time $t$,  $$\code{TotalVol}_n(t) = \sum_{i=1}^n (V^b_i(t)+V^a_i(t)).$$
    \item The \textit{order book imbalance} of the top $n$ levels of the book at time $t$, 
		\begin{equation}
		    \nonumber
			\imb_n(t) = \frac{\sum_{i=1}^n V^b_i(t)}{\sum_{i=1}^n (V^b_i(t) + V^a_i(t))} = \frac{\sum_{i=1}^n V^b_i(t)}{\code{TotalVol}_n(t)}.
		\end{equation}

	\item The \textit{spread} at time $t$, $\code{Spread}(t) = P^a_1(t) -
		P^b_1(t)$.

    \item The \textit{midprice} at time $t$, $P^{\rm mid}_t  = \frac{1}{2}(P^b_1(t) + P^a_1(t))$.
\item The \textit{percentage return} for a time window $[t-\Delta, t]$ is given by
    \begin{equation}\nonumber
        \code{PctReturn}_\Delta(t) = \frac{P^{\rm mid}_t - P^{\rm mid}_{t-\Delta}}{P^{\rm mid}_{t-\Delta}}.
    \end{equation}
	\item Suppose that the order book events occur at times $t_i$ for $i\in\N$.
		Then, the midprice at time $t$ is equal to the midprice at time
		$t_{j^*(t)}$ for $j^*(t) = \sup\{j\in\N: t_j \leq t\}$. After the
		$n^{th}$ order book event, we may then also define the \textit{$n$-event
		percentage return} at time $t$ by 
    \begin{equation}\nonumber
        \code{EventPctReturn}_n(t) = \frac{P^{\rm mid}_{t_{j^*(t)}} - P^{\rm mid}_{t_{j^*(t)-n}}}{P^{\rm mid}_{t_{j^*(t)-n}}}.
    \end{equation}
\end{itemize}

We further define \textit{the touch} to consist of the best bid and the best ask
price levels. 
When we refer to quoting at a distance from the touch, this means that we post a
limit order at that distance away from the best price on the corresponding side
of the order book. 
This is quoted in \textit{ticks} -- the minimum price difference between two
price levels in a LOB.

\paragraph{\textbf{Trade features}} We now introduce a family of \textit{trade
features}. 
Suppose that trades occur at times $s_i$ for $i\in\N$ and let $V^{\rm
trade}_{s_i}$ be the signed volume of the trade; 
if the trade is seller-initiated then it takes a positive sign, if it is
buyer-initiated, it takes a negative sign.\footnote{A trade is seller-initiated
if the sell order arrives at the market after the buy order, and is buyer-initiated
if the buy order arrive after the sell order.} 
Then, for $t\geq \Delta$ one can define the \textit{trade volume imbalance} of
window size $\Delta$ at time $t$ by 
\begin{equation*}
{\code{TradeImbalance}_\Delta(t)} = \frac{
\sum_{t-\Delta \leq s_i \leq t}\1_{V^{\rm trade}_{s_i}\geq0}V^{\rm trade}_{s_i}}{\sum_{t-\Delta \leq s_i \leq t}\left|V^{\rm trade}_{s_i}\right|}.
\end{equation*}

\noindent
\paragraph{\textbf{\lobgan features}}\label{sec:features_cgan} The features which \lobgan conditions upon are:
\begin{itemize}
    \item the \textit{order book imbalance} for $n=1$ and $n=5$ levels;
    \item the \textit{total volume} at the first level and at the top five levels of the book;
    \item the spread;
    \item the \textit{$n$-event midprice percentage return} for $n=1$ and $n=50$ order book events;
    \item the \textit{trade volume imbalance} over the last minute and over the last five minutes.
\end{itemize}
To better capture the market evolution over time, \lobgan concatenates the feature values over the last 
30 seconds.

\subsection{\lobgan features dependence}
\label{sec:cgan_feature_dependence}

The conditional nature of \lobgan is crucial for two main reasons: 
firstly, it makes it easier for the CGAN to learn to produce stable (and therefore more realistic) trajectories; 
secondly, it ensures reactivity when an exogenous agent interacts with the order book. 
In this section, we investigate how the order book dynamics depend upon the
input feature vectors and which features produce the largest changes in the
trajectories of the trained CGAN.

\begin{figure}[h]\centering
\includegraphics[width=\fscale\linewidth]{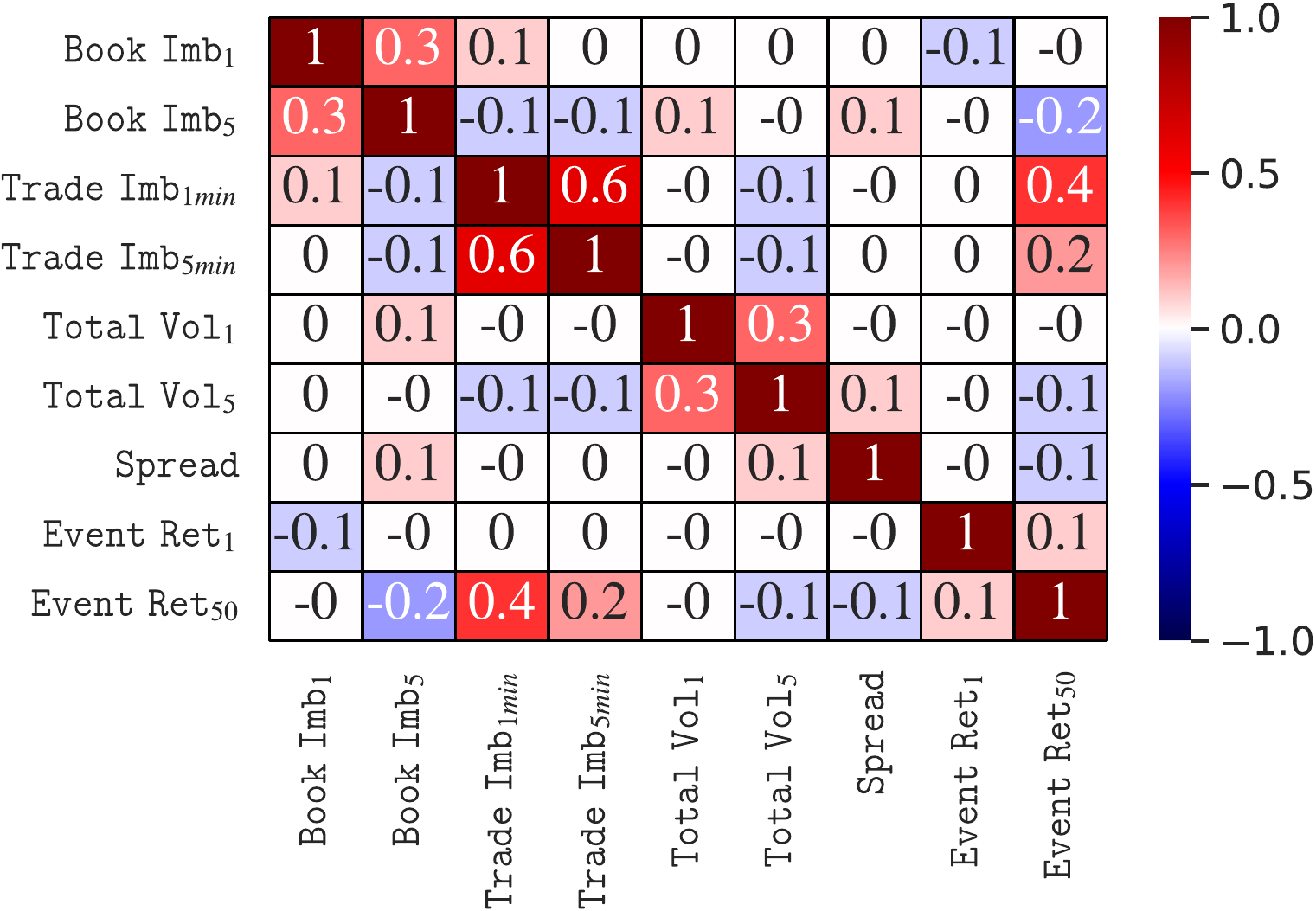}
\caption{The hand-crafted features correlation matrix.}
\label{fig:corr_plot}
\end{figure}

Notice that, analysing the effect of individual features is extremely difficult for a number of reasons: 
firstly, there are cross-correlations between all of the input features (see Figure~\ref{fig:corr_plot}); 
second, we do not only want to investigate the distributional properties of a single output of the CGAN -- 
realistic order flow data must be maintained when the orders generated by the
CGAN are processed and the features are fed back into it which compounds any
errors in the order flow; 
finally, there is also a ``mechanism effect'' that
comes directly from the order book mechanism -- this should ideally be separated
from the effect of the features.

In this section, we introduce a pragmatic approach to investigate the long-run
dependence of the CGAN output upon certain features. 
This approach requires that the correlations between the input features not be
too large. We see in the correlation matrix in Figure~\ref{fig:corr_plot} that
this is indeed a reasonable assumption.

To investigate the effect of various features we first roll out 60 trajectories
each lasting 20 minutes, starting on a lattice of evenly spaced points across
two trading days in January 2021.
These act as baseline trajectories. 
We then repeat the process twice for each input feature of the CGAN, fixing its values 
to be equal to the $5^{th}$ and $95^{th}$ percentiles of the empirical training
distribution; we give these percentiles in Appendix~\ref{sec:featurepercentiles}. 
We allow all of the other features to update as the order book updates. 
Finally, we investigate the
properties of the time-series data for each of these trajectories with the goal
of measuring the effect of fixing these features on certain key properties of
the time-series data that are described in \citet{VyetrenkoBPMDVB20}. In
particular, if the trajectory statistics for these rollouts with extreme values
for the conditioning appear much the same as the baseline trajectories, then
these features are candidates for ablation, and if there is a clear dependence
of any of the output ``stylised facts'' on the feature being perturbed, then
this knowledge provides a partial explainability of the CGAN. It can also be
used to construct "adversarial" strategies for the CGAN.

\paragraph{Discussion of approach.} Our approach allows us to investigate how altering a single input feature of the
financial market effects the output of \lobgan~and, as a consequence, the rolled
out trajectories. 
It is worth noting here that we do not actually change the
\emph{past} orderbook or trades that occur, but rather \lobgan's \textit{perception} of
them. 
This in turn may of course change the future output of \lobgan, which is the whole
point.
This type of A/B testing is simply not possible using historical data, as
it relies on counterfactually altering one component of the financial market and
seeing how the subsequent order flow evolves. 
There are not enough historical
time intervals in which the value for one of the input features remains
unaltered, and so the macro effects of such a change cannot be studied easily.
It is \textit{only} possible to look at the order-level distributions and then
attempt to infer how the changes in distribution affect the dynamics of the
order book throughout time. 
Thus, under the assumption that \lobgan has successfully learnt the conditional
distribution of orders from historical data, this presents a completely novel
approach to investigating LOBs with much more insight into the
causal mechanism. 
In the rest of this subsection, we employ our approach for the different \lobgan
features in turn.
We note that, in this approach the input to the CGAN after having been altered
could correspond to an unrealistic orderbook and historical trade combination.
However, as we see in results below, the dependence of the CGAN output on the
input modified in this manner is actually relatively well-behaved (stable).

\subsubsection{\imbI}\label{sec:book_imb1}

\begin{figure}[h]\centering
\includegraphics[width=\fscale\linewidth]{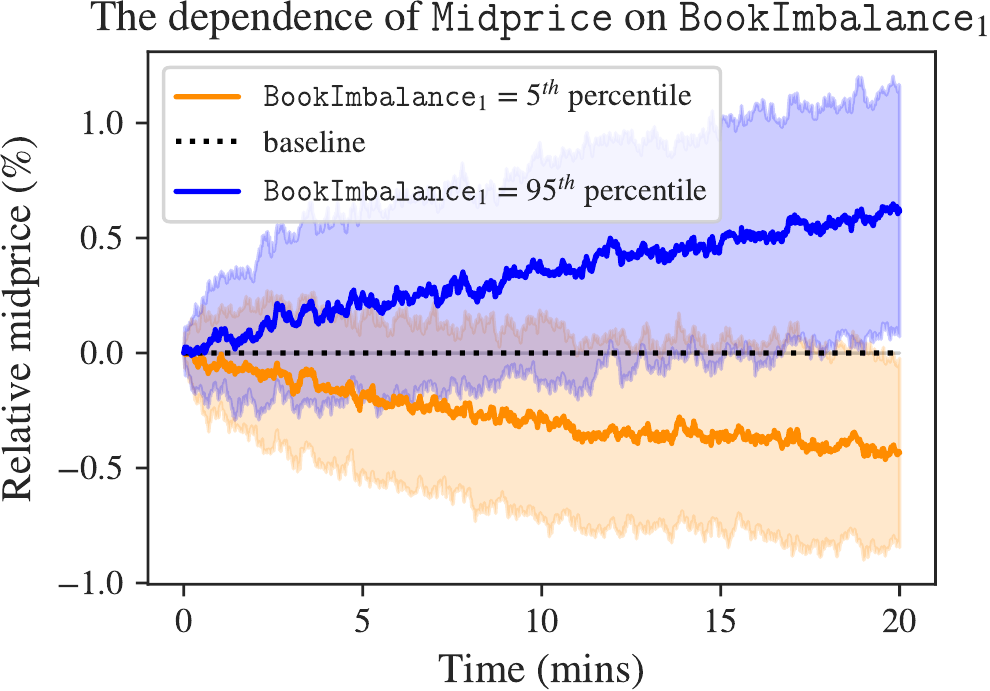}
\caption{The effect of fixing \imbI~on the midprice dynamics. The midprice change is calculated relative to the ``baseline'' case in which the feature \imbI~is unaltered.}
\label{fig:imbalance_and_midprice}
\end{figure}

When the book imbalance at the touch is larger -- due to a larger volume of buy
limit orders than sell limit orders -- the trajectories generated by \lobgan
noticeably trend upwards more on average (see
Figure~\ref{fig:imbalance_and_midprice}). 
This effect has also been frequently observed in analyses of historical data,
where book imbalance is often a predictor of the next midprice move~(see, for
example, \citet{bouchaud2018trades,zheng2012price}). 
This effect could have a variety of causes: there could be an imbalance between
the market buy orders and the market sell orders; on the ask side of the book
there could be an imbalance between limit orders and cancellations; or, on the
bid side, there could be an imbalance between limit orders and cancellations. 
By investigating the properties of the individual orders outputted by the CGAN
during these trajectories, we can answer the question of \textit{how} this trend
appears.

\begin{figure}[h]\centering
\includegraphics[width=\fscales\linewidth]{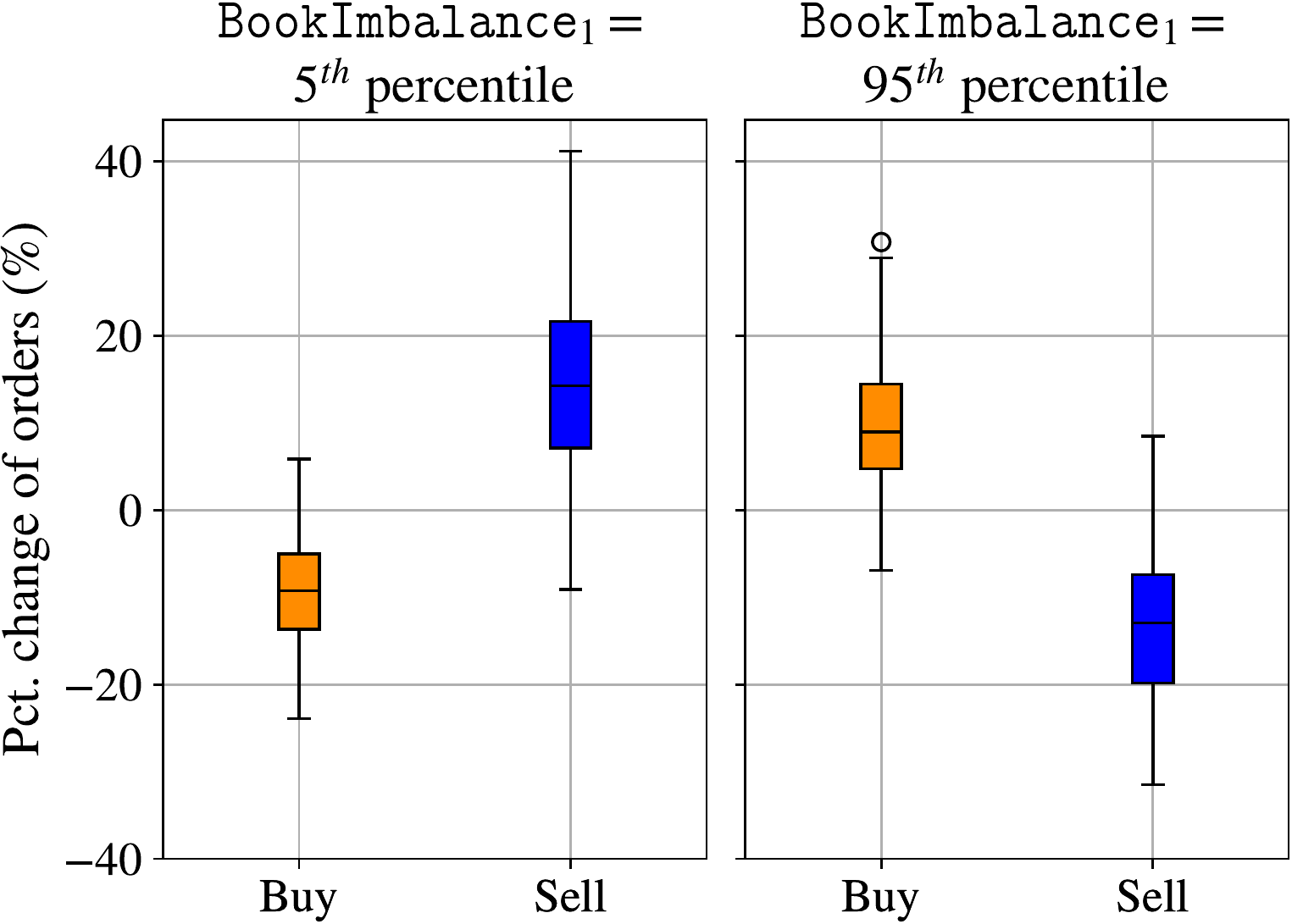}
\caption{The effect of fixing \imbI~on the proportion of market order types.}
\label{fig:imbalance_effect_on market}
\end{figure}

In Figure~\ref{fig:imbalance_effect_on market}, it is clearly shown that one of
the causes of the upwards price trend is the effect on the direction of
outputted market orders. 
Specifically, when the imbalance is lower, \lobgan outputs a larger proportion
of \textit{sell} market orders and vice-versa. This is one of the causes of the
downward price pressure. 
Furthermore, if the input of \imbI to the CGAN is lower, there is an increase in
the quantity $(\textit{cancellation volume} - \textit{limit volume})$ on the sell
side when compared with the baseline trajectory. 
This means that there are relatively much more cancellations than limit orders. 
Similarly, when \imbI takes a large value, there is a decrease on the sell side
of the quantity $\textit{cancellation volume} - \textit{limit volume}$. 
Interestingly, in both cases the difference on the bid side of the book is
unaltered. It seems likely that this is simply due to bias in the training data,
and that the price dynamics w.r.t. \imbI have become overly reliant on the
\textit{sell} side of the book.

\begin{figure}[h]\centering
\includegraphics[width=\fscales\linewidth]{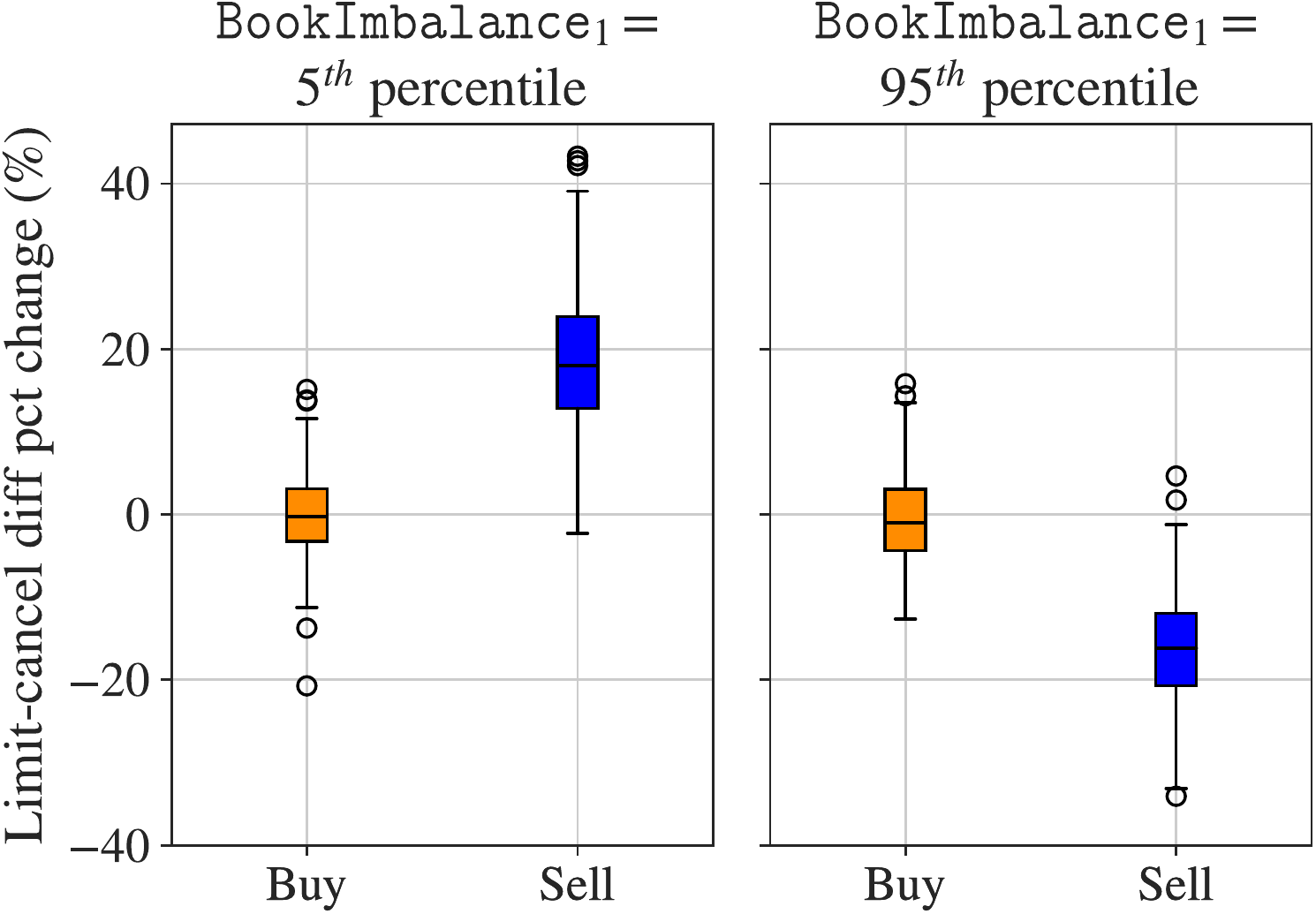}
\caption{The effect of fixing \imbI~on the difference between the volume of cancellation orders and the volume of limit orders. This difference is a key driver of price changes.} 
\label{fig:imbalance_effect_on_diff}
\end{figure}

\subsubsection{\tradeimb}
Both $\tradeimb_1$ and $\tradeimb_5$ have a large effect on the dynamics of the
midprice, with $\tradeimb_1$ playing a particularly big role. 
A plot of the effect can be seen in
Figure~\ref{fig:trade_imbalance_and_midprice} in the appendix. 
In particular, the midprice is increasing in the value of \tradeimb.
It is also worth noting that \tradeimb~is increasing in its own value (which
produces a self-exciting effect). 
Similarly to \imbI, the main causes of the upward (downward) price are the
larger proportion of buy (sell) \textit{market orders} shown in Figure
\ref{fig:trade_imbalance_and_limit_cancel}, and the difference of
(\textit{cancellation volume - limit volume}) on both sides of the book shown in
Figure \ref{fig:trade_imbalance_and_markets}. 
Compared to \imbI, the quantity (\textit{cancellation volume - limit volume}) is
more symmetric, being consistent across both sides of the book. 

\subsubsection{\code{Volume} and \code{PctReturn}}
Upon observing the rollouts when individually fixing \volI, \volV,
$\code{PctReturn}_{1\rm min}$ and $\code{PctReturn}_{5\rm min}$ in turn, it is
clear that none of them has a meaningful effect. 
Therefore, they are candidates for ablation (Section~\ref{sec:ablation}).

\subsubsection{\code{Spread}}
\begin{figure}[h]
\centering
\includegraphics[width=\fscale\linewidth]{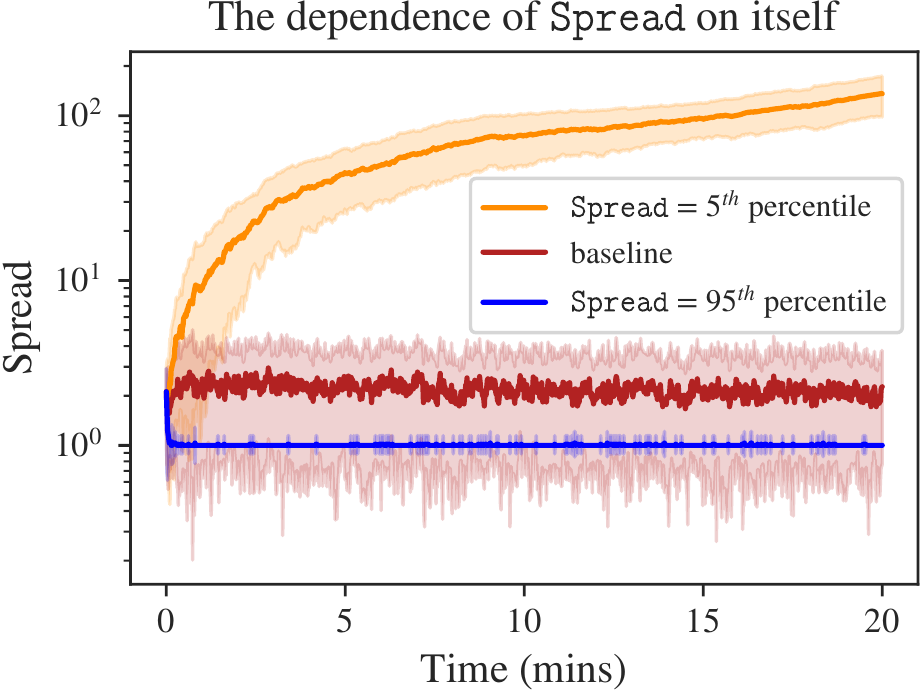}
\caption{The effect of fixing the spread input of \lobgan~on the spread of the subsequent trajectories.
}
\label{fig:spread_trajectories}
\end{figure}

\begin{figure}[h]
\centering
\includegraphics[width=\fscaless\linewidth]{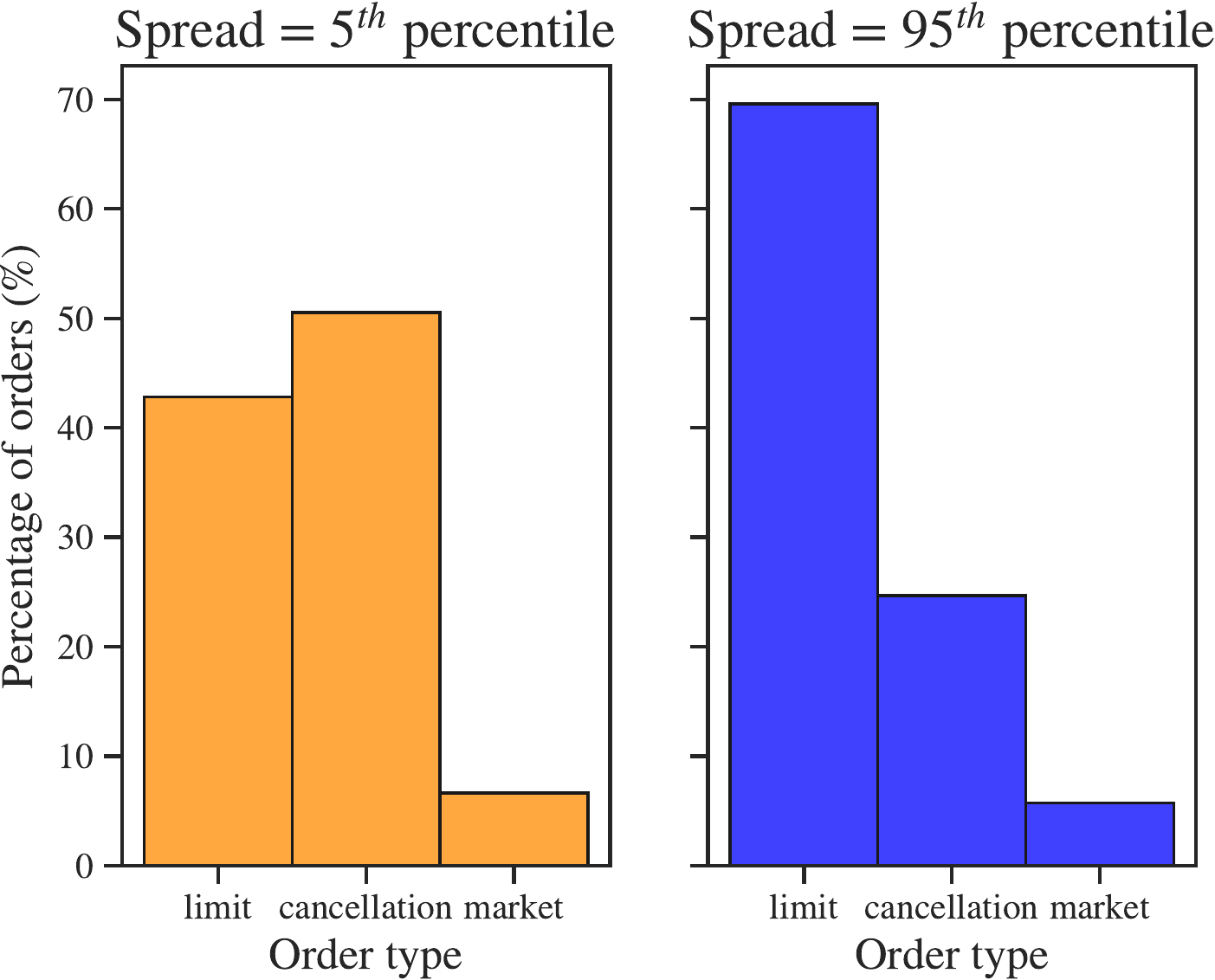}
\caption{The effect of spread on the proportion of different order types.}
\label{fig:spread_effect_on_order_proportions}
\end{figure}

The spread plays a key role in the dynamics of the trajectories generated by
\lobgan. 
In particular, it is highly mean-reverting. 
This is essential for the stability of the outputted market dynamics. 
This can be seen in Figure~\ref{fig:spread_trajectories} -- when \lobgan
perceives the spread to be small, it outputs orders that increase the spread;
when the spread is large, \lobgan tries to close it. 
It does this primarily via the distribution of
\textit{order types} (see Figure~\ref{fig:spread_effect_on_order_proportions}).
When the spread is large (i.e., 95$^{th}$ percentile), \lobgan places more limit
orders so that more liquidity is provided to the market and the spread tightens.
When the spread is small (i.e., 5$^{th}$ percentile), the order type
distribution shifts in favour of deletions and the spread increases massively.

\subsection{Ablation studies for candidate features}
\label{sec:ablation}

To further investigate the \lobgan's features as candidates for inclusion or not
in a potentially improved model, we perform an ablation study in which we train
the model without the features that do not show an observable effect in the
rollouts from Section~\ref{sec:cgan_feature_dependence} (i.e., \volI, \volV, 
$\code{PctReturn}_{1\rm min}$, and $\code{PctReturn}_{5\rm min}$).

In the ablation study we remove features and investigate how time to model convergence and model realism (see Section \ref{sec:realism}) are affected.
Without \volI and \volV the model achieves comparable realism, meaning that it is
able to unconditionally learn the average volume of the orders. 
In fact, we recall that CGAN is trained against a discriminator that rejects
unrealistic volumes.   
Without $\code{PctReturn}_{1\rm min}$ and $\code{PctReturn}_{5\rm min}$ we
observe substantially more training time, i.e., more unrealistic markets in the
early phases, yet comparable performance at the end of the training procedure. 
As the returns are used also by the discriminator, we can conclude that they
help more with rejecting unrealistic markets during training than with
conditioning the generation. 

\section{\lobgan adversarial attacks}
\label{sec:attack}

In this section, we show that certain simple trading strategies are able to
exploit the \lobgan model, completing round trip meta-orders and
turning a profit. 
This means that the agent begins and ends the trajectory with a ``risk-neutral''
inventory of zero and such trades constitute a form of arbitrage. 
Some of these trading strategies, such as market making, are commonly studied in
the literature -- however, these strategies can be implemented so as to realise
unrealistic outsized profits for the adversarial agent. 
Other adversarial attacks have been inspired by Section~\ref{sec:cgan_feature_dependence} to
take advantage of feature dependence in the CGAN model. 
Finally, we introduce trading strategies that ``exploit'' the order creation
mechanism\footnote{By ``order creation mechanism'' we mean the way in which the
output of the CGAN is converted into actual orders in the LOB. This
is explained in detail in Section~\ref{sec:mechanism_attack}.} of the
\lobgan-based simulator.

All of these strategies start with zero inventory and liquidate any terminal
inventory and so -- whilst the profit and loss curves are marked to market (MtM)
-- the terminal profit corresponds purely to an increase in the cash holdings of
the agent from the start to the end of the episode.

\subsection{Market-making}
\label{sec:market_making}
Market makers (i.e., liquidity providers) are a ubiquitous and crucial type of
participant in high-frequency financial markets. 
In orderbook-driven financial markets, they
post limit orders on both sides of the book always offering to buy or sell with
the aim of earn the difference between the bid and the ask whenever they complete a
round trip trade of one order. 
Their main source of risk is \textit{inventory risk} -- the possibility that
they will accumulate a large (positive or negative) inventory by getting filled
unevenly on each side of the book before the price moves against that inventory;
the other source of risk is adverse selection where they are more likely to get
filled by informed traders than uninformed traders.

\begin{figure}[h]\centering
\includegraphics[width=0.468\linewidth]{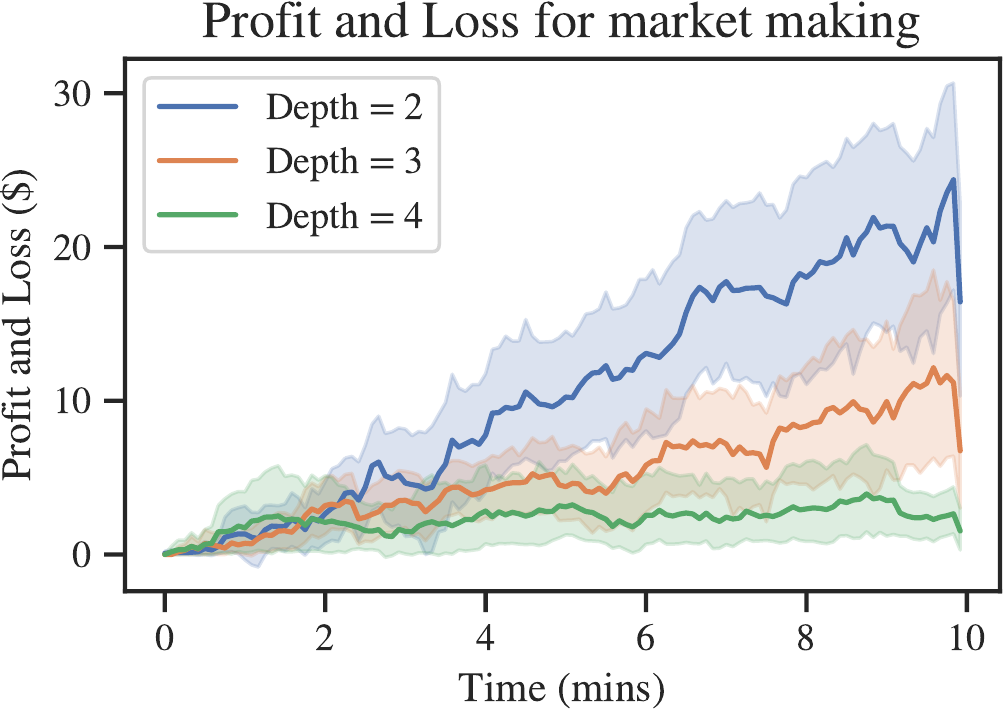}
\includegraphics[width=0.49\linewidth]{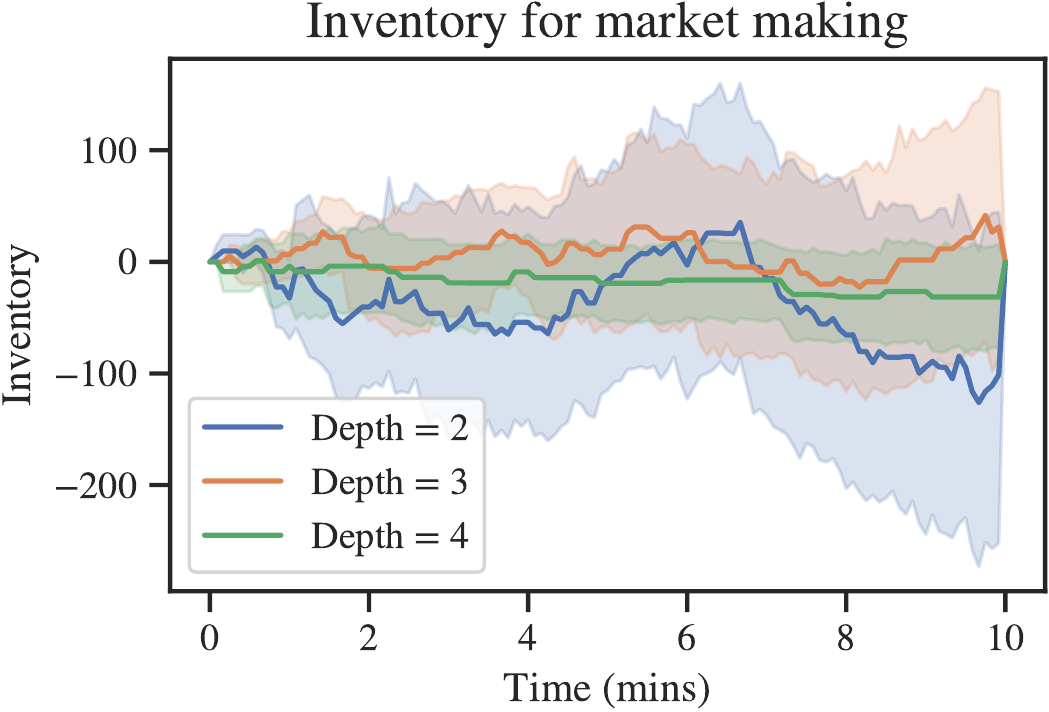}
\caption{\small The left-hand figure shows the profit and loss for a family of market-making strategies that symmetrically posts a (relatively) large volume at a fixed number of levels (its depth) from a target. The right-hand figure shows the associated inventories.}
\label{fig:market_making_rewards}
\end{figure}

In this section, we introduce a naive symmetric market-making strategy that
posts a symmetric volume around the midprice of the asset.  
These strategies update every 5 seconds -- maintaining a fixed volume on each
side the book at a fixed depth.
They do so by cancelling existing orders that after an orderbook update are no
longer at the desired depth, replacing them with orders at the desired depth. 
This simple liquidity provision strategy ends up being consistently profitable
in the trained \lobgan-based simulator (see left hand panel of
Figure~\ref{fig:market_making_rewards}), and is robust across a range of depths. 
While market making is a highly profitable activity in real markets, it is not
realistic for such a simplistic strategy, which uses a fixed distance from the
touch on both sides at all times (i.e., it never skews its spread), to be  so
consistently profitable.

It is worth noting that this strategy is not profitable if it posts exactly at
the touch, as the frequency at which the agent gets filled can cause the agent
to accumulate a overly large signed inventory that needs to be liquidated at the
terminal time (at a cost, through market impact) to satisfy the condition of
being a round-trip meta-order. 
It seems highly likely that a more adaptive market-making strategy (see
\citet{jerome2022market} for example) could avoid this problem by managing
inventory risk better.

In the right hand panel of Figure~\ref{fig:market_making_rewards}, the mean
inventory accumulated by such a market-making strategy is plotted. 
The inventory mean-reverts, causing the inventory risk of the
strategy to be maintained within reasonable bounds. 

\begin{figure}[h]\centering
\end{figure}

\subsection{Accumulating a positive inventory using limit orders and then liquidating}
\label{sec:attack_imbalance}

\begin{figure}[h]\centering
\includegraphics[width=0.42\linewidth]{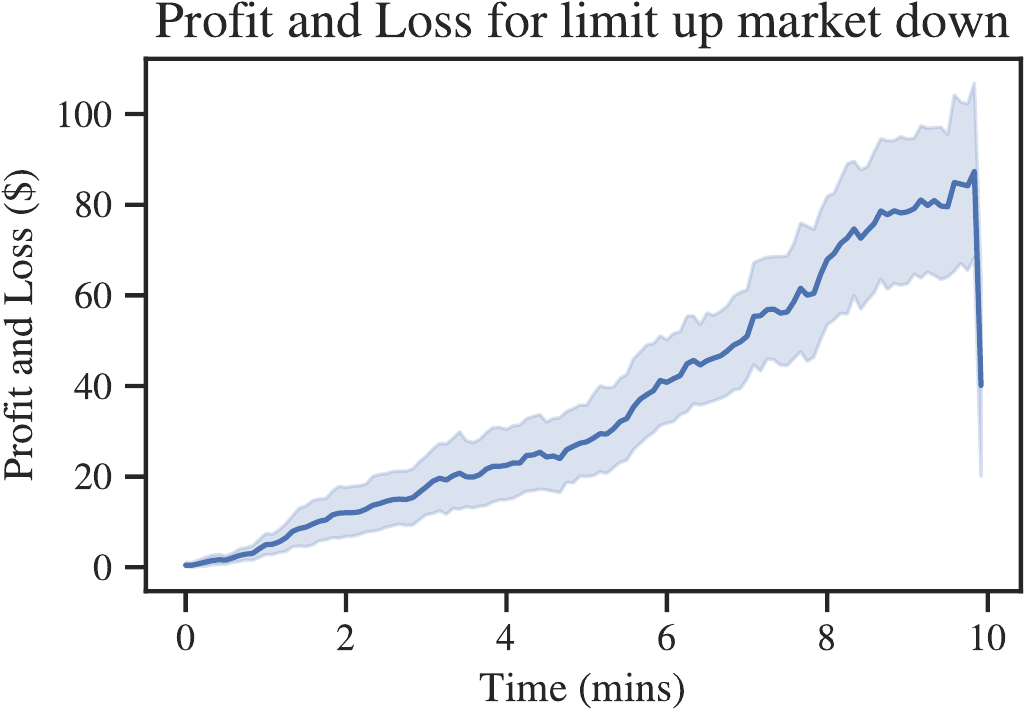}
\includegraphics[width=0.51\linewidth]{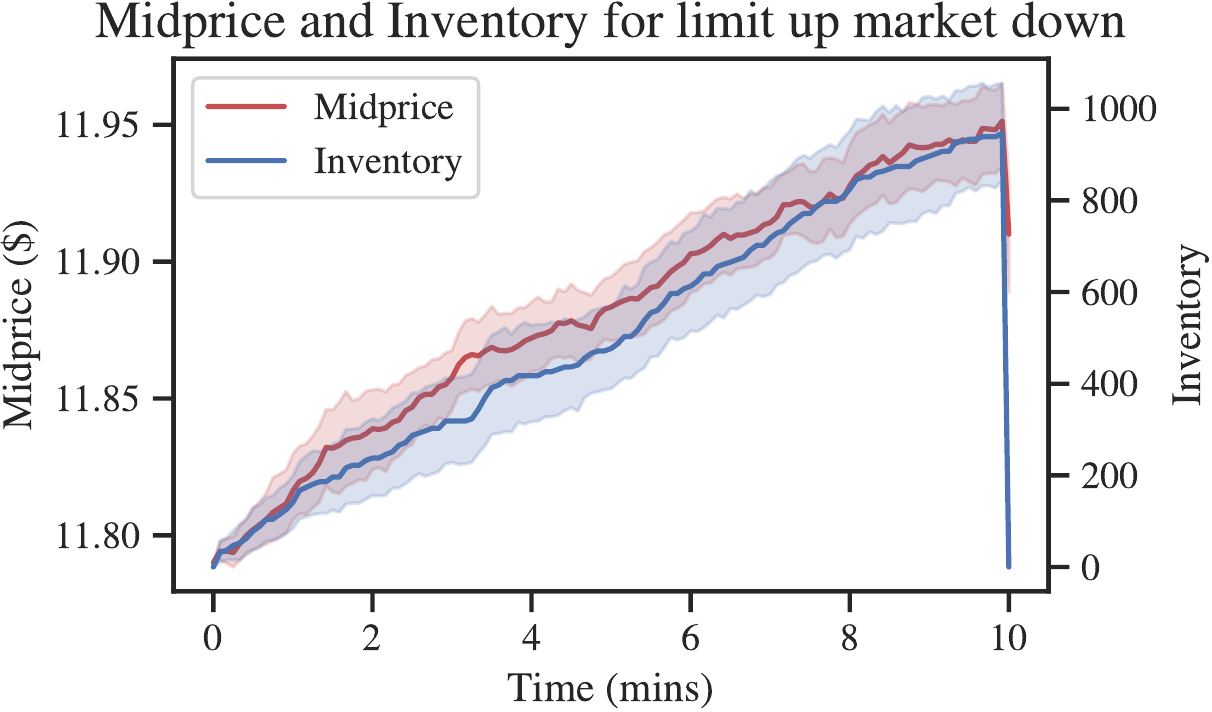}
\caption{The left-hand side shows the profit and loss for a simple strategy that places buy limit orders over a 10 minute period and then liquidates its position with a market order. The right-hand side shows the midprice of the order book and the inventory for this strategy.}
\label{fig:limit_up_market_down}
\end{figure}

As shown in Section~\ref{sec:cgan_feature_dependence}, \lobgan is quite
sensitive to the order book imbalance. 
In this section, we show that a trading strategy can target this feature by
continually placing a large volume of limit orders on one side of the book.
The strategy also places a small volume of orders on the other side of the book.
It accumulates inventory from the larger orders, and after liquidation at the
end of the episode makes a profit from the overall round-trip meta-order.
A specific example is an agent that maintains a limit order of size 200 at one
tick away from the touch on the bid side of the book and an order 10\% of the
magnitude one tick away from the touch on the ask\footnote{The strategy works much less well without the 
ask side, in which case two things happen: 
The agent accumulates a very large inventory. This costs to liquidate at end;
the price moves too quickly, which means that the levels at the touch on
the bid side of the book (when we are pushing price up) don’t get filled up
with external orders as the touch moves too fast, and this also increases
the liquidation cost at the end of the episode.}.
As described in Section~\ref{sec:cgan_feature_dependence}, by making  \imbI
larger, this strategy pushes the price up whilst accumulating inventory as the
price goes up.
The strategy is then able to liquidate at the terminal time for a cost that is
less than the value it gained by pushing the price up, completing a profitable
round-trip meta-order. 
Figure~\ref{fig:limit_up_market_down} shows the profit and loss of this
strategy, along with the effect that this strategy has on the midprice
dynamics as well as the strategy's accumulated inventory.
We recall that this is an adversarial attack, like all of our strategies, created to 
highlight the \lobgan model weaknesses. Moreover, our strategy is not a
form of (illegal) market manipulation, like spoofing, since every limit order we
place has a good chance to be, and often is, executed.

It is worth briefly commenting on a similar strategy -- which pushes the price
up using buy \emph{market} orders. 
This strategy causes price impact, but also importantly increases the values
for \tradeimb. 
Whilst this strategy did manage to have an outsize effect on the midprice of the
market, it was not a profitable round-trip trade as the cost of liquidating the
large accumulated inventory at the terminal time offset its gains. 

\subsection{The weakness of the \lobgan placing mechanism}
\label{sec:mechanism_attack}

\newcommand\wid{0.48}
\begin{figure*}
\centering
\begin{tabular}{cc}
\includegraphics[width=\wid\textwidth]{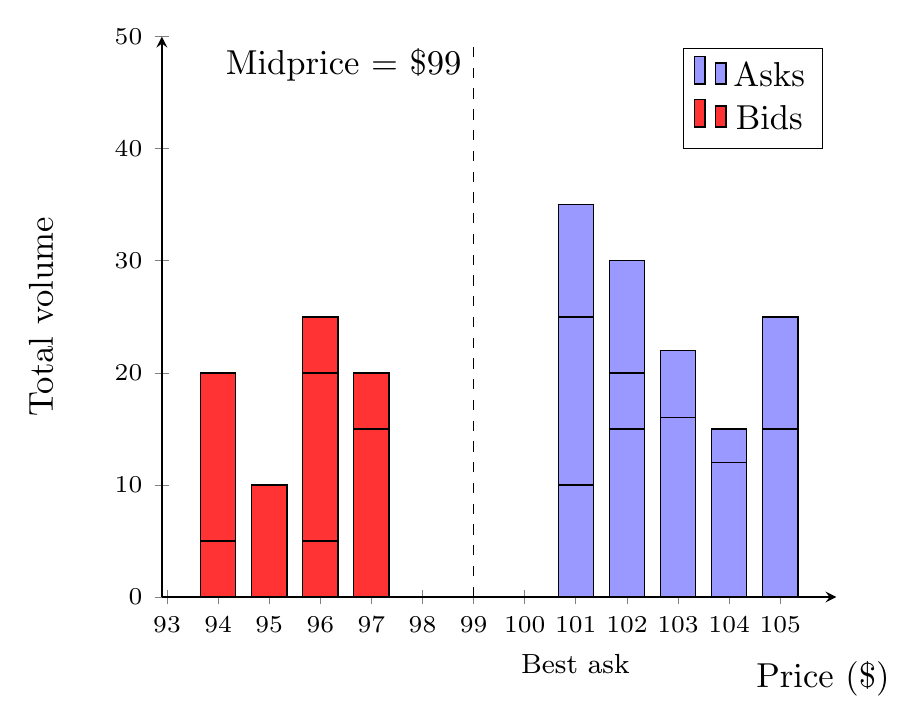} &  
\includegraphics[width=\wid\textwidth]{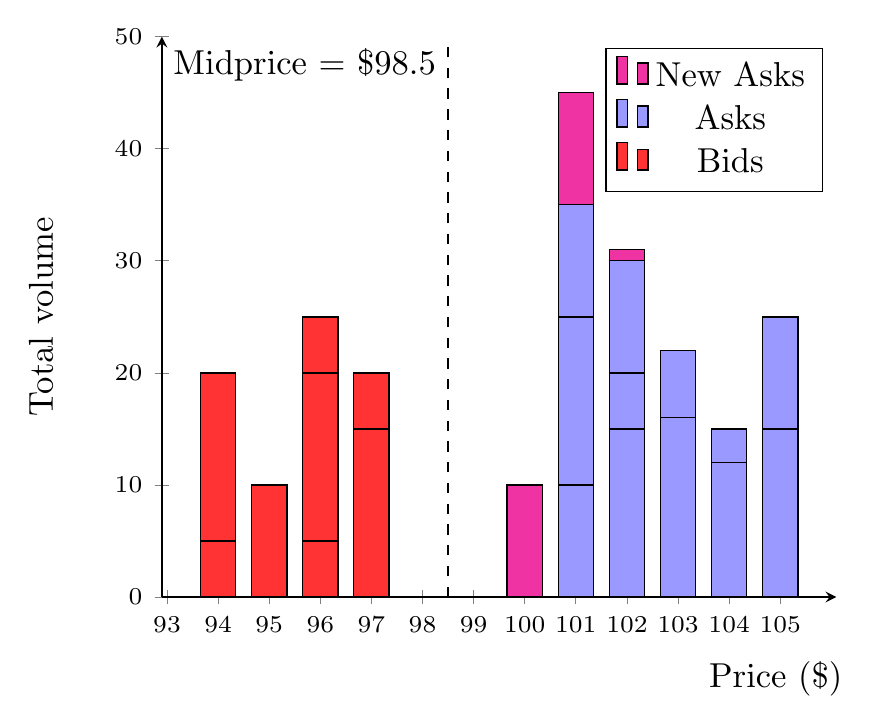} \\
\includegraphics[width=\wid\textwidth]{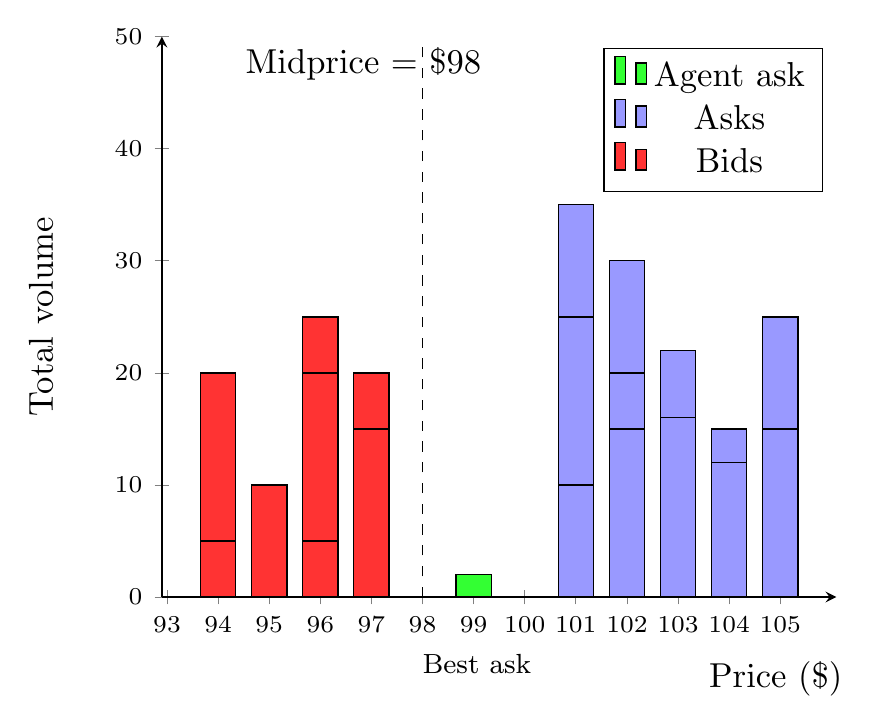} &   
\includegraphics[width=\wid\textwidth]{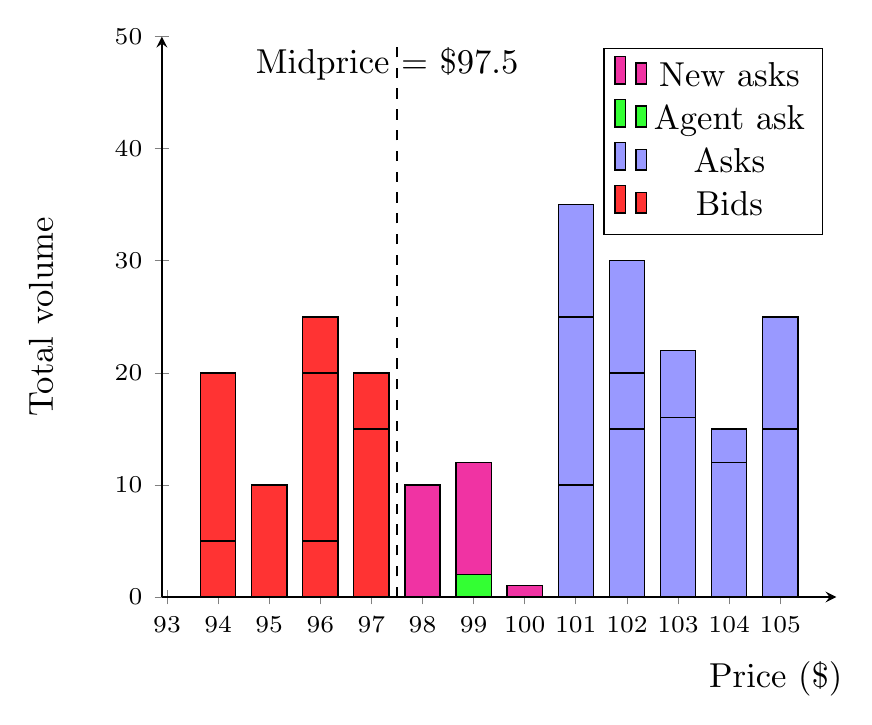} \\
\end{tabular}
\caption{\label{fig:mechanism_attack}
An illustration of a \textit{market-mechanism weakness}. 
Here, the agent places a small aggressive sell order near the original midprice 
(the agent's sell order then becomes the best ask) which puts downwards pressure
on the midprice. 
This is because all of the ask orders generated by \lobgan~-- 
which are assigned a price relative to the best ask -- 
have a lower price assigned to them than if the agent's order was not present.
}	
\end{figure*}

The trained CGAN outputs orders with relative prices called \textit{depths}. For
example, if the CGAN outputs a bid limit order with depth $1$, then it is placed
one tick away from the current touch on the bid side of the book. 
This ensures that the order price distribution is stationary, improving the
performance and convergence of the CGAN compared to training the CGAN using
absolute prices. 
However, as we will show in this section, this rule for assigning prices to
orders can be a exploited by an interacting agent.

A limit order that is placed at a better price than the touch (i.e. the current
best price in the order book) is called an \textit{aggressive} limit order. 
The arrival of aggressive limit orders, the cancellation of orders at the touch,
and the execution of orders at the touch are the ways in which the midprice in a
LOB changes. 
In particular, restricting to the bid side of the book for clarity, the midprice
decreases when sell market orders arrive and are executed, or when the bid touch
is fully cancelled; the midprice increases when aggressive bid limit orders are
placed. 
These same events cause the spread to increase and decrease. 

The adversarial strategy in this section is the following: maintain a single limit order at the
touch on the side of the book that the agent wishes to move inwards. 
For example, we describe the strategy for a case in which the agent wishes to move
the price to move down. 
This strategy is shown in Figure~\ref{fig:mechanism_attack}, where the bottom and
top rows show the market evolution with and without the strategy, respectively. 
In this market, the best ask price is 101\$ (top left chart), thus the agent can
place a new ask limit order at 99\$ being at the touch of the side he wishes to
move inwards (bottom left chart). 
This means that the prices of all new incoming sell limit orders from \lobgan
(which are priced relative to the best ask) are now relatively lower (bottom
right chart) than they would have been had the aggressive order of the agent not
been added to the order book (top right chart). 
In short, the agent's aggressive ask order means that the best ask
is lower than it would have been without and this 
modifies subsequent order placement by \lobgan. 
Then, the agent places an order of a larger size (we chose 300) at a fixed depth
away from the ask touch (to get a better sale price) to accumulate a negative
inventory and profit as the price price goes down further. 

\begin{figure}[h]\centering
\includegraphics[width=\fscale\linewidth]{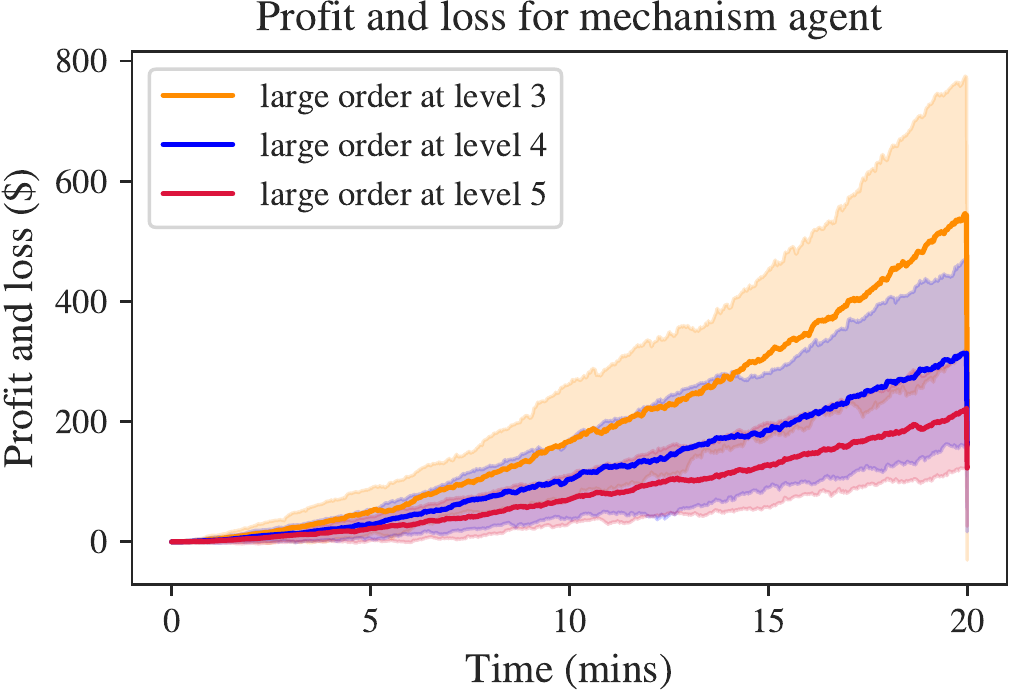}
\caption{The profit and loss for adversarial strategy on the market mechanism  (Section~\ref{sec:mechanism_attack}) for a variety of depths and for a fixed (main) volume of 300.}
\label{fig:mechanism_pnl}
\end{figure}

We recall that also in this case, our adversarial strategy is demonstrating 
a weakness of the \lobgan placing mechanism, which is not representative of a real market. 
Moreover, we never cancel placed orders to make them less likely to be filled.
The resulting strategy is consistently
profitable across a variety of depths for the larger fixed-depth
order (see Figure~\ref{fig:mechanism_pnl}). 

\subsection{Learning liquidity provision with reinforcement learning}

As well as the three hand-crafted strategies in the previous sections, we
trained a simple market-making reinforcement learning agent to act in the
\lobgan environment. 
The state of the agent was simply the agent's inventory and the spread of the
environment,\footnote{By keeping the state of the agent small (in terms of the 
number of features, here just two), the agent
learns more quickly and it doesn't simply learn to overfit the distributional
properties of the data that \lobgan~was trained on.} and the action space was
that considered in \citet{spooner2018market} with one minor alteration and several
extra actions. 
In \citet{spooner2018market}, the agent is allowed to quote symmetrically around
the midprice with a half-spread in $\{1,2,3,4,5\}$; it also has two options in
to skews their orders relative to midprice -- quoting at 1 tick away from the
midprice on one side of the book and 3 ticks away on the opposite side or 2
ticks away from the midprice on one side of the book and 5 ticks away on the
opposite side. 

The first change that we make is to have our market maker agent price relative
to the touch (the best price on the relevant side of the book) instead of the
midprice. 
The second change is to augment these 9 available actions with 6 more: the
ability to post at $1,2$, or $3$ ticks away from the touch on one side of the
book and not quote on the opposite side of the book.

\begin{figure}[h]\centering
\includegraphics[width=\fscale\linewidth, clip]{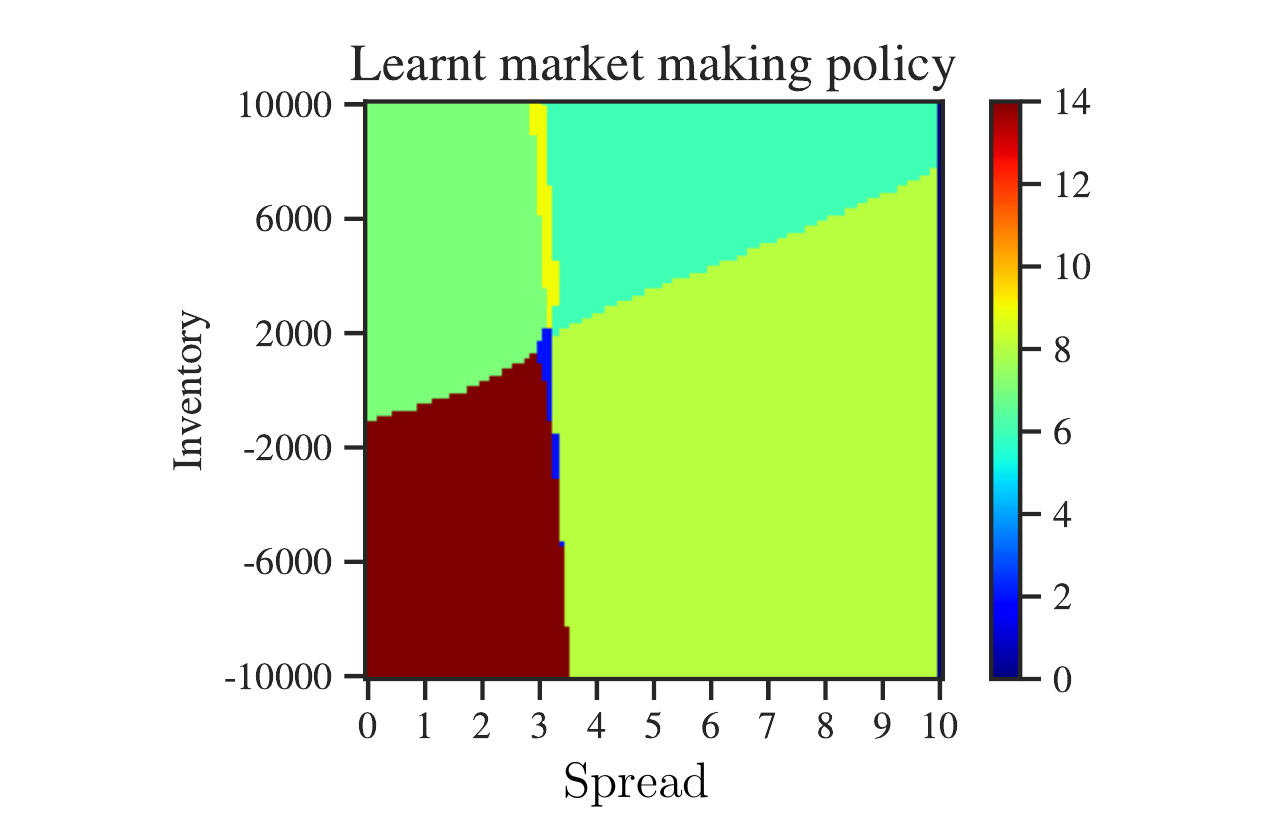}
\caption{The policy learnt by a simple reinforcement learning agent who posts at
the touch. Colours represent the different actions (see Appendix~\ref{sec:rl_actions}).
}
\label{fig:rl_policy}
\end{figure}

After training the reinforcement learning agent, we find that it learns a policy
that is a combination of the strategies found in
Sections~\ref{sec:market_making} and~\ref{sec:attack_imbalance}. 
On the right hand side of the plot -- where the spread is large -- the agent
learns a version of the market making strategy from
Section~\ref{sec:market_making}. 
In the lower left corner of Figure~\ref{fig:rl_policy}, when the agent has a
negative inventory, the agent takes action 14, corresponding to placing a large
order 3 ticks away from the touch on the ask side of book to push the price
down. 
In the top left hand corner, when the agent has a large positive inventory, the
agent learns to place a large buy order 2 ticks away from the touch and a sell
order 5 ticks away from the touch, i.e., it has learnt to emulate the strategy
from Section~\ref{sec:attack_imbalance}.
Thus, the weaknesses of the \lobgan shown in previous sections enable this adversarial strategy, which exploiting the CGAN model, rather than learning a real profitable policy. We recall that, this is an adversarial strategy exploiting the CGAN deep learning model, rather than market manipulation.

\begin{figure}[h]\centering
\includegraphics[width=\fscale\linewidth]{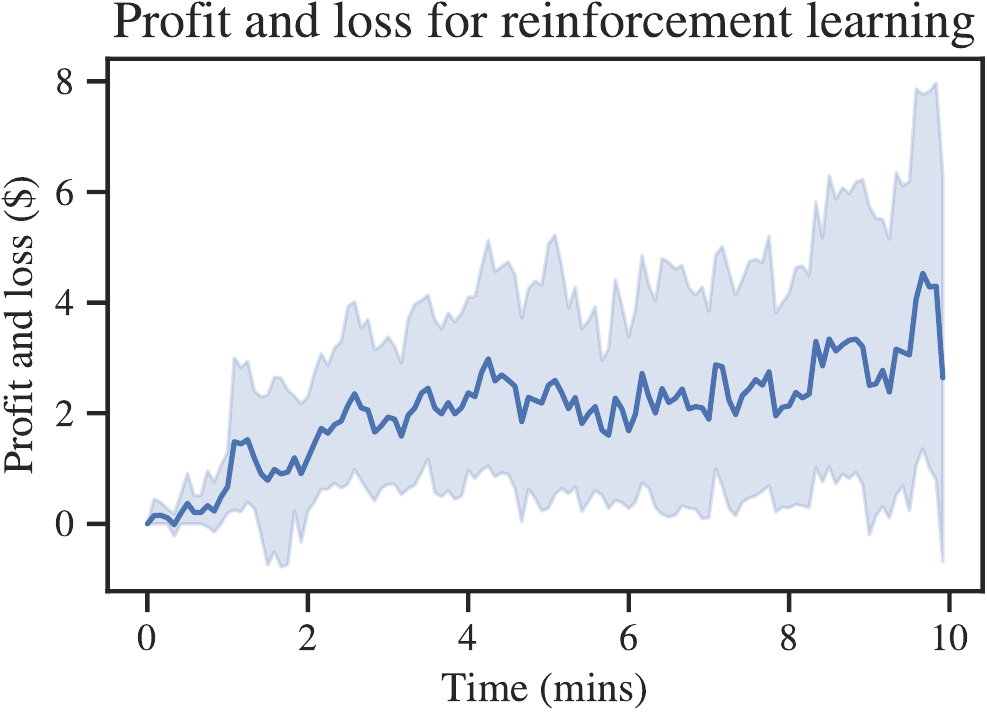}
\caption{The rewards of a reinforcement learning agent.}
\label{fig:rl_rewards}
\end{figure}

\subsection{Insights from the CGAN weaknesses}\label{sec:insights}

From the strategy tested againsts the \lobgan in the previous section, several conclusions emerge,
which we discuss next.

\paragraph{Conservativeness.} The market making strategies are problematic 
because an unsophisticated strategy earns large, consistent profits.
It is important to note that market-making is profitable in real markets,
the main issue is that it should not be so easy to design a profitable trading 
strategy, and such as strategy should not be as simplistic as those in Section~\ref{sec:attack}.
Actually, it is preferable that a simulator be too hard rather than to too easy; that is, we prefer to
be \emph{conservative} when evaluating a trading strategy, and indeed 
this is the typical approach taken in practice.
For example, when backtesting a strategy one would typically assume that trades
incur slippage that hurts profits even if occasionally slippage might help. 
In a similar vein, we would in general like a simulator to provide an
environment in which it is too hard to make profits rather than too
easy~\citep{Pardo08}:
It is common wisdom that strategies practically never outperform their backtests
when traded in reality, and it is desirable to have a ``harder'' simulator that
gives a more accurate picture of expected performance in real markets. 
Thus, our first conclusion is that it is possible to use strategies like ours to
provide a check on how easy it is to produce profit 
and thereby conclude whether or not the generative model simulator 
is suitably ``hard''.

\paragraph{Relative vulnerabiltiy of features.}

The easiest \lobgan~features to exploit were \imb~and \code{Spread}. 
\vol~didn't have as large of an effect on the dynamics of \lobgan as the other
features, as described in Section~\ref{sec:cgan_feature_dependence}.
Among the other \lobgan features, 
$\code{TradeImbalance}$ could be exploited in a similar fashion, 
altering the perception of \lobgan model and generating the desired market regime.
However, creating an adversarial attack on the \tradeimb~requires,
firstly, crossing the spread with market orders, which is costly,
and, secondly, a costly liquidation at the terminal time.

\paragraph{The placing mechanism weakness.} 
This section showed that as well as exploiting the model features, 
the CGAN is also sensible to the relative placing price, which enable 
an aggressive limit order to change the best price on one side of the book (which the CGAN then posts orders in relation to).
We discuss an approach to address the weakness of the placing mechanism in Section~\ref{sec:future_work}.

\section{Improved {\lobgan} models}\label{sec:improved_models}

This section presents various solutions aimed at improving the robustness of
\lobgan and the simulators that are built upon it. 
Based on the the previous section strategies, we can identify four possible
limitations of the current \lobgan-based simulator:

\begin{enumerate}
	\item \textbf{Representativeness} - by using a restricted set of hand-crafted
		features the simulator has a limited view of the financial market which
		could lead to unrealistic behaviour when it conditions on certain market
		regimes. In particular, any predictive feature of the financial market
		that is uncorrelated to the input features of \lobgan 
        will be invisible to it. 
  
	\item \textbf{Overimportance of certain features} - by limiting the number
		of features to a handful of human-interpretable market features, new developed strategies could be ``biased'', i.e., overfitted, 
	on these \lobgan features, and not being profitable in a real market but just on the \lobgan environment. In fact, by only including a limited number of features there is a much higher
		risk of the model becoming over-reliant upon them, thereby facilitating
		the bias in developed strategies.
  
	\item \textbf{Interactiveness} - independently from the chosen features,
		\lobgan is training in a closed-loop: during the training the model
		learns to generate orders from ground truth past states and novel states
		induced by its previous orders. While this training alleviates
		compounding errors (i.e., it reduces the possibility that previous
		suboptimal decisions induce unseen states and failures), the inclusion
		of an interactive agent (i.e., trading agent) during training 
		will almost certainly be able to create novel or adversarial states that are 
		not seen in the current training of \lobgan. In short, lack of 
		interaction with training agents during training is arguably a limitation
		of \lobgan.

	\item \textbf{\lobgan order placement} - by placing orders relative to dynamic
		market features (i.e., best bid/ask) \lobgan enables a trading agent
		to manipulate the next placed \lobgan orders by altering these features,
		as seen in Section~\ref{sec:mechanism_attack}. Thus, using the touch for relative order
		placement is a limitation of \lobgan.
        
\end{enumerate}

Therefore, although hand-crafted features are useful for development and
explainability, they are the first limitations we should address to improve
simulations. 
We now propose three improved {\lobgan} models by addressing the
\textit{representativeness} and \textit{overimportance of features} limitations discussed in
the previous section. 
We discuss a solution to \textit{interactiveness} and \textit{order structure}
in Section \ref{sec:future_work} in the form of a new training procedure and
placing mechanism, respectively.

\subsection{Representativeness}\label{sec:repr} 

The {\lobgan} model uses a set of hand-crafted features introduced in Section
\ref{sec:features_cgan} to create its own representation of the financial
market. 
However, this representation can be limited and cause misleading
behaviour under certain market regimes. 
Learning the market dynamics from raw orderbook observations would be ideal, but
it is difficult and computationally expensive in general~\cite{wu2022robust},
and this is possibly further aggravated by the adversarial training of GANs; we
discuss this further as a future research direction in Section~\ref{sec:future_work}.
Instead, here we show how to improve the \textit{Representativeness} by introducing a
new {\lobgan} model that augments the features detailed in
Section~\ref{sec:features_cgan} with the following new features. 
This allows the CGAN to
have a more detailed view of the current market state. The features that are
added to this new version of \lobgan, which we refer to as \lobgan \texttt{v1}
(with the original \lobgan being denoted by \lobgan \texttt{v0}), are:

\begin{itemize}
	\item the \textit{order book imbalance} for $n=10$  (i.e., for the top 10 levels of the book);
    \item the \textit{total volume} at the top 10 levels of the book;
    \item the \textit{midprice percentage return} over the last $\Delta \in \{1, 5\}$ minutes;
    \item the \textit{trade volume imbalance} over the last 10 minutes;    
    \item the \textit{total execution volume} over the last $\Delta \in \{1, 5, 10\}$ minutes;
\end{itemize}
where the \textit{total execution volume} of window size $\Delta$ minutes at time $t$ is defined by  
\begin{equation}\nonumber
    {\code{TradeVolume}_\Delta(t)} = \sum_{t-\Delta \leq s_i \leq t}\left|V^{\rm trade}_{s_i}\right|,
\end{equation}
where $s_i$ is the time at which the trade occurs, and $V^{\rm trade}_{s_i}$ is the signed volume of the trade.
We also remove the two \textit{$n$-event midprice percentage return} features
to reduce the overlap/correlation of features (given that we now include time-based
midprice returns). 

\begin{figure*}[t]
\centering
\includegraphics[scale=0.45,trim={0 0 0 0}, clip]{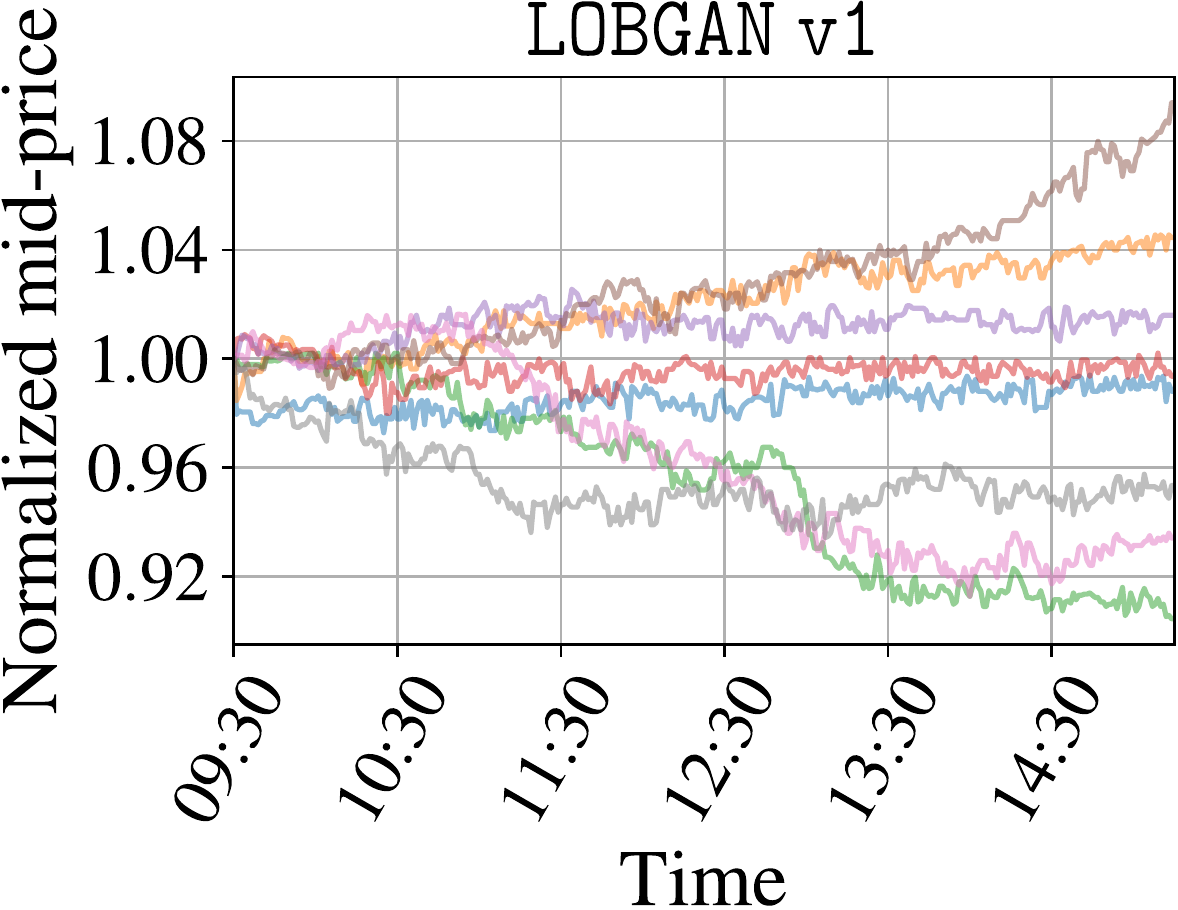}
\hfill
\includegraphics[scale=0.45, trim={0 0 0 0}, clip]{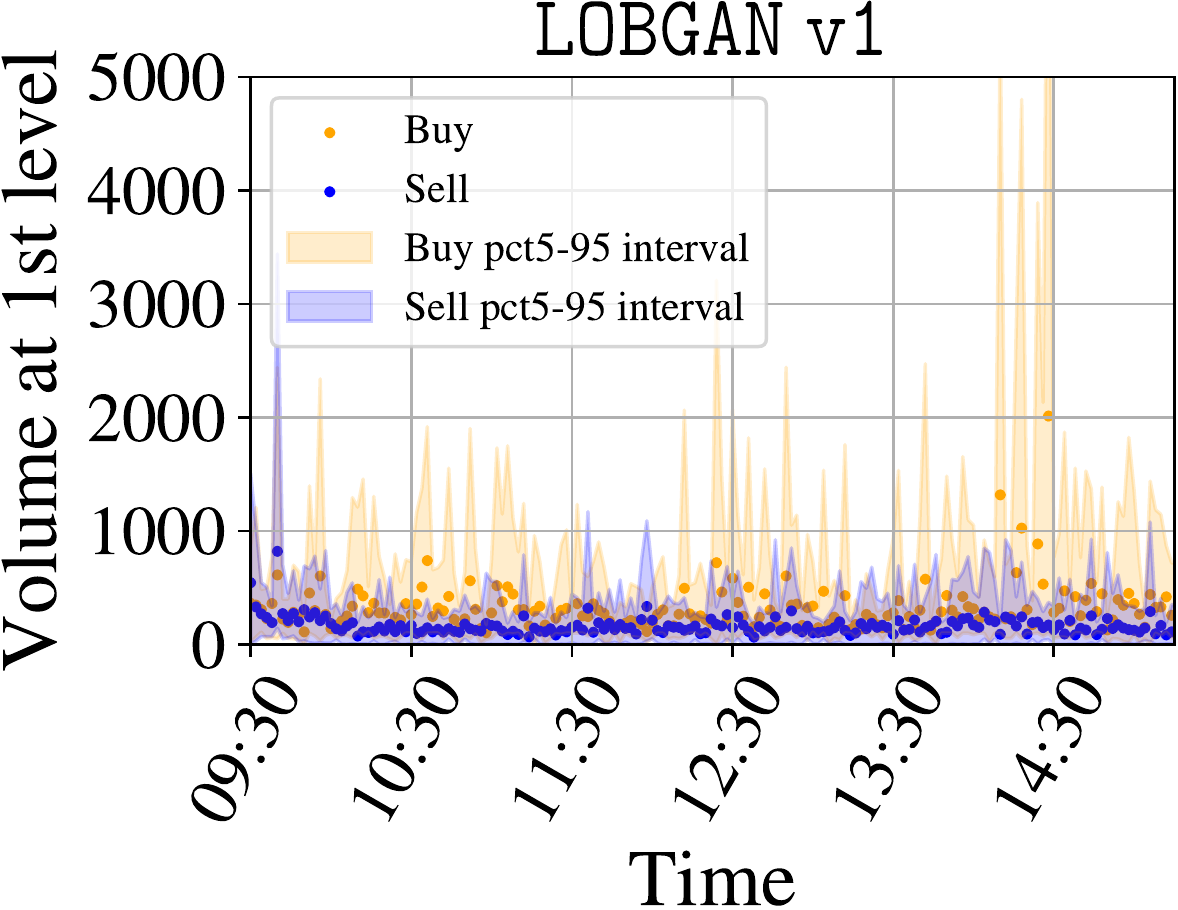}
\hfill
\includegraphics[scale=0.45,trim={0 0 0 0}, clip]{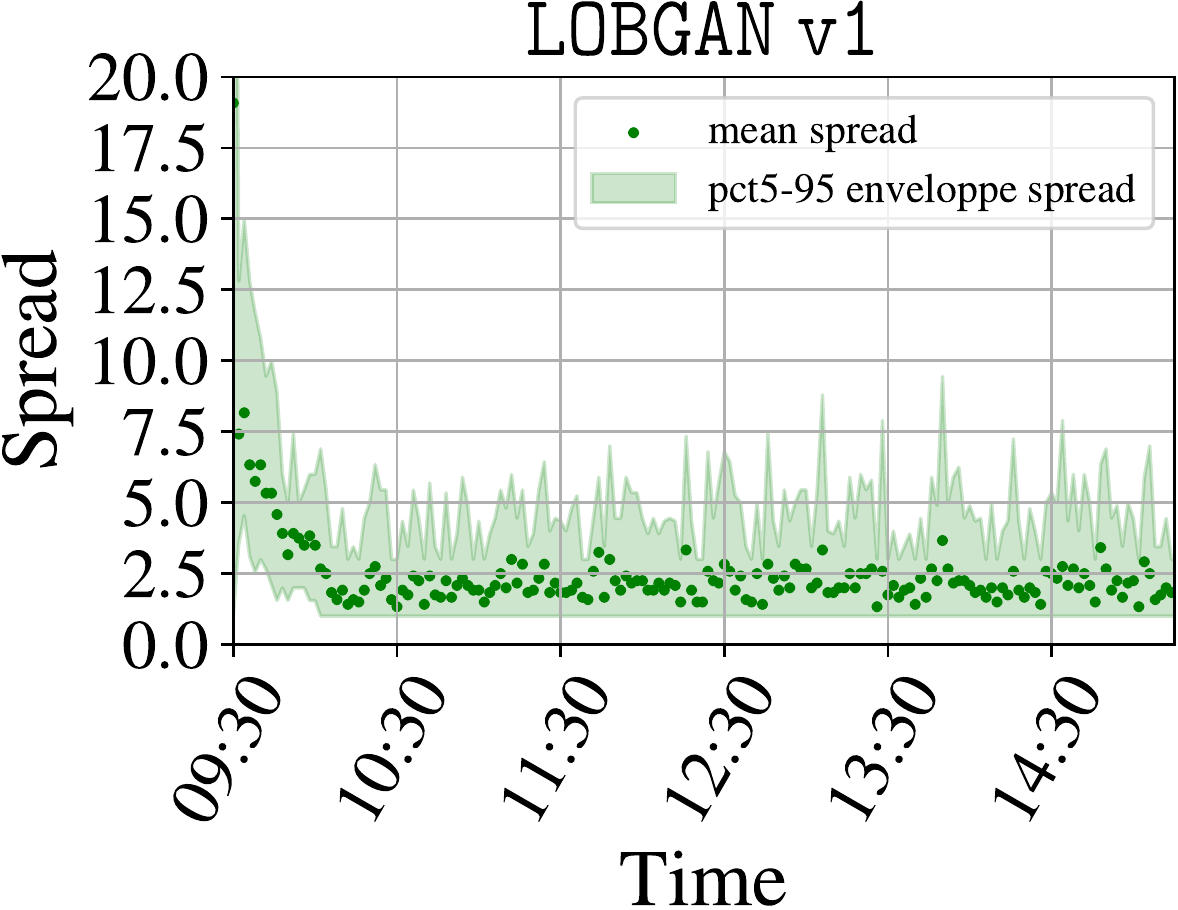}\\
\vspace{0.1cm}
\includegraphics[scale=0.45,trim={0 0 0 0}, clip]{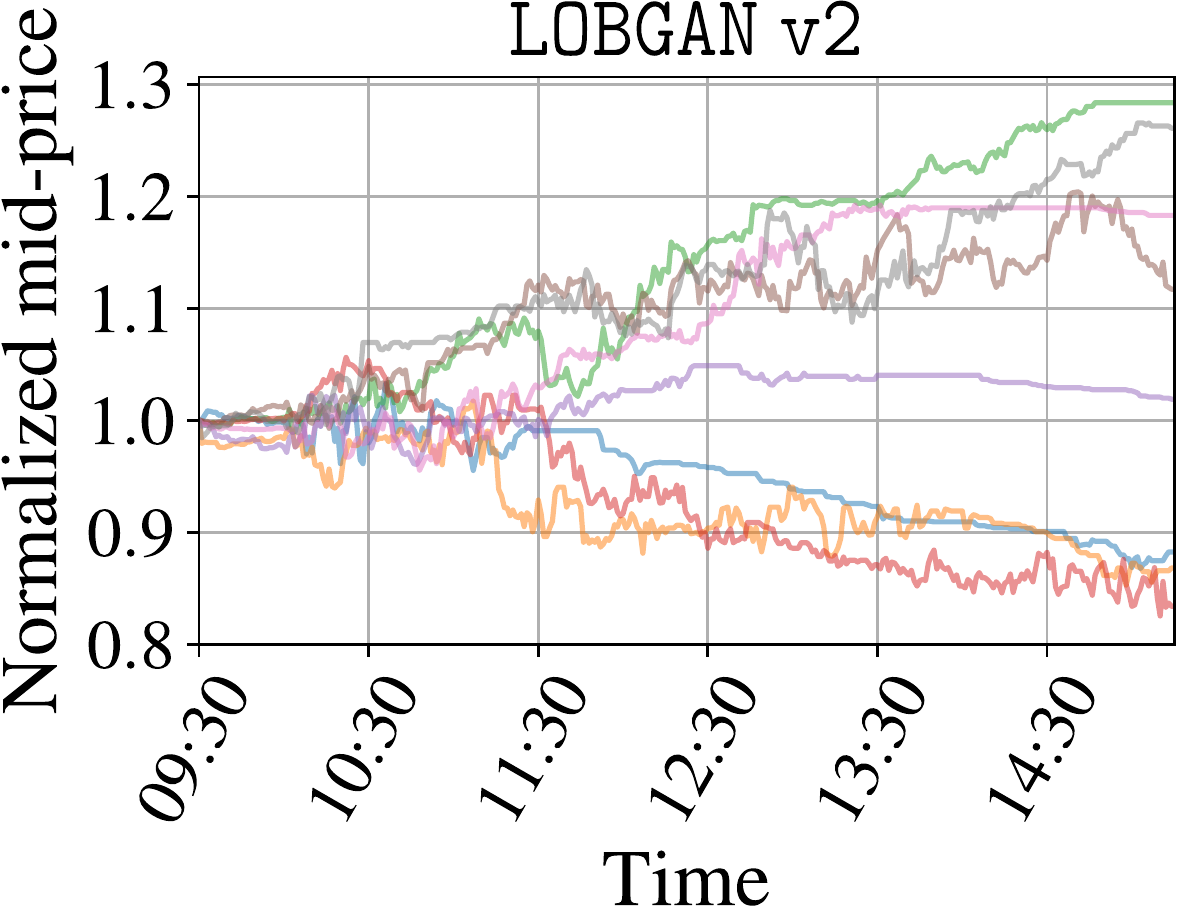}
\hfill
\includegraphics[scale=0.45, trim={0 0 0 0}, clip]{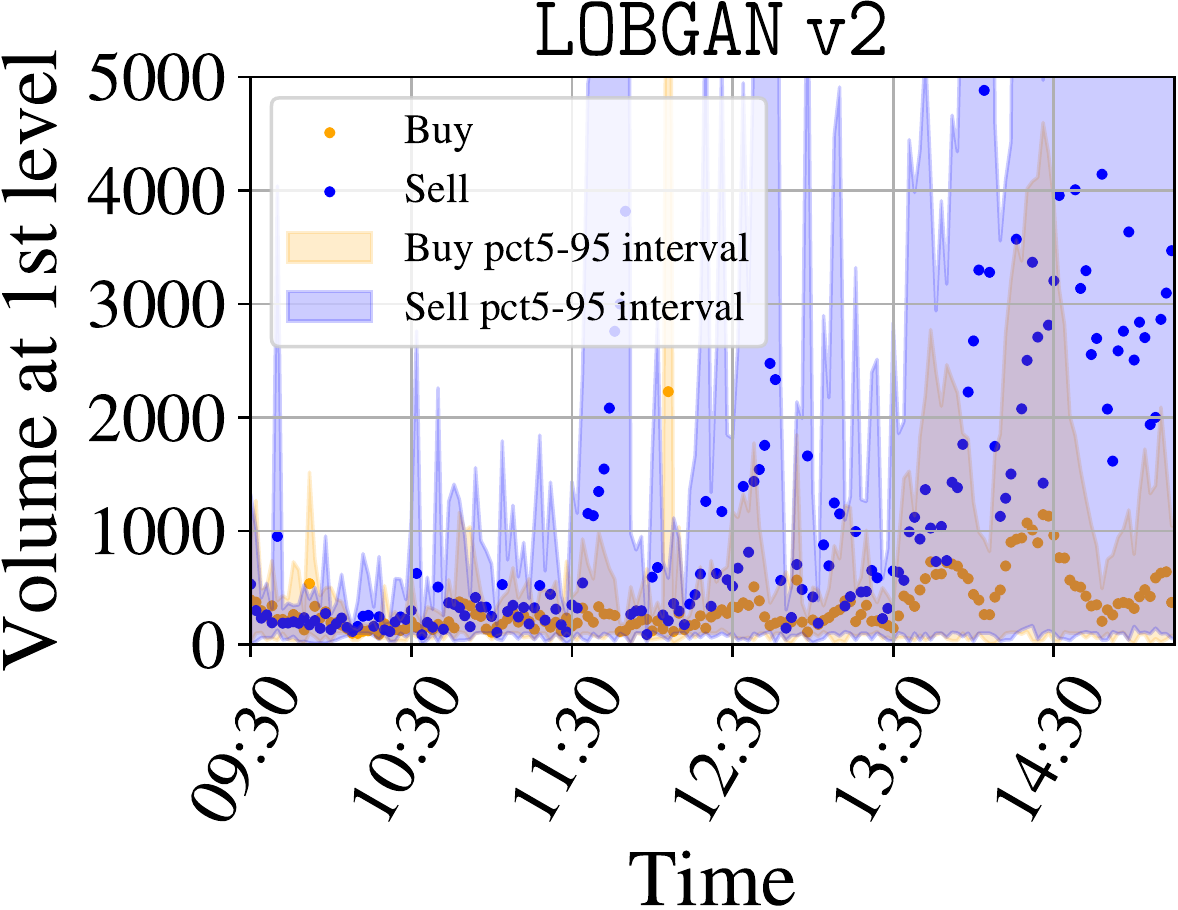}
\hfill
\includegraphics[scale=0.45,trim={0 0 0 0}, clip]{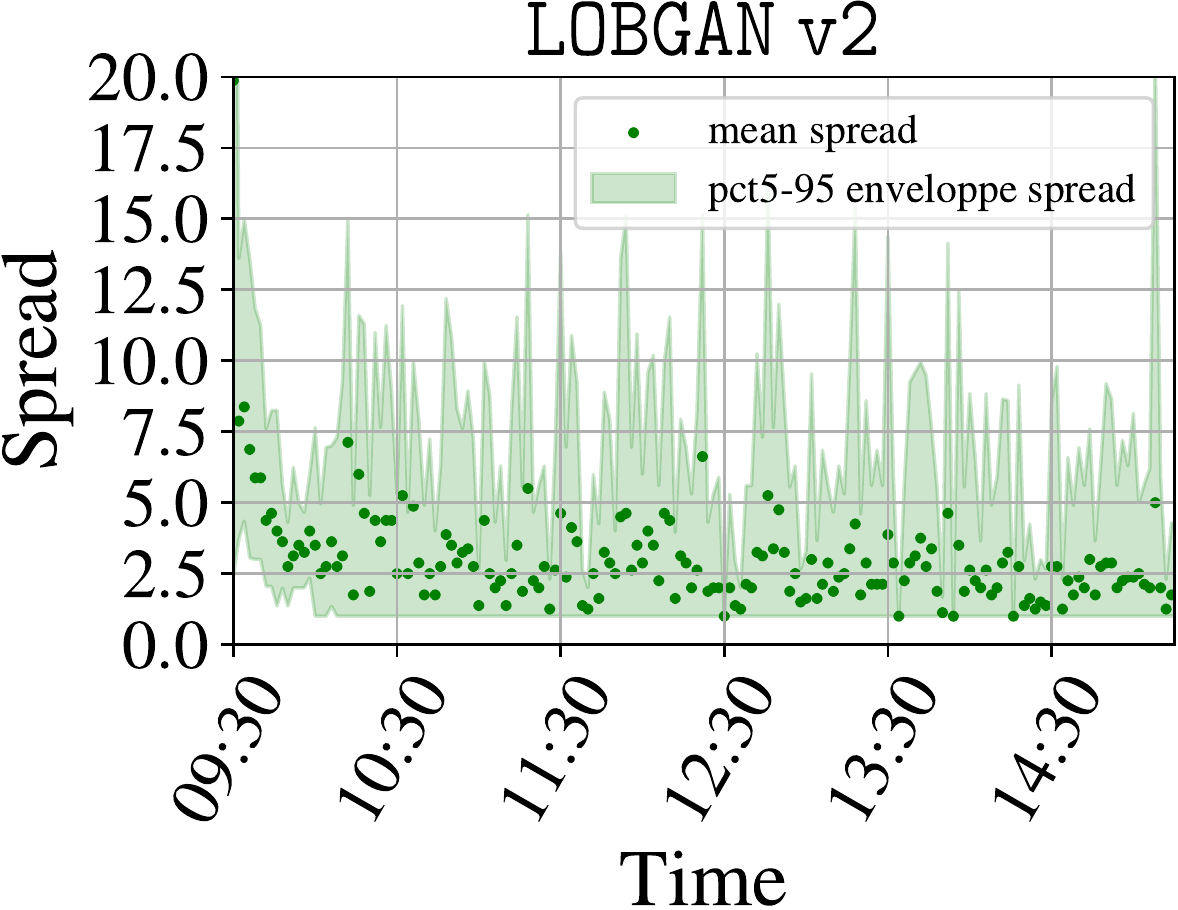}\\
\vspace{0.1cm}
\includegraphics[scale=0.45,trim={0 0 0 0}, clip]{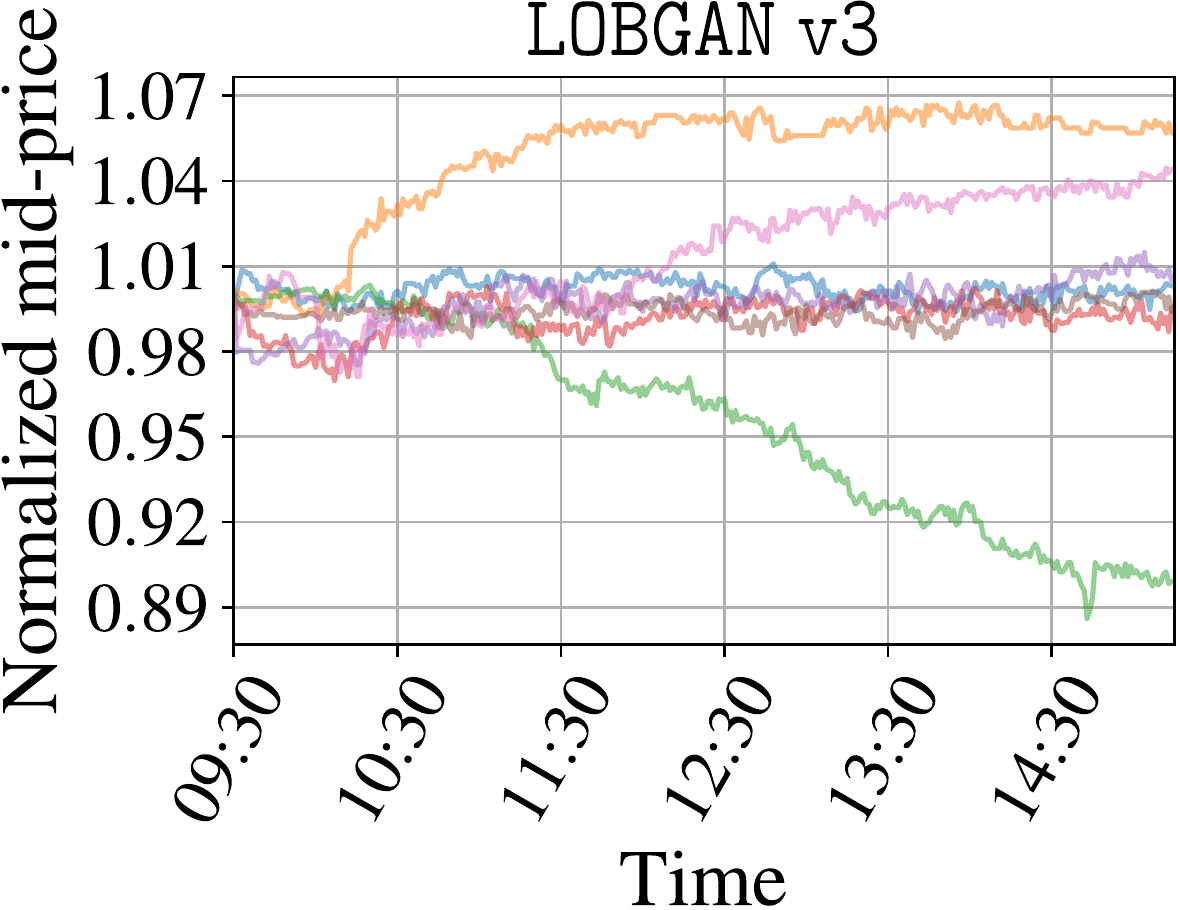}
\hfill
\includegraphics[scale=0.45, trim={0 0 0 0}, clip]{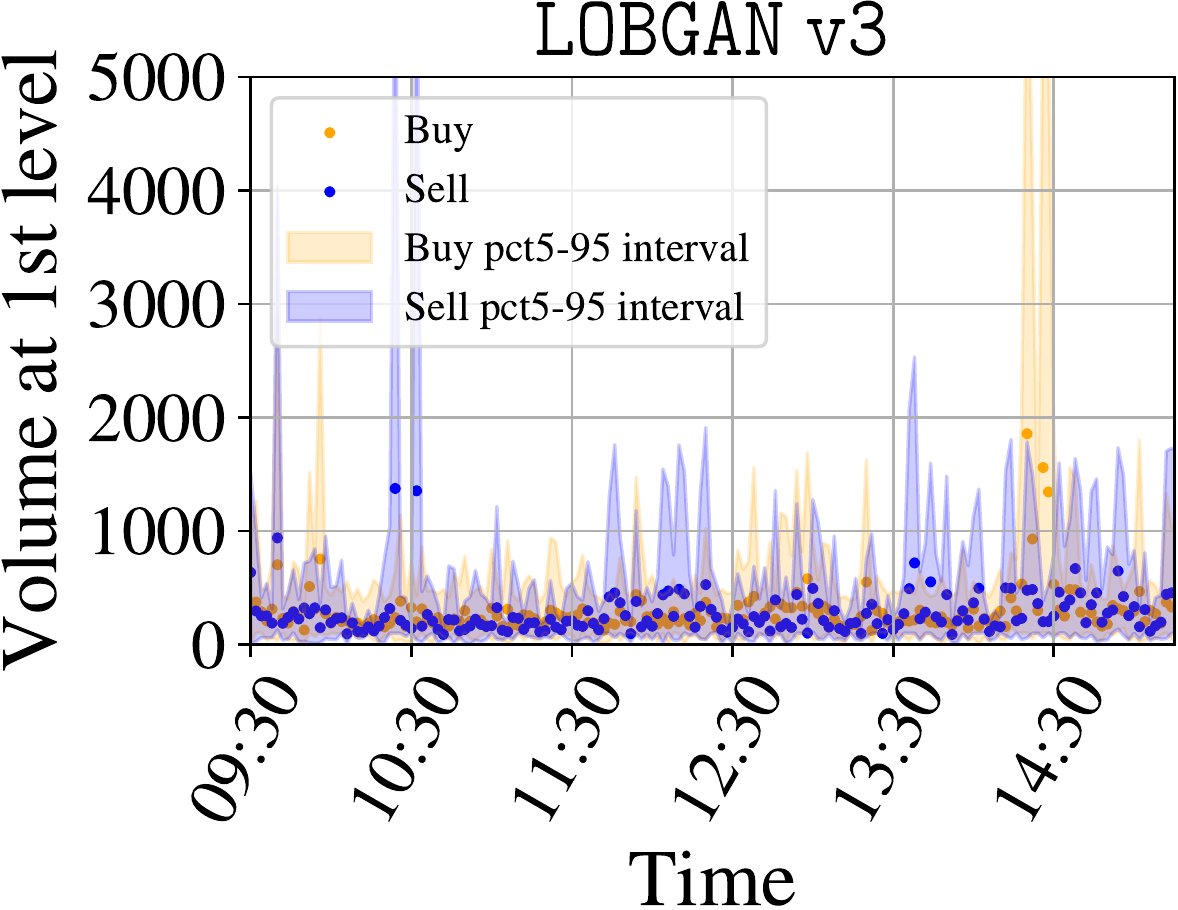}
\hfill
\includegraphics[scale=0.45,trim={0 0 0 0}, clip]{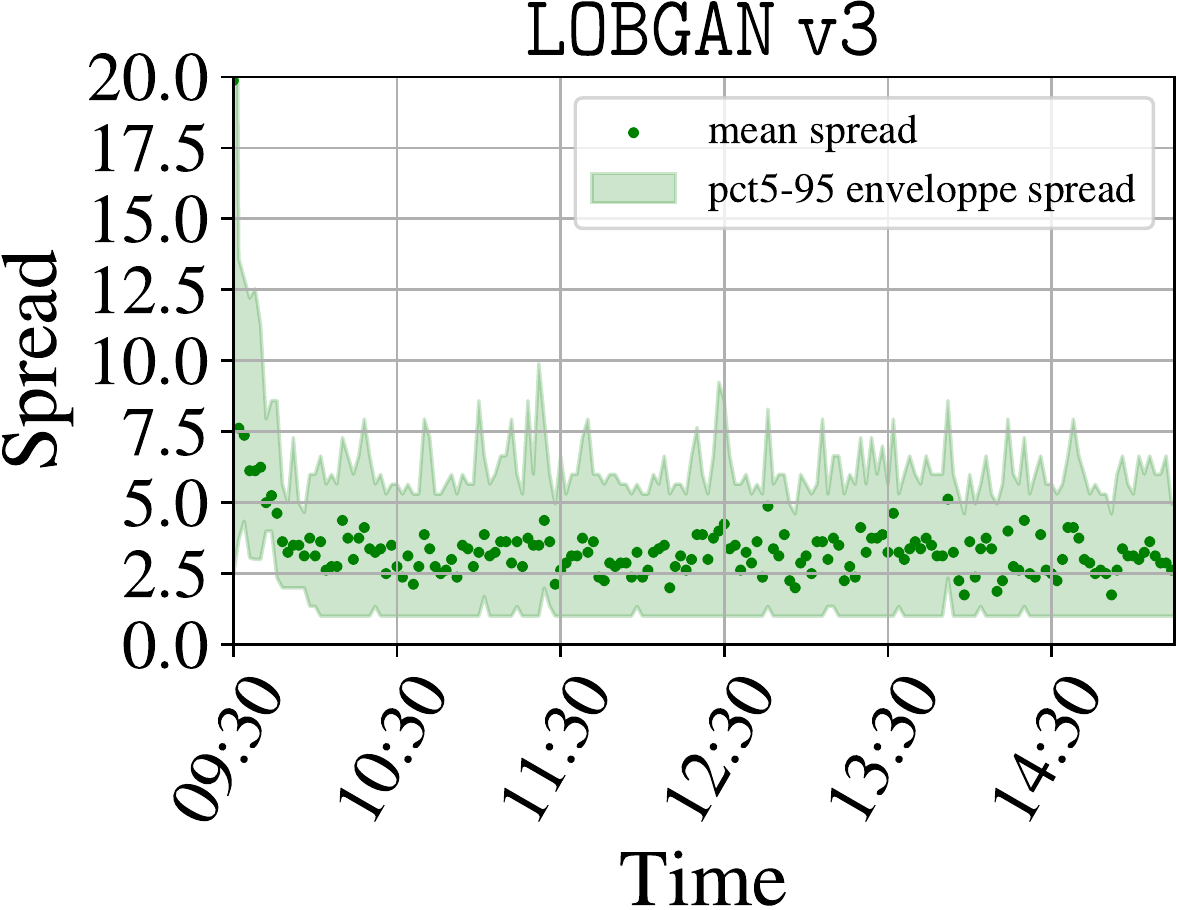}
\caption{Stylized facts of \lobgan \texttt{v1}, \texttt{v2}, and \texttt{v3} }\label{fig:stylized_lobv1v2v3}
\end{figure*}

\paragraph{Realism of \lobgan~\texttt{v1}.}
In the first row of Figure \ref{fig:stylized_lobv1v2v3} we evaluate the realism
of the new proposed model by comparing its stylized facts against those of real
market and \lobgan~\texttt{v0} in Figure~\ref{fig:stylized_lobgan}. 
We note that \lobgan~\texttt{v1} achieves similar realism for the
\textit{volume} and \textit{spread} time-series, with a clear resemblance between
the simulated and real time series. 
Interestingly, the new model has slightly better \textit{price series}: they
exhibit significant diversity and better symmetry. 
We believe the new features, and especially the time-based \textit{midprice
percentage return}, are responsible for the increased realism (in isolation) of
the price series.

\paragraph{Robustness of \lobgan~\texttt{v1}.}
As well as improving realism in isolation, it is
also the case that \lobgan~\texttt{v1} is more robust to the main adversarial strategies 
we introduced in Section~\ref{sec:attack}. 
In Figures~\ref{fig:limit_up_model_comparison} and
\ref{fig:market_making_model_comparison}, we can see that \lobgan~\texttt{v1}
makes improvements in terms of robustness to both the \imbI~strategy and 
the market making strategy. 
While these strategies can still be employed, they will be more expensive
and less profitable: the first strategy just about
breaks even on average, and the second one loses money.  

\subsection{Feature overimportance} 

The introduction of new hand-crafted features in the last section partially 
solved the overly reliance on some features that was outlined in 
Section~\ref{sec:cgan_feature_dependence}. 
In particular, \lobgan \texttt{v1} largely neutralised the imbalance overreliance shown in 
Section~\ref{sec:attack_imbalance}. (see Figure~\ref{fig:limit_up_model_comparison}).
However, in general it is by no means guaranteed that adding new features will 
preclude the model overly relying on just few of them. 
In this section we further investigate strategies to mitigate over-reliance on
certain features.
For simplicity, we continue to focus on \imbI and the related adversarial strategy.
Starting from \lobgan \lobgan \texttt{v1}, we investigate
two approaches to further mitigate potential over-reliance on \imbI.

\subsubsection{Removing \imbI}
\label{sec:remove_imb1}

We first investigate the effect of simply removing \imbI from the feature 
set for \lobgan  \texttt{v1}.

\paragraph{Realism of \lobgan~\texttt{v2}.}
In the middle row of Figure \ref{fig:stylized_lobv1v2v3}, we evaluate the
realism of our second proposed model. 
We recall this model is a first \textit{na\"ive} attempt to show the advantages
and disadvantages of just removing relevant features. 
The figure shows less realism on both \textit{volume} and \textit{spread} with
respect to the real data: volume accumulates over time while spread has higher
variance. 
Most importantly, the proposed model shows \textit{price series} with
strong trends, moving more than 20\% compared to the market open. 
\imbI is a strong indicator of market direction~\citep{bouchaud2018trades}, and
when it is included as a feature, the CGAN uses it to generate more realistic
time-series than it does without~it.

\paragraph{Robustness of \lobgan~\texttt{v2}.}
The goal of \lobgan~\texttt{v2} was primarily to improve the robustness of the
CGAN to the adversarial attacks on \imbI. 
As we can see in Figure~\ref{fig:limit_up_model_comparison}, this goal was
clearly achieved. 
In particular, the adversarial strategy losses its control over the
midprice dynamics. 
Furthermore, as illustrated in Figure \ref{fig:market_making_model_comparison},
\lobgan~\texttt{v2} is also robust to market-making strategies. 
They make a slight loss and have a high variance, both undesirable from a risk
and reward perspective.

\subsubsection{Randomising levels used for the imbalance feature}
\label{sec:randomise_imb} 

We now introduce a more sophisticated attempt to reduce the model dependency on
\textit{order book imbalance at level 1} by using a randomised version of this
feature. 
We remove the \textit{order book imbalance} at levels 1, 5, and 10, and
we add the $\chi$-\textit{level order book imbalance} for a random variable
$\chi$. 
Here, we simply take $\chi$ to be uniformly distributed on the set
$\mathcal{S}=\{1,2,3,4,5\}$ so that at each time period, the agent randomly
samples one of these values to use for that feature. 
Before training, we pre-sample from this distribution so that the
$\chi$-\textit{level order book imbalance} does not change when it is reused as
a data point.

\paragraph{Realism of \lobgan~\texttt{v3}.}
The bottom row of Figure \ref{fig:stylized_lobv1v2v3} shows the stylized facts
of our third model and second attempt to feature overimportance. 
The figure clearly shows an improvement on realism for both \textit{volume} and
\textit{spread} with respect to the previous attempt (the middle row in Figure
\ref{fig:stylized_lobv1v2v3}). 
The stylized facts resemble more those of real data even though the generated
\textit{price series} still have stronger trends, moving around 10\% compared to
the market open. 

\begin{figure}[h]\centering
\includegraphics[width=\fscale\linewidth]{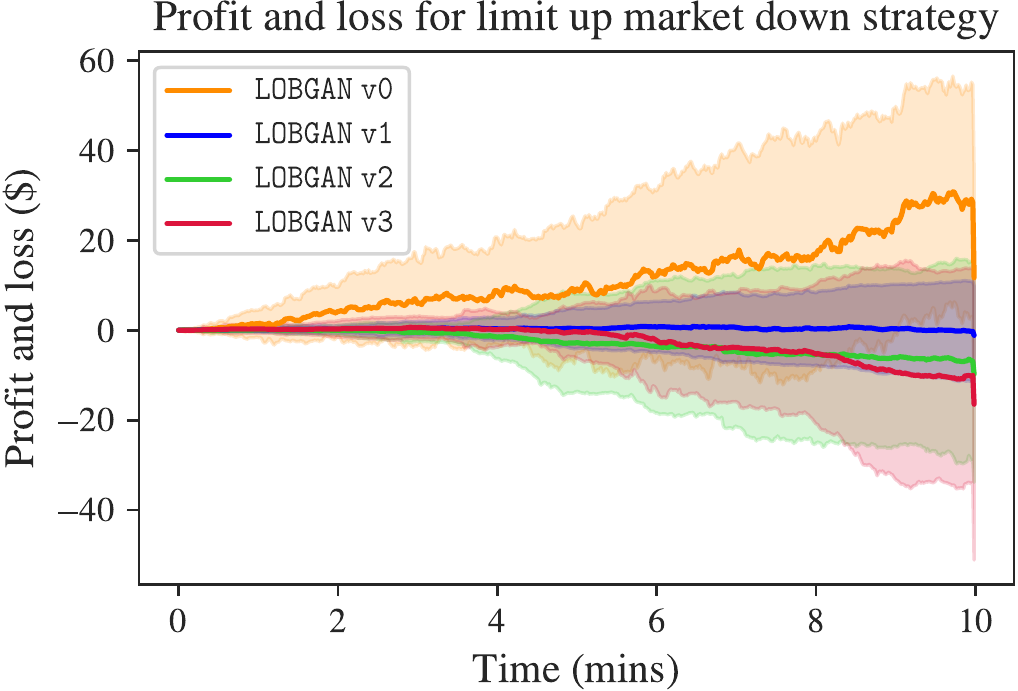}
\caption{The profit and loss trajectories for the \imbI strategy from Section~\ref{sec:attack_imbalance} when used on the new models. 
\lobgan \texttt{v0} is the original version of \lobgan~which is studied throughout Sections 2-5, \lobgan \texttt{v1} is modified to use the features outlines on Section~\ref{sec:repr}, \lobgan \texttt{v2} uses the features in Section~\ref{sec:remove_imb1}, and \lobgan \texttt{v3} uses the features in Section~\ref{sec:randomise_imb}.
}
\label{fig:limit_up_model_comparison}
\end{figure}

\begin{figure}[h]\centering
\includegraphics[width=\fscale\linewidth]{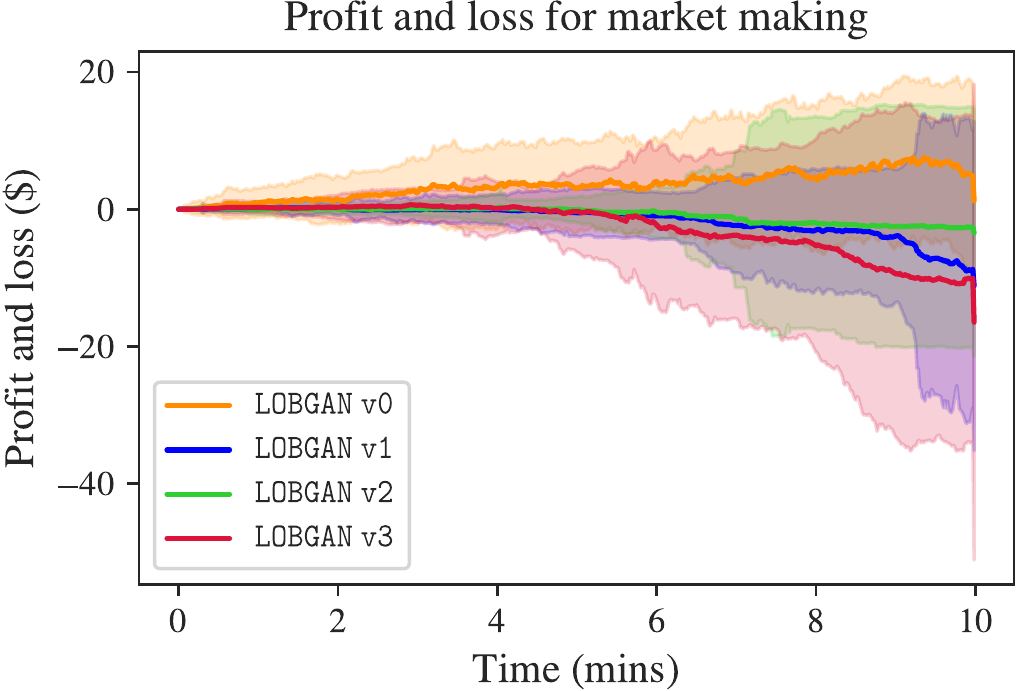}
\caption{The profit and loss trajectories for the market making strategy described in Section~\ref{sec:market_making} when used on the new models.
}
\label{fig:market_making_model_comparison}
\end{figure}

\paragraph{Robustness of \lobgan~\texttt{v3}.}
As shown in Figures~\ref{fig:limit_up_model_comparison} and
\ref{fig:market_making_model_comparison}, we can see that \lobgan~\texttt{v3}
improves upon all of the previous versions of \lobgan in terms of robustness to
both the \imbI adversarial strategy and the market-making strategy. 
Both of these na\"ive trading strategies lose a substantial amount of money on
average and have high variance. 
When considered alongside the improvements in realism over \lobgan~\texttt{v2},
this makes it a clear improvement. 

\subsection{Recommendations for using \lobgan or similar models in practice}
\label{sec:recommendations}

In Section~\ref{sec:future_work}, we provide a roadmap for future research and
development of conditional generative models for LOB environments.
In this subsection, we discuss practical recommendations for how best to use
\lobgan or similar models, given the current state of the art, and in particular
the insights of this paper.
We highlight three main recommendations:

\begin{itemize}

\item \textbf{Training \lobgan with diverse feature sets.}
Our work and prior work has demonstrated that \lobgan models can be generated
with a variety of different features sets, which can result in models that 
have realism and robustness properties.
We recommend to both explore further predictive features that are not highly
correlated with the features already used in versions \code{v0}-\code{v3} of
{\lobgan} and to train a variety of models using different subsets of features. 
The overarching goal is to choose as many uncorrelated and predictive features as possible. 
This will improve realism, as \lobgan is able to form a more nuanced view of the
financial market, and robustness, as \lobgan should then be reliant on each of the
features it uses, rather than overreliant on one or two features, which could
then be easily exploited.
It is further useful to have a wider pool of models to select from (the models
trained with different feature spaces) and, as we discuss next, it can make good
sense to use multiple models.
The features used for \lobgan were handcrafted; we discuss in
Section~\ref{sec:future_work} the, already mentioned, challenging but promising 
direction of using raw market data, rather than hand-crafted features,
to allow the deep conditional generative model to learn its own features.

\item
\textbf{Using multiple models.} 
By using multiple models, one can improve the robustness of trading strategy evaluation.
A natural way to use multiple models is to run a strategy independently in a simulator using each of these models, 
and then use the distribution of performance across the
models to evaluate the strategy.
The evaluation could use a variety of statistics of this distribution; for example,
a worst-case approach would use the worst performance across models.

\item
\textbf{Use existing trading agents for calibration/model selection.}
If the user has access to historical true trade data from a variety of agents,
such as execution and market making agents, then this data could be used to calibrate 
\lobgan to make sure that it gives similar profit as that achieved by these
test agents in historical live trading. 
While this will almost never by feasible for an academic project, it would be 
very natural for a commercial trading entity.
 
\end{itemize} 

\section{Related work}
\label{sec:relatedw}
\paragraph{Conditional generative models.}
Conditional generative adversarial networks (CGANs) were introduced
in~\citet{mirza2014conditional}, where they were demonstrated and evaluated
using the MNIST dataset of images of handwritten digits.
In general, GANs and CGANs have mainly and most prominently been used the
context of image generation tasks, but they have also been explored for
generating a wide variety of other types of data, including tabular
data~\citep{XuSCV19} and time series data~\citep{EstebanHR17}.
Important early works on GANs include~\citet{arjovsky2017wasserstein} (who
introduced the Wasserstein GAN used by Stock-GAN~\citep{LiWLSW20} and
\lobgan~\citep{ColettaMVB22}) and \citet{gulrajani2017improved}, who made
fundamental contributions to improving Wasserstain-GAN training such as gradient
penalties (used in the training of \lobgan).
While conditional generators for LOBs have so far used CGANs, other
important types of generative model exist.
We briefly mention Conditional Variational Autoencoders (CVAEs), which have been
used to for generating lower-frequency financial data (such as hourly or daily
data as opposed to LOB data), e.g.,~\citet{BHLAW20}.
Other generative model approaches that have been applied to time series include
normalizing flows~\citep{RasulSSBV21} and denoising diffusion
models~\citep{ho2020denoising,RasulSSV21}.
It is an interesting direction to explore what these other approaches can offer
for LOB simulation, although the feature exploitation demonstrated in
this paper would likely apply to any conditional model if explicit care to avoid
it is not taken. We discuss denoising diffusion models further in
Section~\ref{sec:future_work}.

\paragraph{Reinforcement Learning.}
\begin{itemize}
\item \emph{Adversarial Reinforcement Learning.}
In ``Robust Adversarial Reinforcement Learning'' (RARL), during the training of
an agent in an RL environment, a second adversarial agent is introduced into the
environment system with opposite rewards, and this adversary is given (limited)
control over (e.g., the transitions of) the environment~\cite{PintoDSG17,spooner2020robust}.
This is closely related to what we propose in Section~\ref{sec:future_work},
namely introducing trading agents during CGAN training.
\item \emph{Learning to simulate.}
There are existing works that use RL in order to learn a simulator, such
as~\citet{RuizSC19}. 
However, these works tend to look at settings where there is a downstream task
that the simulator will be used for, such as image classification, where
accuracy in that task can be used as a reward function. 
In our setting, part of the challenge is that our desiderata for our simulator
are complex and multi-faceted, and to apply RL a key challenge would be design
of the reward function, and dealing with the fact that it would be a
multi-objective problem.
\end{itemize} 

\paragraph{Agent-based models.} 
For a long time, agent-based models (ABMs) have offered the promise of reactive
financial and economic market simulation. 
However, due to a combination of the lack of agent-level historical data, and
the intrinsic difficulty of calibrating ABMs, they are not currently a viable
option for realistic strategy evaluation.

ABMs in principle offer reactivity, similar to conditional generative models,
and in contrast to backtesting (market replay).
In real markets the number of agents is 
huge, and their individual historical actions are not explicit in the historical data, which is anonymous. 
Thus, modelling and calibrating agent-based models (not just in finance, but actually in general) is 
a serious challenge (see, e.g.,~\citet{BaiLBV22,Platt20,LRS18,PG17,coletta2023k}).
While ABMs have been shown to be useful for investigating structural properties
of financial markets \cite{Outkin2012AnAM}, they are difficult to use in
practice for testing trading strategies where a high precision of calibration is
required.
We do note that a combination of approaches is possible, for example, where a
conditional generative model is combined with distinct agents -- in this paper we
investigate the case of various single trading agents operating within a CGAN
environment, but one could include more agents in such an environment
simultaneously.

\section{Future work}\label{sec:future_work}

In this section, we highlight some promising directions for future work.

\paragraph{Using raw market features.} Existing conditional generative models
for LOBs, including \lobgan, use hand-crafted features.
It is an open challenge to effectively train a conditional model using 
raw market features. 
Given the success of deep learning in learning  representations from raw
pixels of images, this does seem like a challenge worth pursuing at this
point in time. 
One piece of work in this direction is~\citet{zhang2019deeplob} which uses
convolutional layers to extract features and create a representation of the
ongoing market using raw order book snapshots.  
However, we found in some preliminary experiments that the training approach
used by \lobgan does not work immediately with raw features, perhaps because
the representation learning that is needed is too challenging in the adversarial
training setup.
We suspect that new innovations
may be needed in terms of a sophisticated training procedure, or an improved
conditional generator architecture, if we are to effectively use raw orderbook
features in a conditional generative LOB model.
For instance, (denoising) diffusion models~\citep{ho2020denoising,RasulSSV21}
could be a good candidate for a different type of generative model
that may be good at processing raw orderbook data, thanks
to their more stable training process and their remarkable results in computer
vision handling high-resolution data. 
However, to simulate LOBs we require to generate tens of thousands of orders which may be computationally expensive for diffusion models (because each sample is generated through a high number of iterations/steps~\citep{croitoru2022diffusion}).

\paragraph{An automated training approach that includes RL-based adversarial attacks.} We
	demonstrated how interactive realism and corresponding adversarial attacks on the
	features and mechanism of \lobgan could be used to build better generative
	models. It is appealing to try and systematize this process, for example, by
	using reinforcement learning (as we did \emph{post-training}) to create adversarial agents \emph{during}
	the CGAN training process. A challenge here is to effectively balance
	the CGAN's training objective between minimising the rewards of adversarial agents,
	and maximising realism in isolation. 

\paragraph{Metrics for model selection.} While a fully automated model selection procedure is
	desirable, it is arguably some way off. Firstly, human input into determining realism as
	in Section~\ref{sec:realism} currently brings an important sanity check to
	the process, and, secondly, there are so many aspects to realism that the
	required step of aggregating these aspects into single realism scores
	requires much more research. 
	Still, this is certainly an important direction and worthy of further
	research as it will open up many possibilities. 
	For example, with a realism
	in isolation metric and a metric for interactive realism one can explore
	whether there is a trade-off between these type of realism. 
	With a metric
	for model quality, which would presumably combine metrics for realism in
	isolation and interactive realism, one could include an automated model
	selection step within the training process. 
	A concrete example of this would
	be to use this metric as a fitness function within a population-based
	evolutionary method for constructing models.

\paragraph{Order placement mechanism.}  
We demonstrated that a fairly simply, yet effective, adversarial strategy can exploit the
order creation mechanism of the \lobgan-based simulator: orders are placed
relative to the best ask/bid. 
We envision the use of a more robust notion of
price to solve this problem: new orders will instead be placed according to a
more sophisticated notion of current price that the model could compute.
This notion should be more resilient to temporary and exogenous trading orders,
than the touch, or even midprice.
Finding such a notion of price is an interesting research direction.
We note, however, that adding sophistication here will increase the 
complexity of the training process.
+
\paragraph{Wider applicability of our work beyond LOBs.}
The use of conditional generative models for simulation has widespread applicability,
and is by no means only relevant for finance.
As three examples, CGANs have been applied for simulation of flows in
aerodynamical systems~\citep{ChenGXCFWW20}, fuel sprays in combustion
engines~\citep{AtesKOKB23}, and sensors in autonomous
vehicles~\citep{ArnelidZM19}.
While some of our work is undoubtedly specific to LOBs, and other
parts to economic/financial systems more generally, significant parts such as 
the distinction between realism in isolation and interactive realism, and our 
techniques for exploring the conditioning of a generative model, apply in general.
In particular, we note that it will often be the case that when a conditional 
generative model is built as a simulator, the initial development will naturally
tend to focus on realism in isolation, but often the end goal will be to allow
interaction with the model.
Our paper shows that is likely important to incorporate interactive realism into
the design process from the start.

\section{Conclusion}

In this paper, we have explored the benefits and challenges of building conditional 
generative models for LOB environments.
In doing so, we have stressed the distinction between realism in isolation
and interactive realism. 
We then explored interactive realism in depth with \lobgan, a specific family of 
generative models.
We analysed \lobgan for its dependence on features, and then we explored the weaknesses of its features
and mechanism.
Using the insights from our analysis and adversarial strategies, we designed new \lobgan models
that are better in terms of both realism and robustness.
We finished by outlining numerous directions for further work on generative models
for LOBs.
All of these directions are at the same time both challenging but also 
potentially extremely fruitful

\section*{Acknowledgments}
Jerome and Savani would like to acknowledge the support of a 2021 J.P.Morgan Chase
AI Research Faculty Research Award. 
Jerome would further like to acknowledge the support of a G-Research Early Career Grant. Finally, we thank Tucker Balch for his valuable support and suggestions.

\section*{Disclaimer} 
This paper was prepared for informational purposes in part by the Artificial Intelligence Research group of JPMorgan Chase \& Co. and its affiliates (``J.P. Morgan''), and is not a product of the Research Department of J.P. Morgan. J.P. Morgan makes no representation and warranty whatsoever and disclaims all liability, for the completeness, accuracy or reliability of the information contained herein. This document is not intended as investment research or investment advice, or a recommendation, offer or solicitation for the purchase or sale of any security, financial instrument, financial product or service, or to be used in any way for evaluating the merits of participating in any transaction, and shall not constitute a solicitation under any jurisdiction or to any person, if such solicitation under such jurisdiction or to such person would be unlawful.

\newpage

\bibliographystyle{named}
\bibliography{papers}

\newpage
\appendix
\section{Appendix}

\subsection{\lobgan features: 5th and 95th percentiles}
\label{sec:featurepercentiles}

\begin{table}[h]
\centering
\begin{tabular}{lrr}
\toprule
Feature  &      5\% &           95\% \\
\midrule
\imbI                    &     0.051 &      0.909 \\
\imbV                     &     0.210 &      0.758 \\
$\tradeimb_{1\rm min}$ &     0.000 &      1.000 \\
$\tradeimb_{5\rm min}$ &     0.000 &      1.000 \\
\volI                           &   200.000 &   1500.000 \\
\volV                           &  2250.000 &   8000.000 \\
\code{Spread}                           &     1.000 &      7.000 \\
$\code{EventPctReturn}_{1}$                    &    -0.0004 &      0.0004 \\
$\code{EventPctReturn}_{50}$                   &    -0.0017 &      0.0017 \\
\bottomrule
\end{tabular}
\caption{5th and 95th percentiles of \lobgan features.}
\end{table}

\subsection{Trade imbalance effect on \lobgan order type}

\begin{table}[h]
\centering
\begin{tabular}{llr}
\toprule
Imbalance$_1$ & OrderType & \%Orders        \\
\midrule
5$^{th}$ Percentile & Cancellation &  0.429407 \\
         & Limit &  0.505465 \\
         & Market &  0.065128 \\ \hline
Baseline & Cancellation &  0.426289 \\
         & Limit &  0.508550 \\
         & Market &  0.065161 \\ \hline
95$^{th}$ Percentile & Cancellation &  0.422418 \\
         & Limit &  0.511946 \\
         & Market &  0.065637 \\ 
\bottomrule
\end{tabular}
\caption{Trade imbalance effect on order type.}
\end{table}

\subsection{RL policy actions}\label{sec:rl_actions}

\setlist[itemize]{leftmargin=*}
\begin{itemize}[noitemsep,nolistsep]
    \item Action 0: place large volume symmetrically at one tick from touch
    \item Action 1: place large volume symmetrically at 2 ticks from touch
    \item Action 2: place large volume symmetrically at 3 ticks from touch
    \item Action 3: place large volume symmetrically at 4 ticks from touch
    \item Action 4: place large volume symmetrically at 5 ticks from touch
    \item Action 5: place large volume 1 tick from bid and 2 ticks from ask
    \item Action 6: place large volume 2 ticks from bid and one tick from ask
    \item Action 7: place large volume 2 tick from bid and 5 ticks from ask
    \item Action 8: place large volume 5 tick from bid and 2 ticks from ask
    \item Action 9: place large order 1 tick away from touch on the bid side, nothing on the ask side
    \item Action 10: place large order 2 ticks away from touch on the bid side, nothing on the ask side
    \item Action 11: place large order 3 ticks away from touch on the bid side, nothing on the ask side
    \item Action 12: place nothing on the bid side, large order 1 tick away from touch on the ask side
    \item Action 13: place nothing on the bid side, large order 2 ticks away from touch on the ask side
    \item Action 14: place nothing on the bid side, large order 3 ticks away from touch on the ask side
\end{itemize}

\subsection{Additional experiments}

\begin{figure}[h]\centering
\includegraphics[width=\fscale\linewidth]{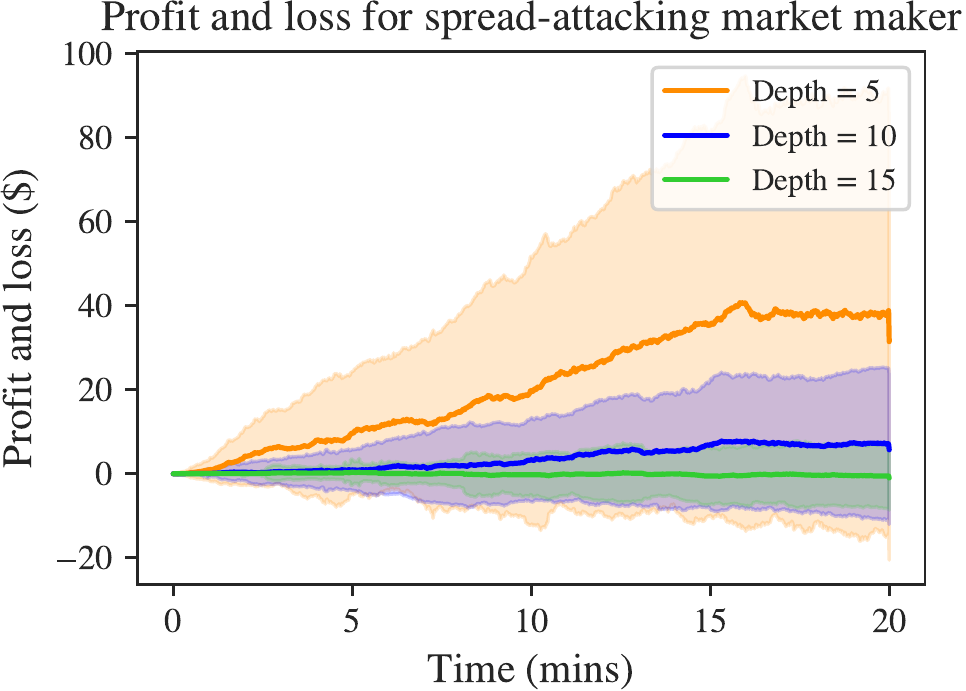}
\caption{A market maker strategy that places a single unit order at the touch to constrain the spread and then a symmetric larger order of volume 300 at a fixed depth away from the touch.}
\label{fig:spread_attack}
\end{figure}

\begin{figure}[h]\centering
\includegraphics[width=\fscale\linewidth]{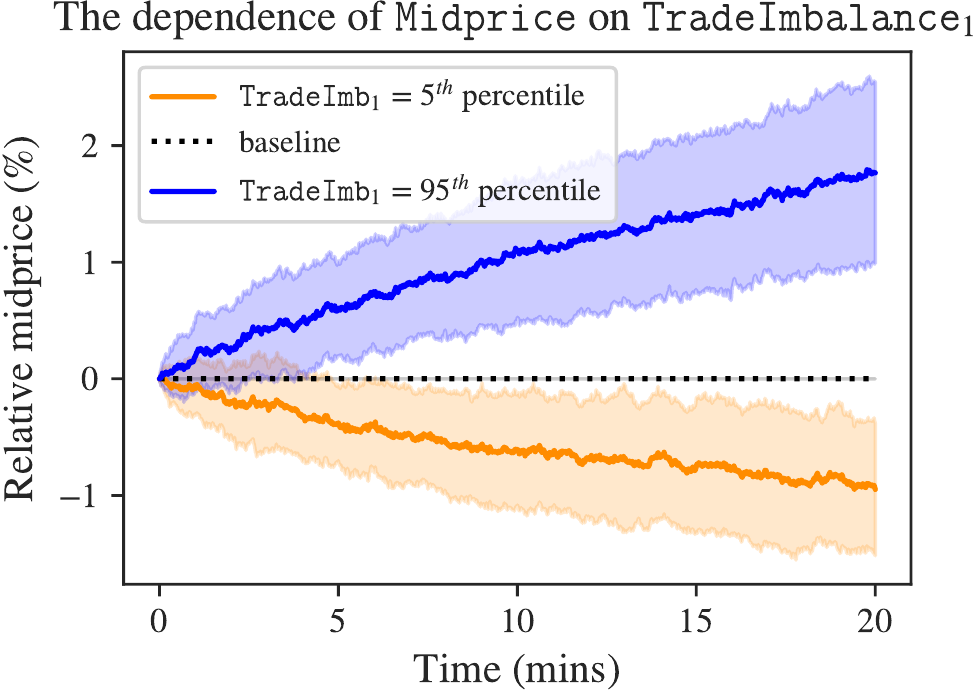}
\caption{The effect of fixing $\tradeimb_1$ on  the midprice dynamics. The midprice change is calculated relative to the ``baseline'' case in which the feature $\tradeimb_1$~is unaltered.}
\label{fig:trade_imbalance_and_midprice}
\end{figure}

\begin{figure}[h]\centering
\includegraphics[width=\fscales\linewidth]{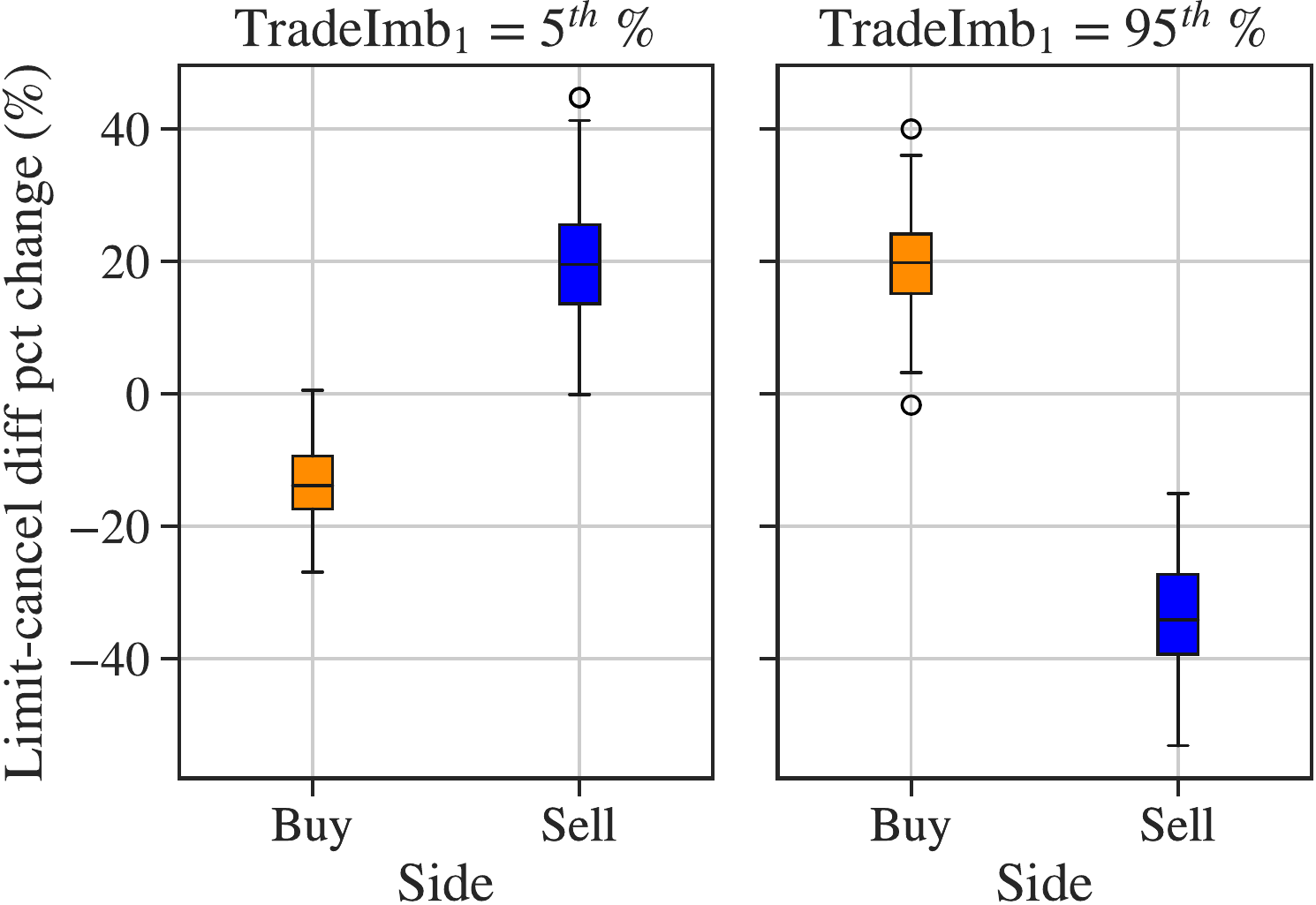}
\caption{The effect of fixing $\tradeimb_1$ on difference between the volume of limit orders and cancellations.}
\label{fig:trade_imbalance_and_limit_cancel}
\end{figure}

\begin{figure}[h]\centering
\includegraphics[width=\fscales\linewidth]{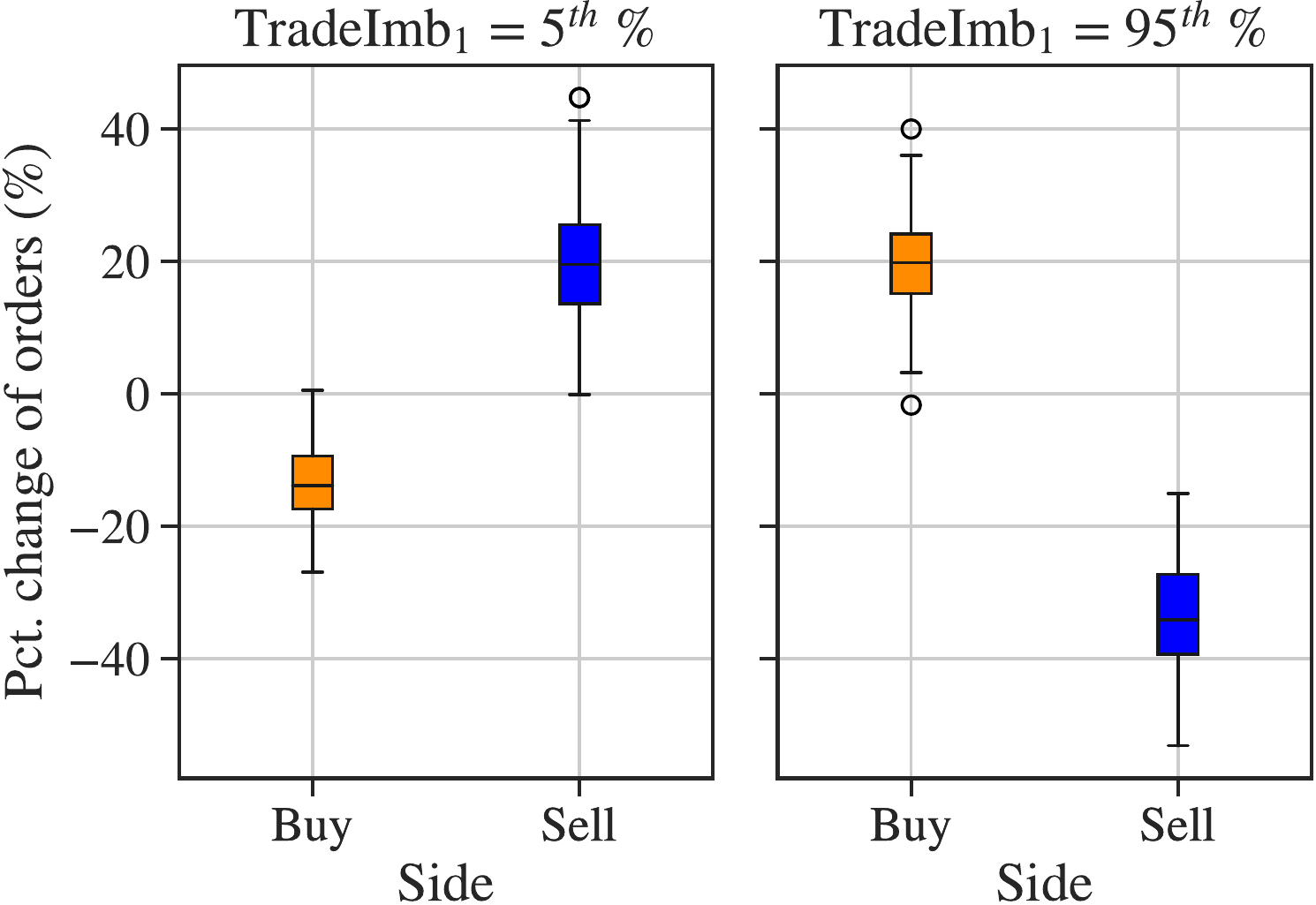}
\caption{The effect of fixing $\tradeimb_1$ on the direction of market orders.}
\label{fig:trade_imbalance_and_markets}
\end{figure}

\end{document}